\documentclass[sigplan,screen]{acmart}

\usepackage{amsmath,amsfonts}
\usepackage{physics}
\usepackage{algorithm}
\usepackage[indLines=false,commentColor=blue]{algpseudocodex} %
\usepackage{graphicx}
\usepackage[]{hyperref}
\usepackage{moresize} %
\usepackage{subcaption} %
\usepackage{multirow}
\usepackage{tabularx} %
\usepackage{booktabs} %
\usepackage{balance}

\algrenewcommand\algorithmicindent{1em}%
\newcommand{\acronym}{RESCQ}

\copyrightyear{2025}
\acmYear{2025}
\setcopyright{cc}
\setcctype{by}
\acmConference[ASPLOS '25]{Proceedings of the 30th ACM International
Conference on Architectural Support for Programming Languages and Operating
Systems, Volume 2}{March 30-April 3, 2025}{Rotterdam, Netherlands}
\acmBooktitle{Proceedings of the 30th ACM International Conference on
Architectural Support for Programming Languages and Operating Systems,
Volume 2 (ASPLOS '25), March 30-April 3, 2025, Rotterdam,
Netherlands}\acmDOI{10.1145/3676641.3716018}
\acmISBN{979-8-4007-1079-7/25/03}

\begin{document}
\title{\acronym{}: \texorpdfstring{\underline{Re}}{Re}altime \texorpdfstring{\underline{S}}{S}cheduling for \texorpdfstring{\underline{C}}{C}ontinuous Angle\\\texorpdfstring{\underline{Q}}{Q}uantum Error Correction Architectures}

\author{Sayam Sethi}
\email{sayams@utexas.edu}
\orcid{0009-0005-3056-5285}
\affiliation{%
    \department{Department of Electrical and Computer Engineering}
    \institution{The University of Texas at Austin}
    \city{Austin}
    \country{USA}
}

\author{Jonathan Mark Baker}
\email{jonathan.baker@austin.utexas.edu}
\orcid{0000-0002-0775-8274}
\affiliation{%
    \department{Department of Electrical and Computer Engineering}
    \institution{The University of Texas at Austin}
    \city{Austin}
    \country{USA}
}

\begin{abstract}

In order to realize large scale quantum error correction (QEC), resource states, such as $\ket{T}$, must be prepared which is expensive in both space and time. In order to circumvent this problem, alternatives have been proposed, such as the production of continuous angle rotation states \cite{akahoshi2023partially, choi2023fault, toshio2024practicalquantumadvantagepartially}.
However, the production of these states is non-deterministic and may require multiple repetitions to succeed. The original proposals suggest architectures which do not account for realtime (or dynamic) management of resources to minimize total execution time. Without a realtime scheduler, a statically generated schedule will be unnecessarily expensive. We propose RESCQ (pronounced rescue), a realtime scheduler for programs compiled onto these continuous angle systems. Our scheme actively minimizes total cycle count by on-demand redistribution of resources based on expected production rates. Depending on the underlying hardware, this can cause excessive classical control overhead. We further address this by dynamically selecting the frequency of our recomputation. RESCQ improves over baseline proposals by an average of $2\times$ in cycle count.

\end{abstract}



\maketitle

\section{Introduction}

\begin{figure*}
    \centering
    \begin{subfigure}[t]{0.32\textwidth}
        \centering
        \resizebox{1.2\linewidth}{!}{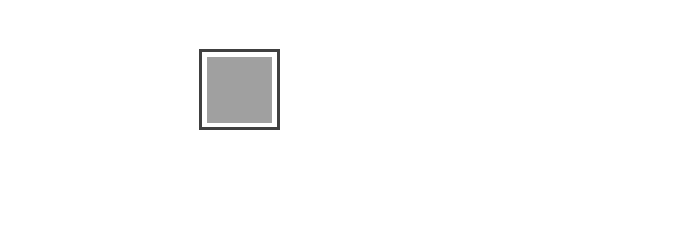}
        \caption{Notation for different surface code tiles}
      \label{fig:intro_a}
    \end{subfigure}
    \hfill
    \begin{subfigure}[t]{0.32\textwidth}
        \centering
        \resizebox{0.5\linewidth}{!}{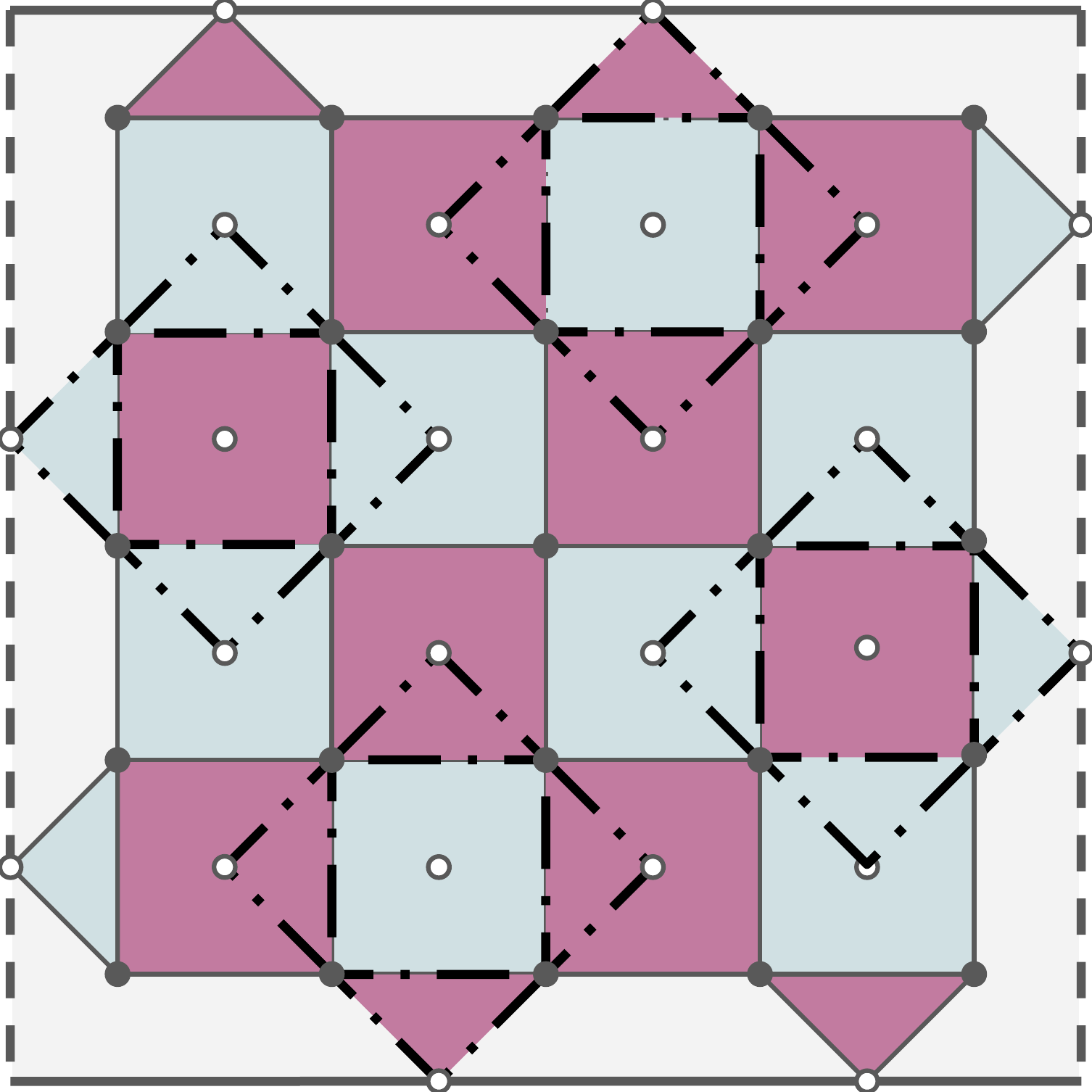}
        \caption{Preparation of $\ket{m_\theta}$ state within an ancilla}
        \label{fig:parallel_prepare}
    \end{subfigure}
    \hfill
    \begin{subfigure}[t]{0.32\textwidth}
        \centering
        \resizebox{0.75\linewidth}{!}{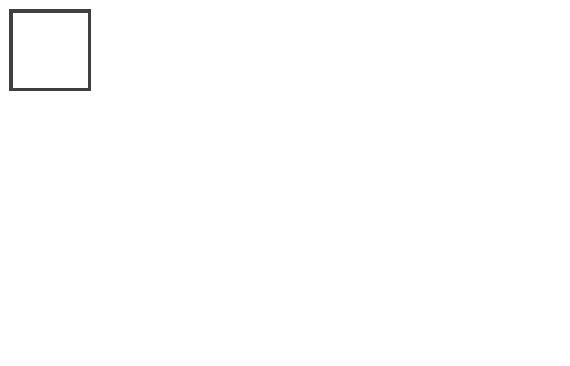}
        \caption{Grid of atomic STAR blocks, defined in \cite{akahoshi2023partially}}
    \end{subfigure}
    \begin{subfigure}{\textwidth}
        \centering
        \resizebox{\linewidth}{!}{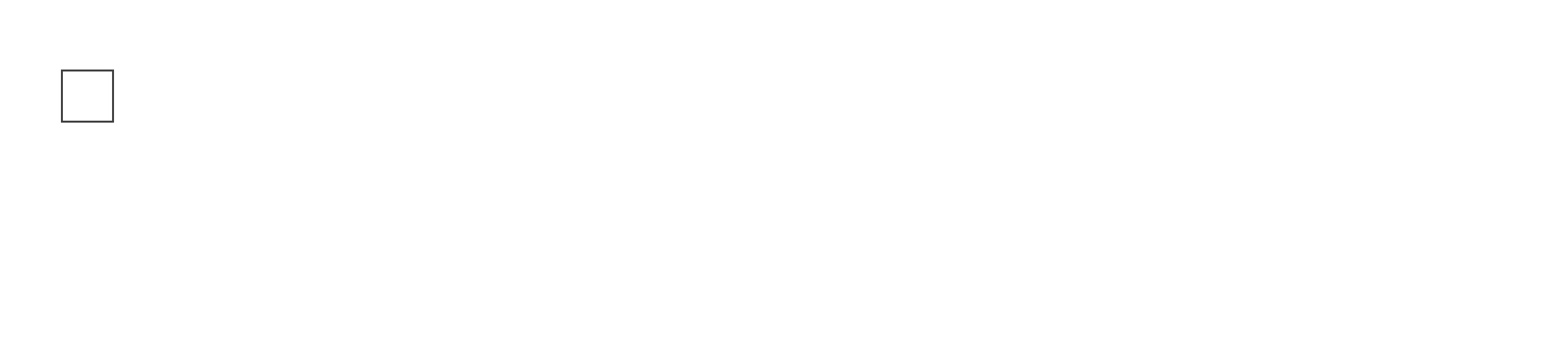}
        \caption{Protocol for the baseline architecture}
      \label{fig:intro_c}
    \end{subfigure}
    \begin{subfigure}{\textwidth}
      \centering
      \resizebox{\linewidth}{!}{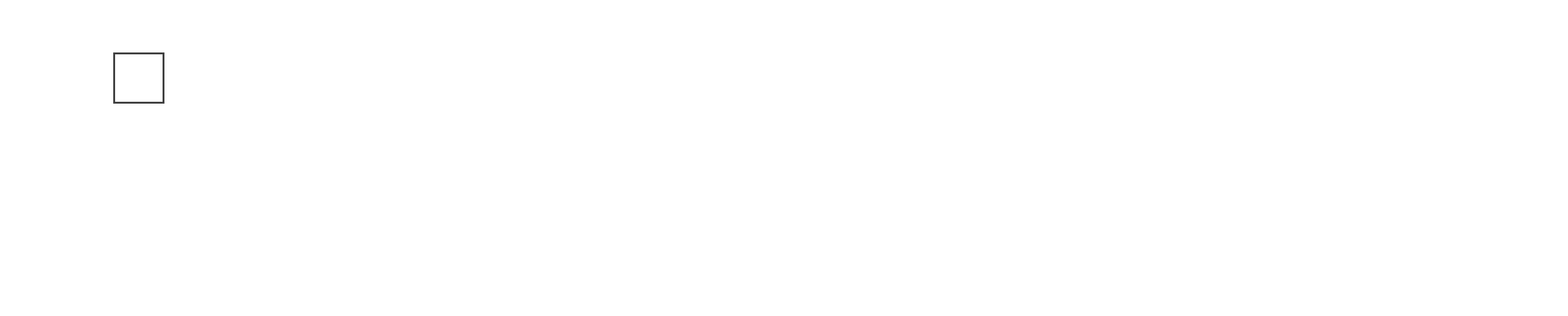}
      \caption{Protocol for the realtime architecture (this work, \acronym)}
      \label{fig:intro_d}
    \end{subfigure}
    \caption{(a, top) Surface Code data qubit with labeled X and Z edges. Gate execution depends on the correct exposure of these edges to ancilla channels. (a, bottom) A prepared $\ket{m_\theta}$ state converted to Surface Code. (b) Each of the four disjoint sub-patches (dotted-and-dashed regions) attempts to prepare a $\ket{m_\theta}$ state in an repeat-until-success (RUS) fashion within a single surface code patch. Whenever any sub-patch succeeds, it expands the state to the entire surface code patch and destroys other sub-patches in the process. If this expansion fails, the entire process is restarted. We discuss this further in Section~\ref{section:relatedwork} and Appendix~\ref{sec:appendix_rus}. (c) Baseline STAR architecture from \cite{akahoshi2023partially} with three ancilla (white) for every data qubit (dark grey) with prepared rotation states labeled (green). Four star patches are outlined in red. (d) The baseline RUS protocol which always attempts preparation of $\ket{m_\theta}$ in the upper right ancilla of the atomic STAR patch. On success, injection can proceed, otherwise, we prepare a $\ket{m_{2\theta}}$ fixup and start from the beginning. (e) In our realtime scheduling scheme we attempt multiple preparations in parallel reducing the number of restarts (to maximise the probability of successful preparation). During injection, we preemptively prepare fixups to prevent restarting the entire process.
    }
    \label{fig:intro}
\end{figure*}
Quantum error correction is necessary for \textit{fault tolerant quantum computation} (FTQC). As we scale, demand for quantum resources will grow rapidly and it is critical to develop realtime management of available resources. These resources include classical bandwidth for decoding \cite{ravi2023better}, ancilla management for communication and decompositions \cite{ding2020square}, and the creation and use of special resource states called \textit{magic states} \cite{ding2018magic, litinski2019magic}. The study of these resources for small to intermediate scale quantum hardware \cite{litinski2019game, gidney2021factor} has focused on the quantum error correcting code known as the surface code due to its limited hardware connectivity, high threshold, and well studied decoding procedures. 

Providing support for the creation and consumption of \textit{magic states} (or resource states) is critical to the success of any quantum error correcting code. Since no code natively supports a transversal and universal gate set \cite{eastin2009restrictions}, there exist gates that need to be executed in \textit{another} code and \textit{distilled} into the original code. For example, the surface code natively supports the Clifford gates (e.g. CNOT, X, Z, and H) which are not universal. This gate set is commonly made universal by producing T states and inject them as necessary. While most physical quantum computers support $Rz(\theta)$ for any $\theta$, the surface code only supports a discrete set. Combining the T gate with Clifford set we get a universal gate set since we can approximate any $Rz(\theta)$ gate to an arbitrary precision \cite{ross2014optimal}. However, this synthesis results in an increase in circuit depth by two orders of magnitude compared to a Clifford + Rz compilation and requires a large numbers of bulky \textit{factories} to produce the magic states; both space and time requirements for any circuit become dominated by T state production. This is significant as, for near-term FTQC, where we have about 1-2 orders of magnitude improvement in the logical error rates, this leads to many orders of magnitude drop in program fidelity.%

Recently, several alternative approaches have been proposed for near-term FTQC \cite{akahoshi2023partially, choi2023fault, toshio2024practicalquantumadvantagepartially}, which instead propose methods to create analog rotation states $\ket{m_\theta} = Rz(\theta)\ket{+}$, which enables arbitrary Rz rotations when injected. This is especially powerful because it reduces the space requirement for producing non-Clifford gates. We can use ancilla qubits local to the injection site to prepare the necessary $\ket{m_\theta}$ in contrast to distilling hundreds or thousands of T gates per $Rz$ rotation \cite{ross2016gridsynth}. There is limited architectural support for this strategy specifically for the realtime management of the nondeterministic behavior of the $\ket{m_\theta}$ production. In \cite{akahoshi2023partially}, the STAR architecture provides a basic structure to demonstrate this technique's efficacy, but 1. limits state production to atomic STAR patches (see Figure \ref{fig:intro}) which limits parallel production despite additional ancilla availability and 2. uses a static schedule which does not account for non-deterministic failures during runtime.%

In this work, we provide an improved scheduling scheme, \acronym{}, for operations on continuous angle rotation architectures in the near-term FTQC regime. In our technique, we consider the ancillas independent of the data qubit, allowing for sharing of resources across multiple qubits and gates. This necessitates efficient resource allocation during runtime. Allocating excessive ancilla for a single gate operation will starve ancillas for neighbouring gate operations. To counter this problem, we propose a mechanism to manage the ancilla allocation in realtime for different gate operations, including but not limited to $Rz$ and CNOT gates, allowing us to flexibly allocate ancilla. We also propose a space-time efficient scheduling technique for long-distance gate operations that minimizes wait time. We achieve an average improvement of $2\times$ in cycle count over a statically compiled and scheduled execution. %
Our approach can be directly incorporated in any quantum architecture involving non-deterministic execution and/or variable ancilla availability.

The primary contributions of this work are
\begin{enumerate}
    \item \acronym: An open source \cite{this_work} efficient realtime scheduling protocol for QEC systems which support native continuous angle resource states. We improve over baseline proposals by an average of $2\times$ in cycle count.
    \item Scheduler which accounts for inherent non-deterministic behaviour of continuous angle resource state production; we introduce the notion of real-time rescheduling depending on the prior success of production and consumption of these states.
    \item An improved architecture for local resource state production which reduces total space (ancilla) requirements while simultaneously reducing the runtime of programs on these systems. Even in the most constrained architectures, \acronym{} results in an average $1.65\times$ improvement in cycle time.
    \item Measurement latencies restrict the frequency of classical recomputation that can occur without incurring stalls on the quantum execution; our scheme easily adapts to any hardware platform and dynamically selects the frequency of realtime updates (recomputation) of classical data structures used: to our knowledge, the first of its kind to do so. This real-time control consideration is extensible to other QEC systems.
\end{enumerate}

\section{Background and Related Work}\label{section:background}

\begin{figure}
    \centering
    \begin{subfigure}[t]{0.24\columnwidth}
        \centering
        \def\svgwidth{0.9\linewidth}
        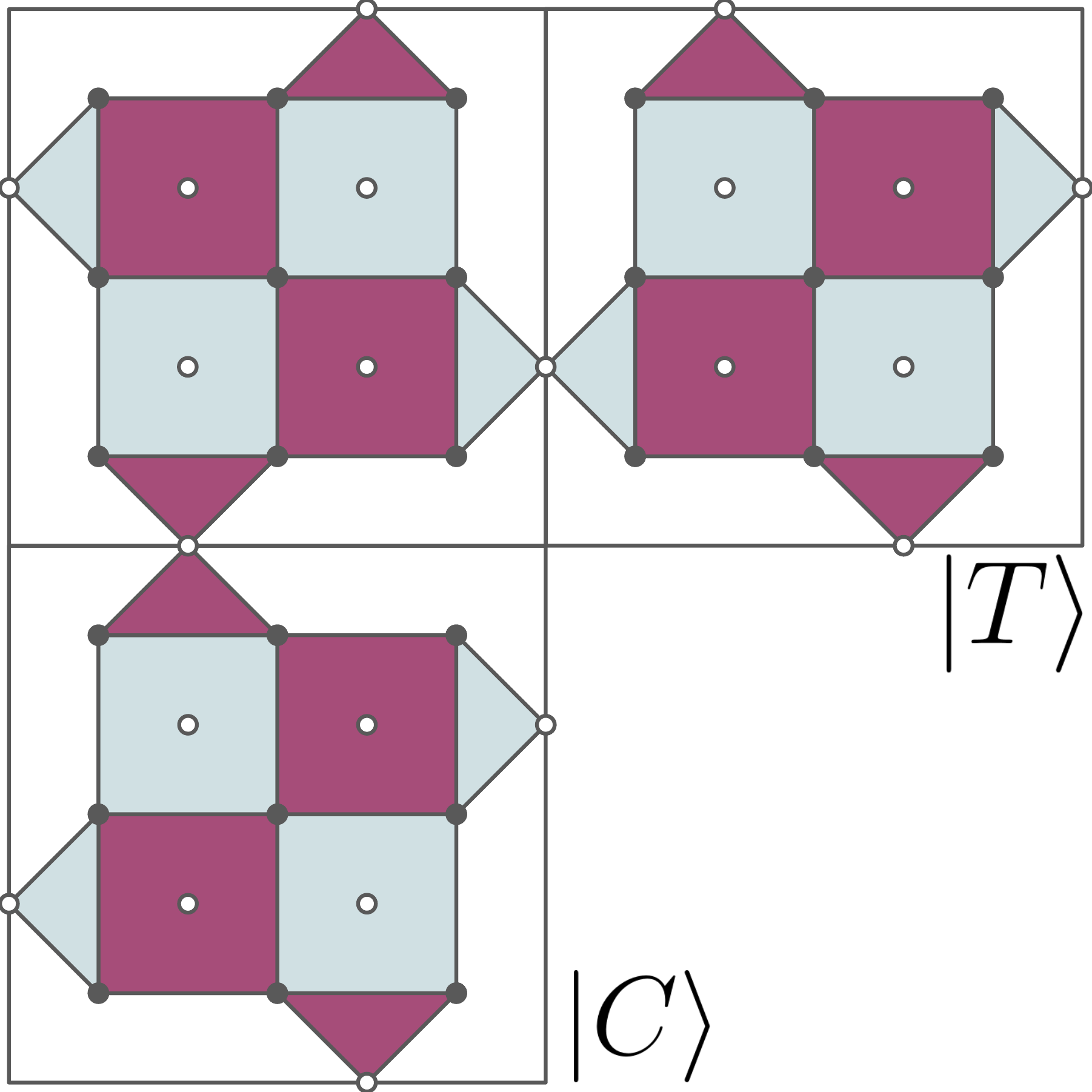
        \caption{Initial state}
    \end{subfigure}
    \hfill
    \begin{subfigure}[t]{0.24\columnwidth}
        \centering
        \def\svgwidth{0.9\linewidth}
        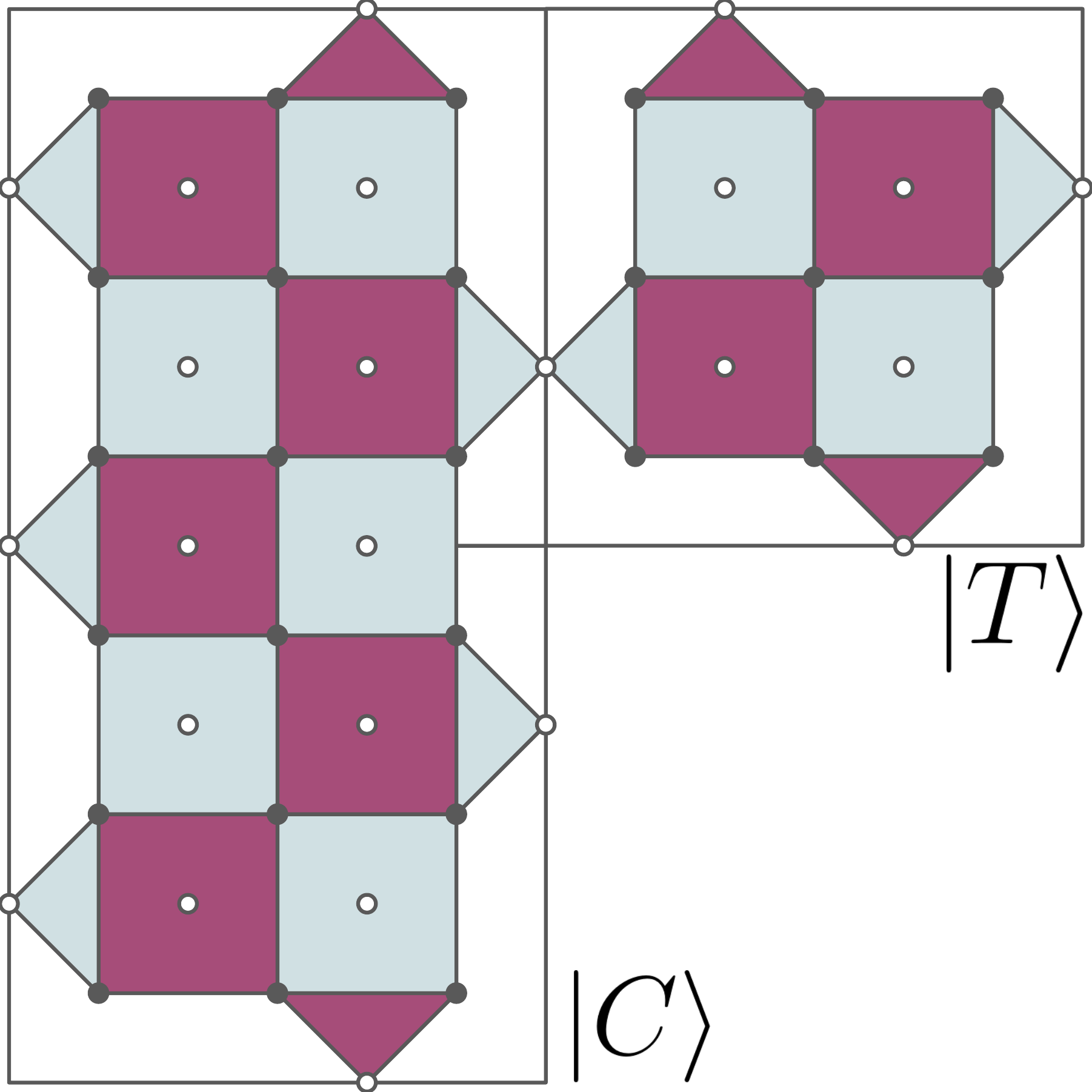
        \caption{Merge\\(1 cycle)}
    \end{subfigure}
    \hfill
    \begin{subfigure}[t]{0.24\columnwidth}
        \centering
        \def\svgwidth{0.9\linewidth}
        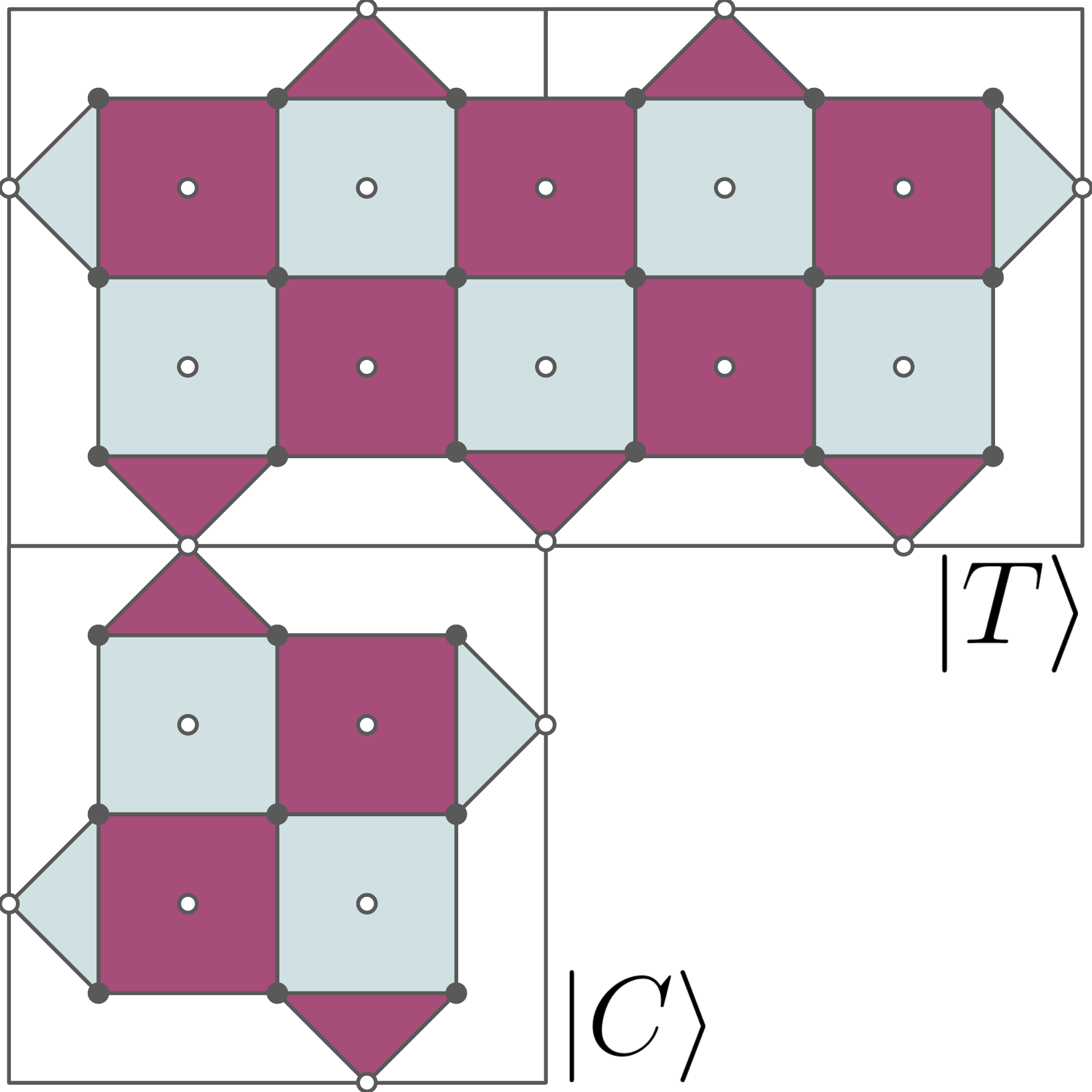
        \caption{Split+Merge\\(1 cycle)}
    \end{subfigure}
    \hfill
    \begin{subfigure}[t]{0.24\columnwidth}
        \centering
        \def\svgwidth{0.9\linewidth}
        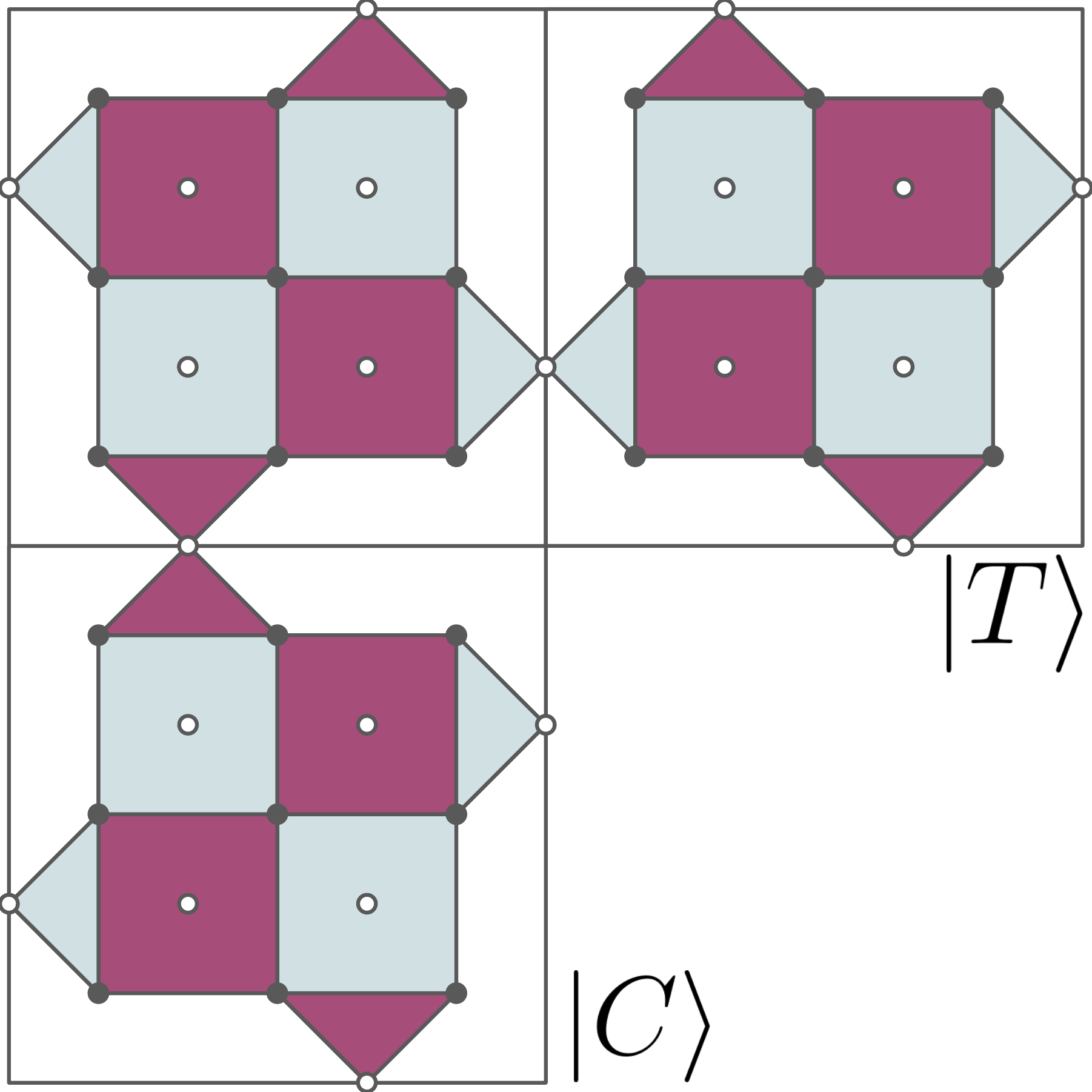
        \caption{Split\\(0 cycles)}
    \end{subfigure}
    \caption{Execution of a CNOT gate taking 2 cycles on surface code qubits (of $d=3$) indicated by the squares. Here the CNOT is performed between the control ($\ket{C}$) and the target ($\ket{T}$) with a single ancilla qubit in between. The horizontal edges are the $Z$ edges (all purple) and the vertical edges are the $X$ edges (all blue). This patch can be arbitrarily long and shaped as long as the correct boundaries are adjacent. We assume ancilla preparation is done ahead of time.}
    \label{fig:lattice_cnot}
\end{figure}

Scalable quantum computation requires error correction in order to achieve sufficiently low \textit{logical error rates} (LER) for successful program execution. We focus on Surface Codes: a code with many attractive properties for available or soon-to-be available hardware platforms. Surface codes require only nearest neighbour connectivity between physical qubits, its parity checks require only 4 two qubit gates, it has a high threshold and it has well-studied decoders \cite{Fowler_2012, higgott2022pymatching}. Surface code architectures are also fairly well studied with the most popular being the rotated surface code with lattice surgery operations \cite{litinski2019game, horsman2012surface}. The rotated surface code is a $d \times d$ patch of qubits with $O(d^2)$ data qubits and ancilla qubits, where $d$ is the distance of the code. The patch is composed of $X$ and $Z$ checks which collect syndromes about $X$ and $Z$ errors, respectively. The boundaries of the square patch are either $X$ or $Z$ edges which determine how the logical qubit can interact with other logical qubits. Figure~\ref{fig:lattice_cnot} shows three surface code tiles for $d=3$.

Surface code architectures are a fabric of $d \times d$ sized tiles onto which program qubits are mapped. Tiles can be deformed as long as the distance of the shortest path between any pair of corresponding edges (e.g. $X$ edge to another $X$ edge) is at least $d$. Deformation, is one of the many available operations in a surface code architecture and can be used to interact logical qubits at a distance. To do so, the grid of tiles must allocate some number of logical \textit{ancilla}.  To interact two logical qubits, we employ lattice surgery using intermediate ancilla. We refer readers to \cite{litinski2019game} for a complete coverage of these operations.

We summarize the necessary information for this work:
\begin{itemize}
    \item Program qubits are assigned to a single $d \times d$ tile in the larger fabric, e.g. using AutoBraid \cite{hua2021autobraid}.
    \item The logical operation CNOT (Figure~\ref{fig:lattice_cnot}) occurs in \textbf{two} steps: a) measure the $ZZ$ operator between the control and the ancilla followed by b) measure the $XX$ operator between the ancilla and the target.
    \item Interactions between logical qubits use a contiguous path of ancilla qubits which must touch every interacting qubit.
    \item A special type of multi-qubit interaction, or \textit{Pauli-product measurement}\cite{litinski2019game}, can be executed by interacting the $P$-edge of each involved logical qubit with an intermediate contiguous ancilla channel, where $P$ is the specific Pauli. %
    This can be done in \textbf{one} step regardless of distance between the qubits, so long as the ancilla channel is contiguous.
\end{itemize}

\subsection{Resource State Distillation}
To make universal gate sets for QEC codes, additional resource states (for example $\ket{T}$ states are used to perform logical T gates) are prepared in a different code which admits a transversal implementation. Preparation of these states can fail with failure probabilities ranging between $0.1\%$ to $11\%$ for some commonly used protocols\cite{litinski2019game}. This preparation has a fixed logical error rate (LER) which differs from the LER of the base code. More error prone states are \textit{distilled} to generate higher fidelity instances of these states. Resource states are prepared remotely in \textit{distillation factories} and teleported on demand. The total space-time cost of factory distillation can take upwards of 90\% of the total space-time volume of the program's execution \cite{holmes2020nisq+}.

\subsection{Continuous Angle Rotation Architectures}\label{section:relatedwork}
\begin{figure}
    \centering
    \def\svgwidth{0.9\linewidth}
    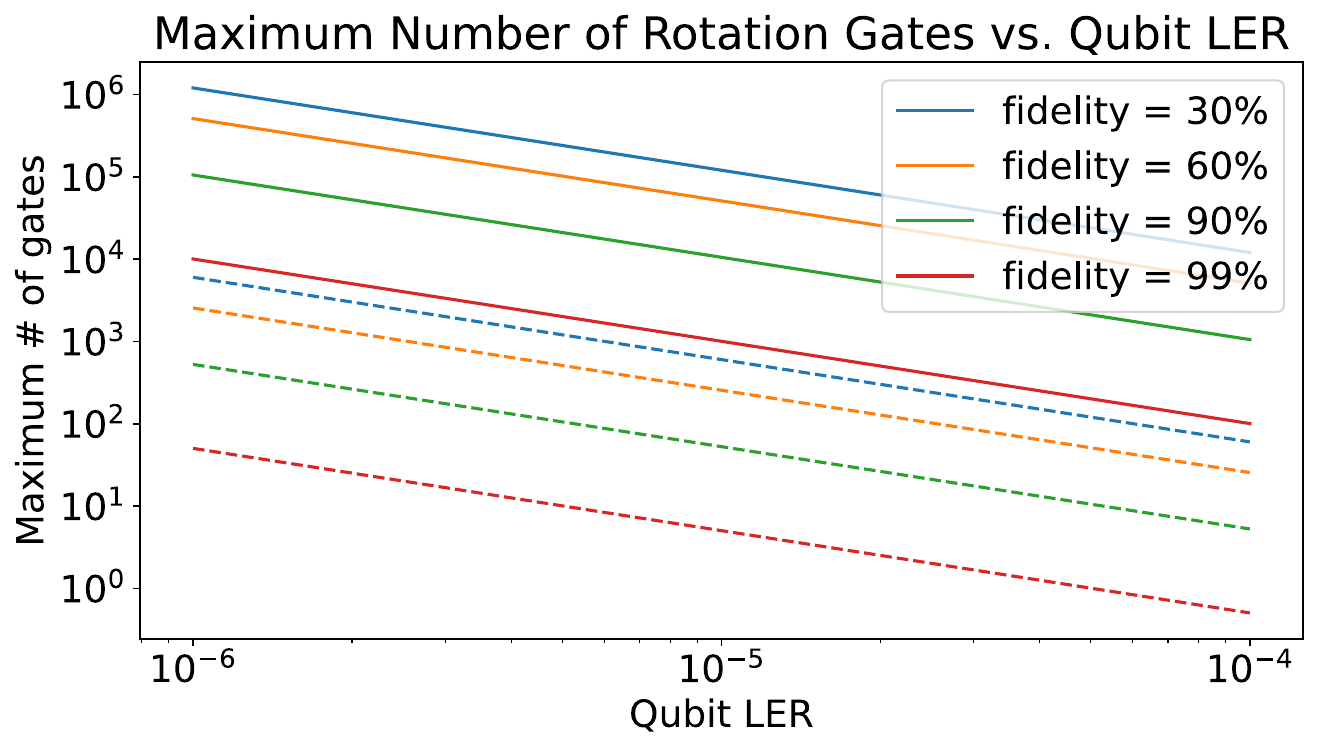
    \caption{A qualitative plot approximating
    the maximum number of rotation gates that can be executed for target program fidelities in the Clifford+Rz (solid lines) vs the Clifford+T (dashed lines) compilation.}
    \label{fig:fidvsler}
\end{figure}
Recent works \cite{akahoshi2023partially, choi2023fault, toshio2024practicalquantumadvantagepartially} have proposed FTQC architectures to prepare arbitrary small angle rotation gates with low overhead and low LER. The typical architecture for QEC systems is to have the primary \textit{computational} code which maintains all data qubits and external factory region(s) which produce special resource states of a single variety, usually $\ket{T}$ for surface codes. $T$ count and depth dominates resource costs in both space and time \cite{litinski2019magic}. Some alternative proposals include code switching \cite{anderson2014fault}, higher dimensional codes \cite{kubica2018abcs} and most recently ``continuous-angle'' synthesis procedures \cite{akahoshi2023partially, choi2023fault, toshio2024practicalquantumadvantagepartially} which use repeat-until-success (RUS) procedures for the production of \textit{arbitrary} magic states $\ket{m_\theta}$. These can be injected to perform $Rz(\theta)$ (Figure~\ref{fig:intro_d}) directly without expensive decompositions or distillation, but with varying success. The space-time cost of these procedures is appealing for near and intermediate term demonstrations of QEC by significantly reducing the physical qubit overhead required for distillation. Unlike Clifford+T, compilation in the Clifford+Rz does not suffer from an increase in circuit depth, increasing the maximum number of gates we can execute for a target program fidelity (Figure~\ref{fig:fidvsler}). These works propose simple architectures which support RUS strategies, but are limited in scale and lack a full scheduler to optimize for overheads caused by non-deterministic behaviour of their protocols, which has gone largely ignored even in other literature. Our work \acronym{} and other baselines do not modify the non-deterministic behaviour of these protocols but address the overheads incurred by modifying the schedules of program operations.

We focus on the technique and architecture proposed in \cite{akahoshi2023partially} which uses a [[4, 1, 1, 2]] error \textit{detection} code to produce $\ket{m_\theta}$ which can be embedded into the larger surface code architecture multiple times (Figure \ref{fig:parallel_prepare}). This can be abstracted out as a non-deterministic preparation with appropriate probability of success (Figure~\ref{fig:intro}).  We discuss the inner workings the preparation scheme in Appendix~\ref{sec:appendix_rus} and also compare it with the Clifford+T compilation scheme.
\cite{akahoshi2023partially} gives three examples of simple architectures which localize the production of these states: 1. STAR block, a $2\times 2$ grid of surface code tiles 1 of which is data, and 1 of which produces the resource state, and 2 ancilla used for communication, 2. Compact STAR block, a $3\times 1$ grid with 1 data and 2 ancilla and 3. Compressed STAR block, a $2\times 1$ grid with 1 data and 1 ancilla. This atomic abstraction leads to wasted ancilla resources when qubits are idling and \acronym{} addresses this by better management of ancilla resources (Section~\ref{section:ancillamanage}).

\subsection{Prior Compilation Work}
There are two primary categories for compilation for quantum systems: 1) \textit{physical} compilation focused on the implementation of physical gates and 2) \textit{logical} compilation focused on managing \textit{logical} qubits. The first is typically defined by the constraints of the underlying hardware. Several works have focused on tailoring efficient execution of physical syndrome extraction circuits to the hardware properties e.g. superconducting qubits \cite{wu2022synthesis}, trapped ions \cite{leblond2023tiscc} and neutral atoms \cite{viszlai2023architecture, jandura2024surface}. These are often distinct from generalized compilers \cite{tannu2019not, li2019tackling, murali2019noise} because of the simple and repetitive structure of the underlying circuits.\par
We focus on the second type of compilation. As QEC codes, and surface codes in particular, become more realistically implementable, several general purpose compilers \cite{comp1, comp2, comp3, litinski2019game, hua2021autobraid, javadi-abhari2017optimized} have emerged, focusing on problems related to path finding and logical qubit routing. However, these suffer from the pitfalls of schedules generated from static compilation. These are also distinct from the synthesis problem designed to convert circuits into versions using only gates from the limited universal gate set, such as in \cite{ross2016gridsynth}. While prior works focus on Clifford+T synthesis and reducing T requirements\cite{camp1, camp2} we prioritize continuous-angle rotation synthesis.

\section{Realtime Scheduling and Compilation for Continuous Angle Rotation Architectures}
We discuss the compilation of high level programs and scheduling gate operations onto surface code architectures which support Clifford+Rz gates as opposed to the traditional Clifford+T. We assume all programs have already been synthesized into the appropriate gate set. 
In contrast to typical distillation factory architectures, in continuous angle rotation architectures, $\ket{m_{\theta}}$ states are prepared and expanded locally into logical surface code qubits. It takes at most a single logical patch for preparation at the cost of additional uncertainty in preparation time. In prior work \cite{akahoshi2023partially}, atomic units of data qubits and ancilla for the preparation of single qubit gates are used. We propose a much more versatile architecture which allows ancilla to be reused and allocated for various rotations in realtime.

\subsection{Execution of CNOT Gate}\label{section:cnotexecution}
\begin{figure}
    \centering
    \begin{subfigure}[t]{\columnwidth}
        \centering
        \resizebox{0.35\linewidth}{!}{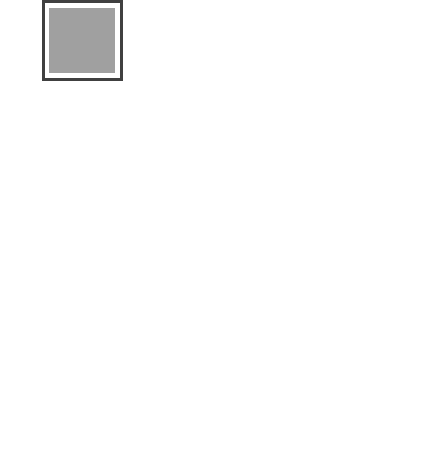}
        \caption{Path exists}
    \end{subfigure}
    \hfill
    \begin{subfigure}[t]{\columnwidth}
        \centering
        \resizebox{0.95\linewidth}{!}{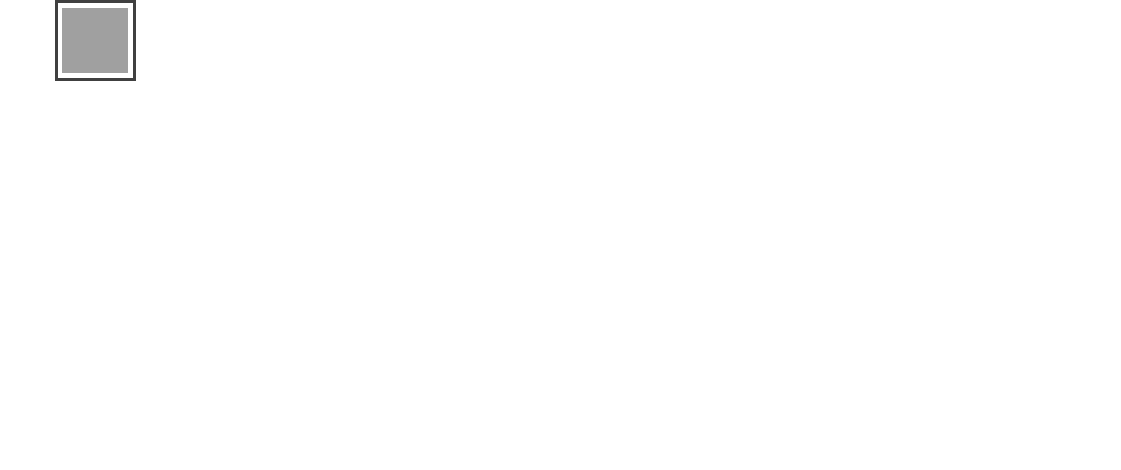}
        \caption{No path exists}
        \label{fig:cnot_execution_b}
    \end{subfigure}%
    \caption{Two different possible scenarios for CNOT execution (from $Z$ edge of $\ket{C}$ $\to$ $X$ edge of $\ket{T}$) in the lattice surgery grid. The qubits colored in red and marked with a cross ($\times$) are busy and/or unavailable for routing the CNOT gate. Recall from Figure~\ref{fig:intro_a} that the solid edges are the $Z$ edges and the dotted edges are the $X$ edges. CNOT execution in (a) requires only $2$ lattice surgery cycles in contrast with (b) which requires a total of $5$ cycles.} %
    \label{fig:cnot_execution}
\end{figure}

We can perform logical CNOT gates on data qubits that are arbitrarily far apart using lattice surgery. %
Lattice surgery requires a contiguous path of (non-empty) ancilla tiles between the Z edge of the control qubit and the X edge of the target qubit. Such a path may not exist, either due to the ancilla being occupied for another gate operation, or the required edges not having any ancilla neighbours. In either case, we need to insert an edge-rotation gate to expose the required edge of the qubit onto the chosen path. This gate operation requires a free neighbouring ancilla tile and takes $3$ lattice surgery cycles (Figure~\ref{fig:cnot_execution}).\par
The routing algorithm and path selection are crucial to an optimal scheduler. Multiple path finding techniques have been proposed in literature, such as shortest path selection \cite{javadi-abhari2017optimized} and AutoBraid \cite{hua2021autobraid}, however, they don't account for non-deterministic ancilla activity and edge-rotation gates. As a result, these techniques do not permit the next set of gates to begin execution early if any gate in the current layer takes longer to execute. More specifically, execution of the next layer is stalled until the gate with the highest execution time of the current layer is completed. For instance, if two CNOTs are scheduled to execute simultaneously with one CNOT taking $2$ cycles and the other taking $5$ cycles (due to requiring an edge rotation), the next set of gates will only be scheduled after the CNOT gate taking $5$ cycles is finished. However, some gates in subsequent layers can start after the shorter CNOT completes. This problem is amplified when we have non-deterministic $Rz$ gates, which we discuss in more detail in Section~\ref{section:rzexecution}.\par
\begin{figure}
    \centering
    \begin{subfigure}{\columnwidth}
        \centering
        \resizebox{\linewidth}{!}{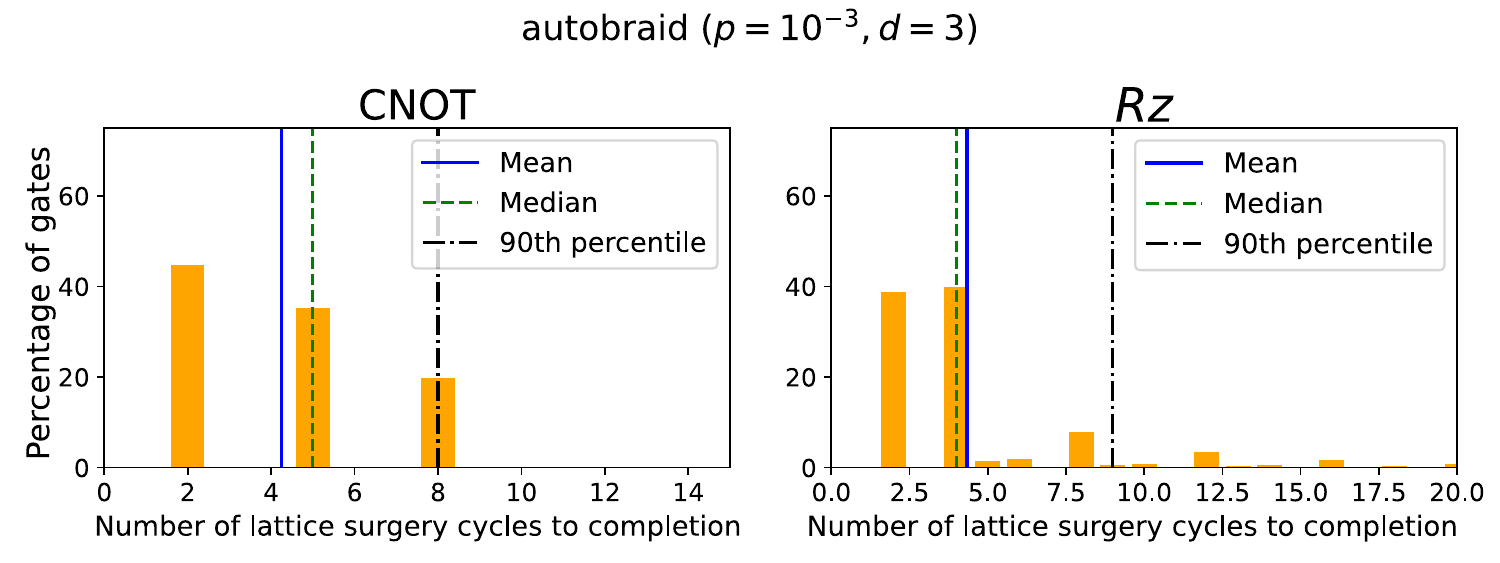}
    \end{subfigure}
    \vspace{1mm}
    \begin{subfigure}{\columnwidth}
        \centering
        \resizebox{\linewidth}{!}{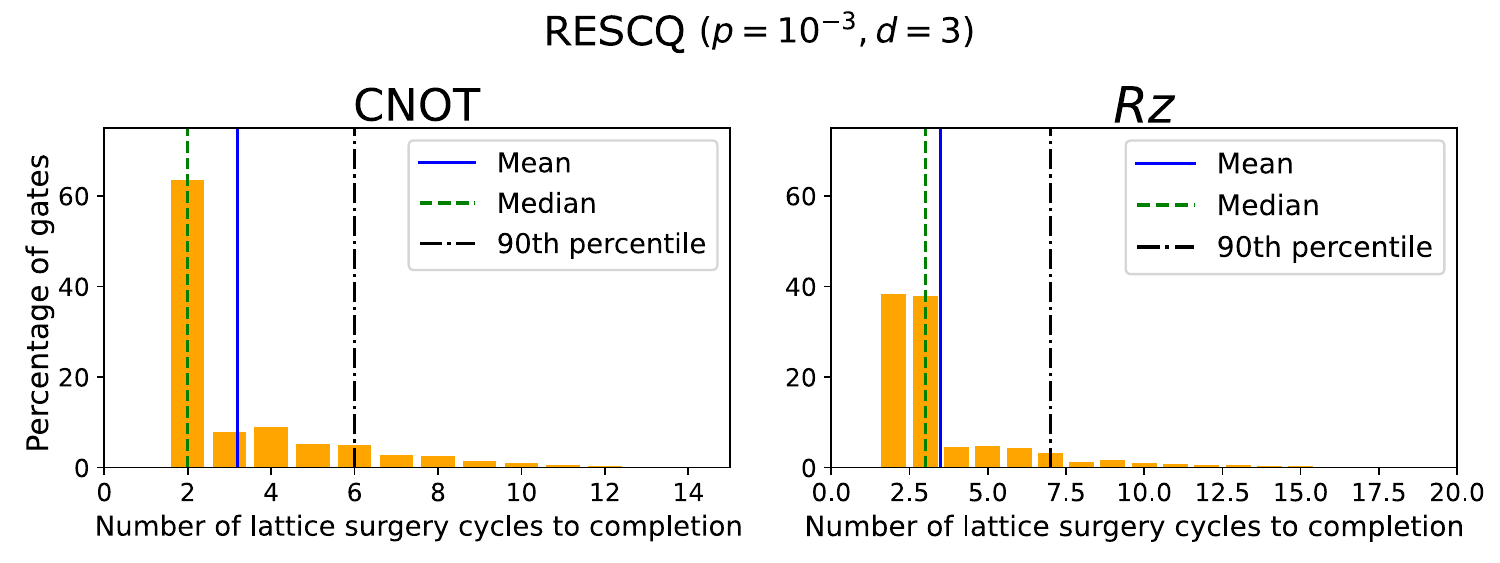}
    \end{subfigure}
    \caption{Histograms showing the time taken for CNOT and $Rz$ gates (including all correction gates) to complete \textit{after they are scheduled} (accumulated over all benchmarks).} %
    \label{fig:histograms}
\end{figure}

As evident from Figure~\ref{fig:histograms}, a large percentage of CNOTs in AutoBraid\cite{hua2021autobraid} take $5$ and $8$ cycles to execute. $8$ cycles are required when a single ancilla is initialized between incorrect edges, thus requiring $3+3+2$ cycles: edge-rotation of the control followed by the target followed by the CNOT. Since there is only one ancilla available, edge-rotation on control and target will be sequential. In contrast more than $50\%$ of the CNOT gates in \acronym{} take $2$ cycles and more than $90\%$ of CNOTs take $6$ or less cycles. Here, we only consider the time taken after the gate is scheduled. 

Because static schedulers such as AutoBraid only schedule the next set of gates once all gates in the current layer are scheduled, this incurs unecessary idle time on the program qubits. Static schedulers suffer from this drawback because we cannot know the state of non-deterministic procedures ahead of time. In \acronym{}, we attempt to schedule the next gate operation immediately when the previous gate on the data qubit completes. Thus, even though CNOT gates may have been scheduled, they cannot start immediately since the ancillas required for the CNOT might be busy executing a different gate, as seen from the continuous distribution of the cycles taken by \acronym{}. We stall execution of a multi-qubit gate if and only if one or more of the involved qubits is currently executing another gate (i.e. is busy).%

\subsection{Execution of \texorpdfstring{$Rz(\theta)$}{Rz(theta)} Rotation Gate}\label{section:rzexecution}
\begin{figure}
    \centering
    \begin{subfigure}[b]{0.52\columnwidth}
        \centering
        \resizebox{\linewidth}{!}{\includegraphics{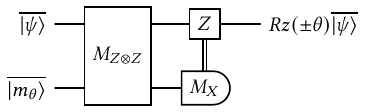}}
        \caption{$Z\otimes Z$ Pauli measurement}
        \label{fig:pauliprod}
    \end{subfigure}
    \begin{subfigure}[b]{0.47\columnwidth}
        \centering
        \resizebox{\linewidth}{!}{\includegraphics{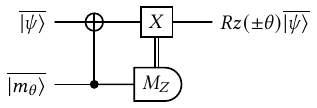}}
        \caption{Rotation gate injection}
        \label{fig:rzrotation}
    \end{subfigure}
    \caption{Circuits for different injection strategies.}
    \label{fig:injstrats}
\end{figure}

\begin{table}
    \centering
    \begin{tabular}{lcc}
        \toprule
        \textbf{Parameter} & \textbf{CNOT} & $\mathbf{ZZ}$ \\
        \midrule
        Exposed edge & $X$ & $Z$ \\
        Number of ancillas required & 2 & 1 \\
        Lattice surgery cycles needed for injection & 2 & 1 \\
        \bottomrule
    \end{tabular}
    \caption{Difference between the two injection strategies. The injection in Figure~\ref{fig:intro_c} is an example of a CNOT injection and in Figure~\ref{fig:intro_d} is an example of a ZZ injection.}
    \label{table:LvsZZ}
\end{table}
The $Rz(\theta)$ rotation gate is executed in two steps, 1. preparation of an ancilla qubit in the state $\ket{m_\theta}$ (Figure~\ref{fig:parallel_prepare}), and 2. injection of the $\ket{m_\theta}$ state into the data qubit and measurement. As discussed in \cite{akahoshi2023partially}, there are two potential injection strategies, shown in Figure~\ref{fig:pauliprod} and Figure~\ref{fig:rzrotation} which we refer to as the ZZ injection and CNOT injection strategies, respectively. The differences between the two strategies is shown in Table~\ref{table:LvsZZ}. Both these injection strategies involve a measurement which outputs +1 and -1 with \textbf{equal} probability. We refer to an output of -1 as a failure.\par
If the measurement output signifies a failure (with fixed probability $1/2$), a correction $Rz(2\theta)$ is required. If this correction fails, another correction gate $Rz(4\theta)$ is required and so on, as Repeat-Until-Success (RUS). Failed injection means an $Rz(-\theta)$ gate was executed so executing an $Rz(2\theta)$ correction gate would yield the proper rotation. However, since $Rz(2\theta)$ is likely a non-Clifford, we must repeat. In general, if an $Rz(2^k\theta)$ injection fails we require $Rz(2^{k+1}\theta)$ correction. Every injection fails with probability $1/2$, hence
\begin{equation}
\mathbb{E}[\text{\#Injections}] = \sum_{k=1}^{\infty} k \cdot \text{Pr[k-1 fail, 1 success]} = \sum_{k=1}^{\infty} \frac{k}{2^k} = 2
\label{eq:exp_inj_attempts}
\end{equation}

If $Rz(2^k\theta)$ is a Clifford for some $k$ (for example consider T or $\sqrt{T}$), this expectation will be $< 2$ since it will no longer require an injection step. %
For either of the two injection strategies (CNOT or ZZ), the scheduler must allocate a contiguous block of ancilla. Our scheduler reserves ancilla and generates a schedule for each ancilla in realtime, determining when it's free. We simply require that some path between $\ket{m_\theta}$ and target all be free at some time.

The preparation of the rotation states $\ket{m_\theta}$ is itself non-deterministic and is prepared in an ancilla patch. Consequently, assigned ancilla are claimed for an indeterminate number of cycles until the state is prepared correctly. Different states $\ket{m_\alpha}$ and $\ket{m_\beta}$ are prepared independently in disjoint ancillas. Multiple ancilla can be assigned for the preparation of any individual state and any additional successful preparations can be discarded if necessary. The number of ancilla dedicated to the production of a particular state $\ket{m_\theta}$ can dynamically change. For example, if $n$ ancilla are assigned in cycle 1 and each fails, wherein some $m$ of these ancilla are needed for other operations, we can reclaim them and try to prepare the state using $n - m > 0$ ancilla in the next cycle. Non-deterministic preparation implies that the exact cycle in which consumption can occur at is unknown ahead of time, motivating eager preparation. Our realtime scheduler \acronym{} decides 1. which ancilla are good candidates for preparation (and injection), and 2. when to start preparation, since beginning preparation too early would prevent the ancilla from being used for other gates while waiting for the data qubit. Conversely, starting preparation too late would stall the gate execution. Making informed and intelligent decisions is clearly beneficial as seen in Figure~\ref{fig:histograms}. The baseline scheme only attempts to prepare a single $\ket{m_\theta}$ state and does not perform eager preparation of the correction state. \acronym{} attempts to prepare multiple $\ket{m_\theta}$ states in parallel and begins eager preparation of the $\ket{m_{2\theta}}$ correction state during the injection of the $\ket{m_\theta}$ state.

\section{\acronym{}: Realtime Scheduling Framework}

Our proposed solution relies on two dynamically constructed and modified data structures throughout execution: 1. a dynamically \underline{re}computed minimum \underline{s}panning tree (MST) weighted by historical use for selecting routing paths based on expected availability, \underline{c}ombined with 2. a \underline{q}ueue for every ancilla qubit in the system; which again abbreviates to \acronym{}.%

For each ancilla, we construct a queue to track not only which operation the ancilla is participating in but what its action or role is in that operation. For example, an ancilla may be used for a routing path or a rotation state preparation. Operations may request many of the same ancilla and compete for shared resources; orderly allocation of resources is important to prevent wasted resources or race conditions. 
We maintain a queue for each ancilla qubit wherein each enqueued element contains metadata about the ancilla for an associated operation (Section~\ref{section:ancillamanage}).\par
Another key component of \acronym{} is the routing protocol. Choosing an optimal path for each CNOT that minimizes the program execution time is computationally intensive, even if done statically and all gates take deterministic execution times \cite{hua2021autobraid}. We maintain an MST of ancilla throughout computation and update its weights in realtime. To determine the best path for each CNOT in realtime, we query the MST and compute the set of ancilla which is most likely free earliest (Section~\ref{section:efficientmst}).

\subsection{Managing Ancilla Operations}\label{section:ancillamanage}
\begin{table}[!hbtp]
    \setlength\tabcolsep{0pt}
    \centering
    \begin{tabular*}{\columnwidth}{@{\extracolsep{\fill}}p{14mm}p{24mm}p{48mm}}
        \toprule
        \textbf{Variable} &\textbf{Purpose} & \textbf{Possible Values} \\
        \midrule
        \multirow{4}{*}{$gate$} & \multirow{4}{0.9\linewidth}{the gate that the ancilla qubit will help to execute} & $\mathsf{C}_{p,q}$: CNOT gate from qubit $p$ (control) onto qubit $q$ (target)\\
        \cline{3-3}
        &&$\mathsf{R_\theta}_p$: $Rz(\theta)$ rotation gate on qubit $p$\\
        \cline{3-3}
        &&$\mathsf{H}_p$: Hadamard gate on qubit $p$\\
        \cline{3-3}
        &&$\mathsf{E}_p$: edge rotation gate on qubit $p$\\
        \hline
        \multirow{2}{*}{$helper$} & \multirow{2}{0.9\linewidth}{other ancilla qubits required} & $\phi$: none\\
        \cline{3-3}
        &&$i$: ancilla qubit $i\in\mathbb{A}$ (required when injecting $\ket{m_\theta}$ via a CNOT)\\
        \hline
        \multirow{4}{\linewidth}{$status$} & \multirow{4}{0.9\linewidth}{the current status of the ancilla} & $\mathcal{R}$: ready to execute the next gate\\
        \cline{3-3}
        && $\mathcal{E}$: executing the top of queue\\
        \cline{3-3}
        && $\mathcal{P}$: preparing $\ket{m_\theta}$ state for the $Rz(\theta)$ gate at top of queue\\
        \cline{3-3}
        && $\mathcal{D}$: done preparing $\ket{m_\theta}$ state and ready to execute $Rz(\theta)$ gate at top of queue\\
        \cline{3-3}
        && $\mathcal{F}$: finished executing the gate at top of the queue\\
        \bottomrule
    \end{tabular*}
    \caption{Variables stored in each node of the queue that contains information about the associated gate. The $status$ is associated only with the top element of the queue.}
    \label{table:queuevars}
\end{table}

\begin{figure*}
    \centering
    \resizebox{\linewidth}{!}{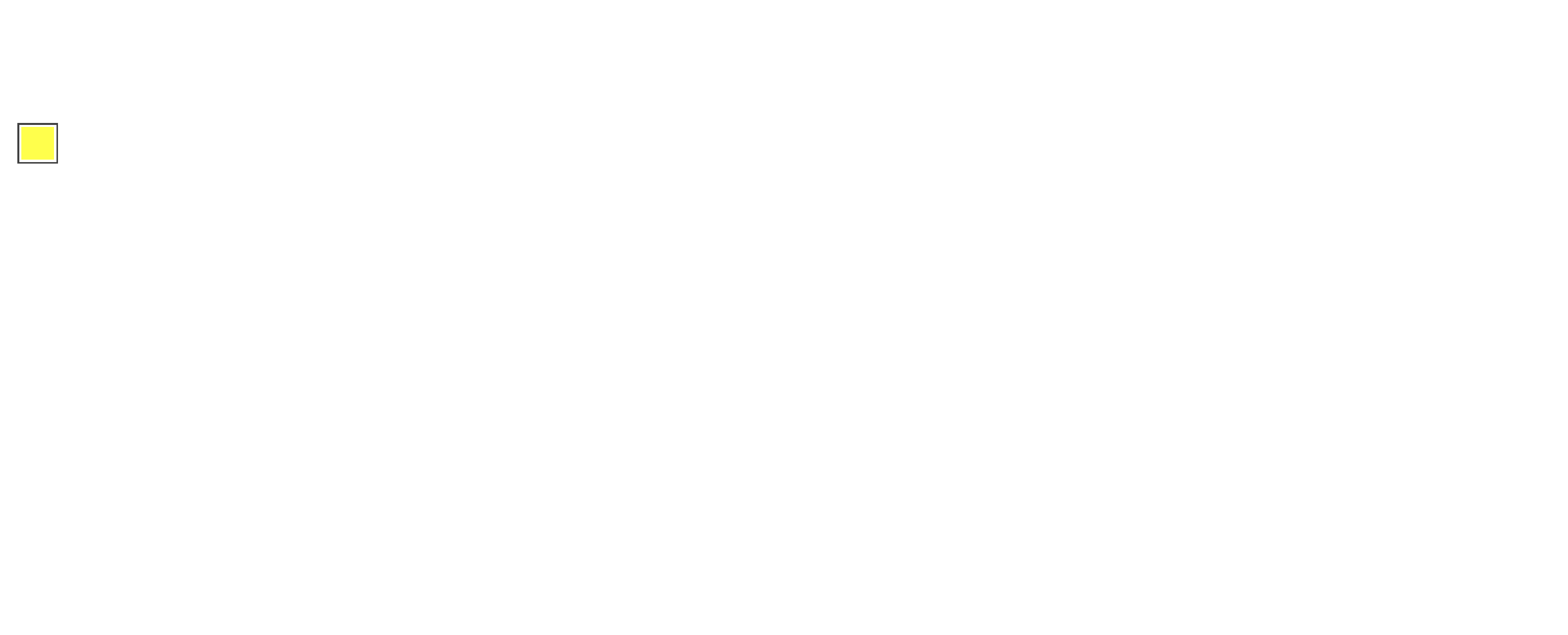}
    \caption{An example of program execution using the queue by showing a small section of the larger surface code grid. Yellow indicates operations in progress and green indicates completion. The CNOT gate is with a target qubit $\ket{B'}$ (not present). The ancillas coloured in red and marked by $\times$ are unavailable to simplify the example. For this example, the CNOT gate is enqueued before the Rz gate. In our actual implementation, however, if multiple gates must be scheduled simultaneously, gates on qubits with larger circuit depth are prioritised (since they are more likely to be on the critical path).}
    \label{fig:queue_execution}
\end{figure*}

Table~\ref{table:queuevars} contains the list of variables necessary to determine the operations the ancilla qubit will perform for a gate's execution. Along with storing the gate and the data qubit it acts upon, we store a helper ancilla if needed for the gate execution. To execute a $\mathsf{R_\theta}_p$ gate, we enqueue the gate into the queues of all ancillas adjacent to the $Z$ edge of $p$ (for $ZZ$ type injection). We also enqueue the gate into the queues of ancilla that are diagonally adjacent to $p$. Ancillas along the $X$ edge of the data qubit $p$ are also reserved for potential execution of a CNOT injection. This gate is enqueued preemptively to reduce the time the data qubit spends waiting for the $\ket{m_\theta}$ state after executing the previous gate. In the example execution shown in Figure~\ref{fig:queue_execution}, for the $\mathsf{R_\theta}_A$ gate, the corresponding neighbouring ancillas are $1, 2, 3$. Ancilla $2$ is adjacent to the $Z$ edge of data qubit $A$. Ancillas $1$ and $3$ can be connected to $A$ via ancillas $4$ and $5$ respectively, which are both adjacent to the $X$ edge of $A$. We enqueue $\mathsf{R_\theta}_p$ into the queue of all three preparing ancillas ($1$, $2$ and $3$) and both routing ancillas ($4$ and $5$).\par
Intuitively, one might think that enqueuing $\mathsf{R_\theta}_p$ into all \textit{valid} neighbouring ancillas would reduce ancilla availability for other gate operations. However, this does not happen. Specifically, when $\mathsf{R_\theta}_p$ is enqueued into all \textit{valid} neighbouring ancilla, there will already be some gates in the queue that are pending execution and this gate will enter the queue at a different position for each ancilla. The ancilla that has the least gates in its queues will start preparing $\ket{m_\theta}$ first. Having fewer gates enqueued is an indication that fewer gates are competing for this ancilla and hence it will execute $\mathsf{R_\theta}_p$ sooner. Using the queue also ensures that the priority of the gates is decided by \textit{seniority}, i.e., gates that have already been added to the queue must have been scheduled earlier and thus are executed before more recent gates.\par
When any of the preparing ancilla succeed, the gate for the other ancillas is modified in-place within their queues from $\mathsf{R_\theta}_p$ to $\mathsf{R_{2\theta}}_p$, in anticipation of an injection failure of $\ket{m_\theta}$ into $p$. An in-place update allows the other ancillas to start preparing the correction state as early as their queue permits, minimizing the execution time for the original $Rz(\theta)$ gate. %
We attempt to maximize parallel preparation since both the preparation and the injection are done in a RUS fashion, and hence more attempts leads to reduced expectation times. This has also be seen from the smaller mean in Figure~\ref{fig:histograms}.

\subsection{Efficient Path Finding}\label{section:efficientmst}

We use a greedy strategy to initialize ancilla for the execution of CNOT gates. To minimize the total program execution time, we choose the path that will finish the earliest. Even if a path exists that starts earlier, it might not \textit{finish} the earliest if it requires an edge-rotation gate. An ideal scheduler would use the exact earliest time each ancilla qubit could be used to route the CNOT gate. However, this cannot be computed due to the non-deterministic nature of the $Rz$ gates. We instead use the $activity$ of the ancilla qubits as a metric to determine which ancilla qubits are less \textit{likely} to be available in the near future. The activity of each ancilla qubit is defined as the ratio of the number of cycles the ancilla qubit was active in the last $c$ cycles, i.e., \[activity = \frac{\#\text{cycles active in last }c\text{ cycles}}{c}.\]
Using $activity$ as the metric, we choose the path that has the smallest maximum activity, since such a path is most likely to have all its ancilla qubits freed the earliest. This can be computed by constructing a weighted undirected graph of the grid with the qubits as the nodes, and the edge weights as the maximum of the activity between the neighbouring qubits. We then compute the Minimum Spanning Tree (MST) of the grid and choose the path between the control and target that lies on this MST. The Minimum Spanning Tree is guaranteed to contain the path between \textbf{\textit{every pair of vertices}} that has the least maximum weight out of every possible path between each pair of vertices \cite{clrs2009algorithms}. Therefore, we can use the same MST to route all CNOT gates of the current layer between any pairs of qubits.

\begin{figure}
    \centering
    \resizebox{\linewidth}{!}{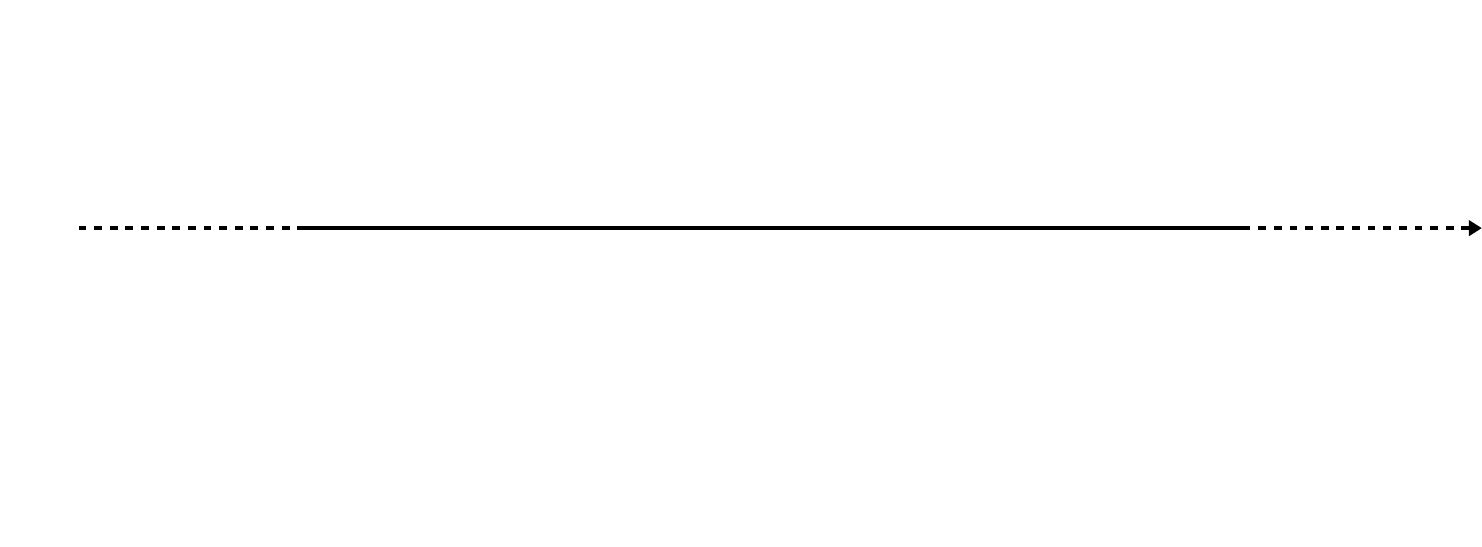}
    \caption{An example showing the MST computation timeline mid-execution. A new MST computation begins every $k = 25$ cycles and each MST computation takes $\tau_{MST} = 50$ cycles. All CNOT operations that are scheduled between time $t_2$ and $t_3$ will use the MST $\mathcal{T}_2$, whose computation started at $t_0$ and corresponds to the activity in the grid at time $t_0$. Note that this avoids any stalling of the quantum program at the cost of using an MST with \textit{stale} information.}
    \label{fig:mst_timeline}
\end{figure}

\par Computing the MST takes $O(n\log{n})$ time and cannot be computed in constant time for every CNOT gate, scaling with the size of the grid. We instead compute the MST of the grid every $k$ cycles and use the latest computed MST. For example, say computing the MST takes $\tau_{MST} = 50$ cycles and $k$ is $25$. An MST computation which began at $t = 0$, would be available only by time $t = 50$. Therefore, the ancilla activity information obtained from the MST would be \textit{stale} by $50$ cycles. In addition to this, two more MST computations would be running in parallel, one that started at $t = 25$ and another that began at $t = 50$. Figure~\ref{fig:mst_timeline} gives an example execution of the MST computation process. We show in Section~\ref{section:evaluationk}, this delay information has negligible effect on the performance of \acronym{}. We also provide a more detailed overhead analysis in Section~\ref{section:mstcomplexityanalysis}.\par

\begin{figure}
    \centering
    \resizebox{0.9\columnwidth}{!}{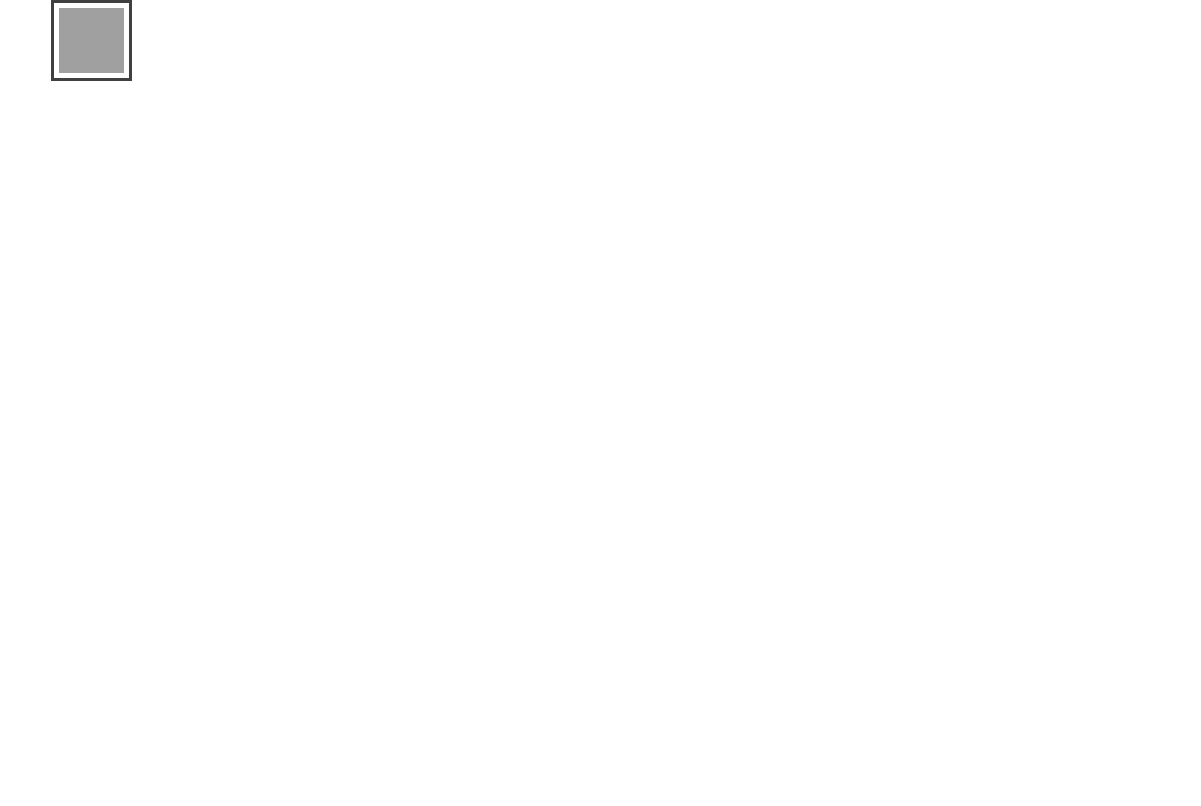}
    \caption{MST construction protocol and comparison of paths chosen by different schedulers. $a_i$ denotes activity of ancilla $i$. The path chosen by Algorithm~\ref{algorithm:cnotexecution} requires an edge rotation gate, but takes lesser time to execute (in expectation) since it chooses the ancillas with less activity. The edge rotation gate does not require all ancillas to be free, only the ancilla adjacent to the control needs to be freed.}
    \label{fig:mst_construction}
\end{figure}
\begin{algorithm}
    \scriptsize
    \caption{CNOT Execution Algorithm}
    \label{algorithm:cnotexecution}
    \begin{algorithmic}[1]
        \Procedure{ExecuteCNOTGate}{$\mathsf{QUEUE}, \mathsf{Control}, \mathsf{Target}$}
            \State{MST $\gets$ \Call{GetLatestComputedMSTOfAncillas}{\null}}
            \State{bestStartTime $\gets$ $\infty$}
            \State{bestPath $\gets$ $\phi$}
            \ForAll{edges $e_C\in$ \Call{NeighbouringAncillas}{$\mathsf{Control}$}}
                \ForAll{edges $e_T\in$ \Call{NeighbouringAncillas}{$\mathsf{Target}$}}
                    \State{startTime $\gets\ 0$}
                    \State{ancilla$_C$ $\gets$ \Call{GetAncilla}{$e_C$}}
                    \State{ancilla$_T$ $\gets$ \Call{GetAncilla}{$e_T$}}
                    \If{\textbf{not} \Call{IsXEdge}{$e_C$}}\Comment{add edge rotation time at ancilla$_C$}
                        \State{expFreeTime $\gets$ \Call{GetExpectedFreeTime}{ancilla$_C$}}
                        \State{startTime $\gets$ \Call{max}{startTime, expFreeTime + 3}}
                    \EndIf{}
                    \If{\textbf{not} \Call{IsZEdge}{$e_T$}}\Comment{add edge rotation time at ancilla$_T$}
                        \State{expFreeTime $\gets$ \Call{GetExpectedFreeTime}{ancilla$_T$}}
                        \State{startTime $\gets$ \Call{max}{startTime, expFreeTime + 3}}
                    \EndIf{}
                    \State{path $\gets$ \Call{FindPath}{MST, ancilla$_C$, ancilla$_T$}}
                    \ForAll{ancilla $\in$ path}
                        \State{expFreeTime $\gets$ \Call{GetExpectedFreeTime}{ancilla}}
                        \State{startTime $\gets$ \Call{max}{startTime, expFreeTime}}
                    \EndFor{}
                    \If{startTime $<$ bestStartTime}
                        \State{bestStartTime $\gets$ startTime}
                        \State{bestPath $\gets$ path $\cup\ \{e_C, e_T\}$}
                    \EndIf{}
                \EndFor{}
            \EndFor{}
            \State{gate $\gets$ \Call{CNOTGate}{$\mathsf{Control}$, $\mathsf{Target}$, bestPath}}
            \State{$\mathsf{QUEUE}$ $\gets$ \Call{AddGateToQueue}{$\mathsf{QUEUE}$, gate}}
            \Statex{\Comment{this also adds EdgeRotationGate(s) if needed}}
            \State{\Return{$\mathsf{QUEUE}$}}
        \EndProcedure{}
    \end{algorithmic}
\end{algorithm}

We want to minimize the \textit{finish} time of the CNOT gate but computing the MST only guarantees that the \textit{start} time is minimized. For this, we consider the expected completion time of each of the 16 paths, $\mathscr{p}_i$, from control to target (4 neighbors each = 16 paths using the MST of ancilla).  For each ancilla $a \in \mathscr{p}_i$ with queue $Q_a$, it's expected free time is given as $$\mathbb{E}[f_a] = \sum_{o \in Q_a} \mathbb{E}[\tau_o]$$, where $f_a$ is the free time of ancilla $a$ and $\tau_o$ is the execution time of operation $o$. Then the expected completion time for a given path $\mathscr{p}_i$ is $$\mathbb{E}[\mathscr{p}_i \text{ completes}] = 3r_C + 3r_T + \mathbb{E}[\tau_{CNOT}] + \max_{a \in \mathscr{p}_i}\mathbb{E}[f_a]$$ where $\mathbb{E}[\tau_{CNOT}] = 2$, $r_C, r_T \in \{0, 1\}$. Here, $r_C, r_T = 1$ if an edge rotation is necessary for the control or target, respectively. We choose $\min_{\mathscr{p}_i}\mathbb{E}[\mathscr{p}_i \text{ completes}]$. An example of this in practice is found in Figure \ref{fig:mst_construction} and corresponding pseudo-code in Algorithm~\ref{algorithm:cnotexecution}.

\section{Evaluation}\label{section:evaluation}

\subsection{Benchmarks and Simulator Setup}
\begin{table}[!hbtp]
    \setlength\tabcolsep{0pt}
    \centering
    \begin{tabular*}{\columnwidth}{@{\extracolsep{\fill}}llccc}
    \toprule
    \textbf{Suite} & \textbf{Benchmarks} & \textbf{\#Qubits} & \textbf{\#Rz} & \textbf{\#CNOT} \\
    \midrule
    \multirow{13}{*}{large} & \multirow{5}{*}{ising} & 34 & 83 & 66 \\
    \cline{3-5}
     &  & 42 & 103 & 82 \\
    \cline{3-5}
     &  & 66 & 163 & 130 \\
    \cline{3-5}
     &  & 98 & 243 & 194 \\
    \cline{3-5}
     &  & 420 & 1048 & 838 \\
    \cline{2-5}
     & \multirow{2}{*}{multiplier} & 45 & 2237 & 2286 \\
    \cline{3-5}
     &  & 75 & 6384 & 6510 \\
    \cline{2-5}
     & \multirow{3}{*}{qft} & 29 & 708 & 680 \\
    \cline{3-5}
     &  & 63 & 1898 & 1836 \\
    \cline{3-5}
     &  & 160 & 5293 & 5134 \\
    \cline{2-5}
     & \multirow{3}{*}{qugan} & 39 & 411 & 296 \\
    \cline{3-5}
     &  & 71 & 763 & 552 \\
    \cline{3-5}
     &  & 111 & 1203 & 872 \\
    \hline
    \multirow{4}{*}{medium} & gcm & 13 & 1528 & 762 \\
    \cline{2-5}
     & dnn & 16 & 2432 & 384 \\
    \cline{2-5}
     & qft & 18 & 323 & 306 \\
    \cline{2-5}
     & wstate & 27 & 156 & 52 \\
    \hline
    \multirow{6}{*}{supermarq} & \multirow{3}{*}{HamiltonianSimulation} & 25 & 49 & 48 \\
    \cline{3-5}
     &  & 50 & 99 & 98 \\
    \cline{3-5}
     &  & 75 & 149 & 148 \\
    \cline{2-5}
     & QAOAFermionicSwap & 15 & 120 & 315 \\
    \cline{2-5}
     & QAOAVanilla & 15 & 120 & 210 \\
    \cline{2-5}
     & VQE & 13 & 78 & 12 \\
    \bottomrule
    \end{tabular*}
    \caption{List of the benchmarks for which we evaluate \acronym{} against the baseline implementations. The large and medium benchmark suites are from QASMBench \cite{li2022qasmbench}. We also evaluate on benchmarks from SupermarQ \cite{tomesh2022supermarq}.}
    \label{table:benchmarks}
\end{table}

\begin{figure*}
    \centering
    \resizebox{\linewidth}{!}{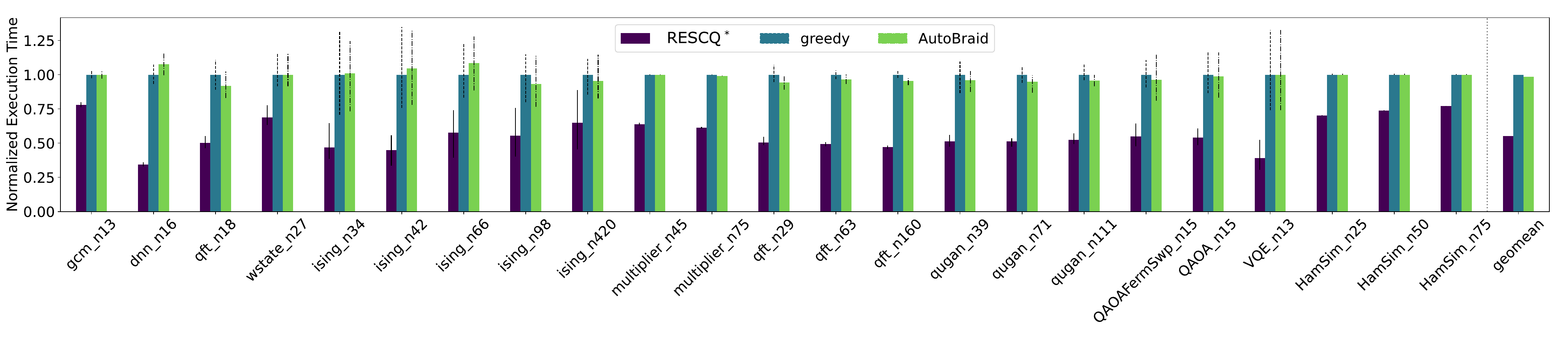}
    \caption{Normalized average execution time for \acronym{} versus the baseline schemes. The above plot is for $d = 7$ and p $= 10^{-4}$. We compute execution time for $k \in \{25, 50 , 100, 200\}$ and report best average as $RESCQ^*$. Error bars show minimum and maximum execution times in black. We report the geometric mean across all the benchmarks.}
    \label{fig:plot}
\end{figure*}

Table~\ref{table:benchmarks} lists the benchmarks (from QASMBench \cite{li2022qasmbench} and SupermarQ \cite{tomesh2022supermarq}) that we evaluate \acronym{} on. These benchmarks were chosen as a representative set from QASMBench (\textit{medium} and \textit{large} suites) and Supermarq since these benchmarks span a large range of Rz gate to CNOT gate ratios ($\approx 1$ to $\approx 6.5$) and number of qubits (from $13$ qubits to $420$ qubits). These benchmarks also represent a wide range of gate densities. For example, \textit{wstate} and \textit{qft} circuits are largely sequential while \textit{ising} circuits are largely parallel. Most importantly, these benchmarks are chosen because they contain continuous angle rotations in the given representation. Other benchmarks which contain only deterministically prepared gates, e.g. programs with only Cliffords will behave identically in the static and realtime cases.

We compile each benchmark into the basis gate set of $Rz$, $H$, $X$ and CNOT using Qiskit \cite{Qiskit}. After compilation, we report the total number of gates, the number of Rz Gates, and the number of CNOT gates in Table~\ref{table:benchmarks}. We perform a one-to-one mapping of program qubits to logical qubits since gates in the benchmarks were between qubits of numerically close indices. We use the greedy path selection \cite{javadi-abhari2017optimized} and AutoBraid\cite{hua2021autobraid} as the baseline schedulers, and add edge-rotation gates if required. We augment both schemes with the naive $Rz$ gate preparation protocol; exactly one ancilla is reserved for preparing the $\ket{m_\theta}$ state and early preparations are not done. We simulate each quantum program by choosing a random seed to model the preparation and injection probabilities, as well as the execution times, as a function of the code distance and physical qubit error rate\cite{akahoshi2023partially}. We then perform symbolic execution of the program to simulate total execution time, accounting for delays due to non-deterministic failures, qubit stalls (both by ancilla and data qubits) and routing congestion. Each benchmark is executed multiple times, each time with a unique seed. For \acronym{}, we fix $c$, the number of cycles used to determine the ancilla activity, to be $100$ and evaluate the execution for the MST computation frequency $k \in \{25, 50, 100, 200\}$ cycles. Based on realistic grid sizes and estimates from MST computation time on modern CPUs (MacBook Air machine with M2 chip), it takes about $100\mu s$ (refer Section~\ref{section:mstcomplexityanalysis}) to compute the MST. Since a lattice surgery cycle takes about $1\mu s$ \cite{litinski2019game}, the time taken to compute the MST is $\tau_{MST} = 100$ lattice surgery cycles. We compare the performance of \acronym{} versus the baseline schedulers in Figure~\ref{fig:plot} and observe considerable performance improvements with a geomean of $2\times$ speedup.

\subsection{Sensitivity Analysis}\label{section:sensitivity}
\begin{figure*}
    \centering
    \begin{subfigure}{0.24\textwidth}
        \centering
        \resizebox{\linewidth}{!}{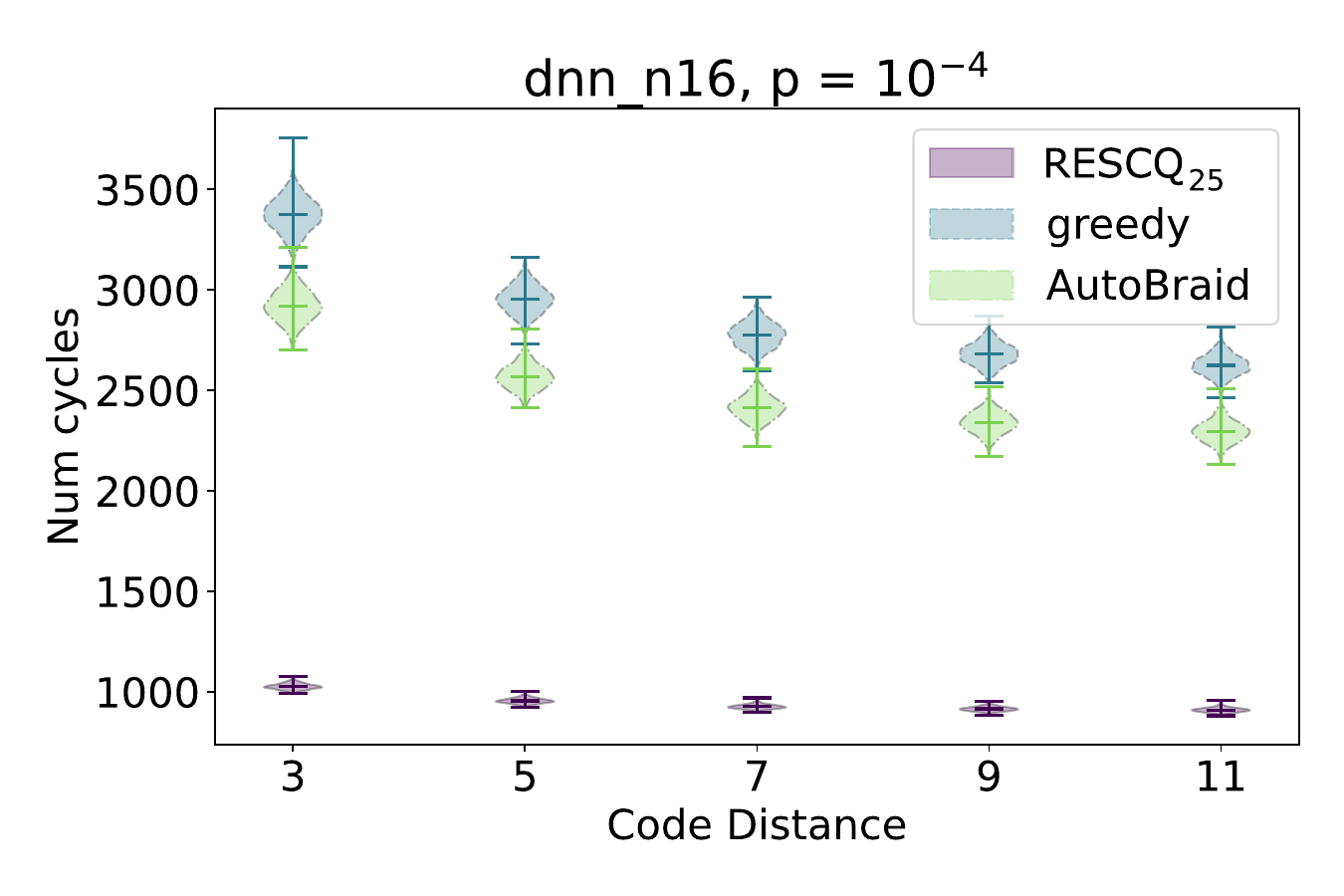}
    \end{subfigure}
    \begin{subfigure}{0.24\textwidth}
        \centering
        \resizebox{\linewidth}{!}{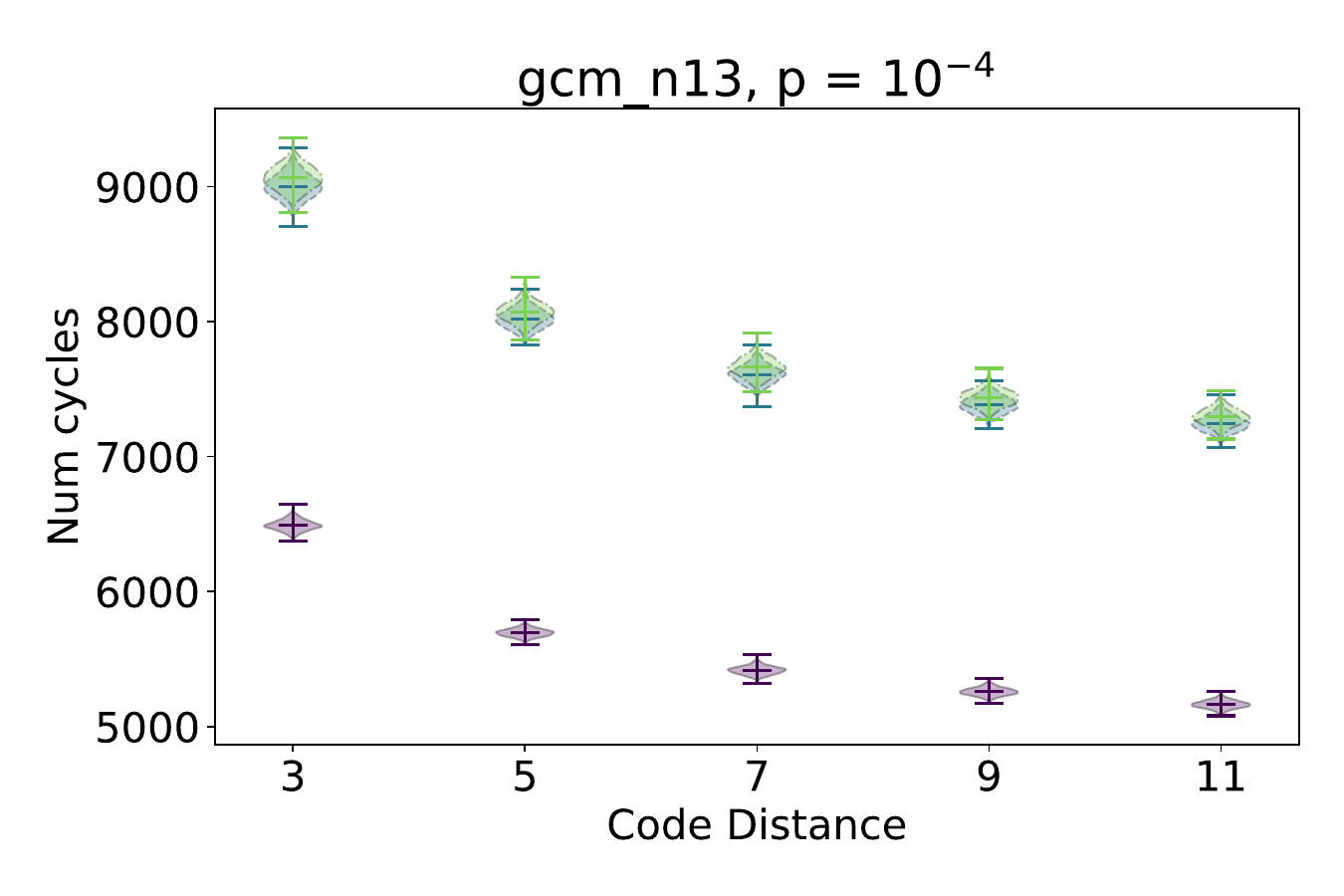}
    \end{subfigure}
    \begin{subfigure}{0.24\textwidth}
        \centering
        \resizebox{\linewidth}{!}{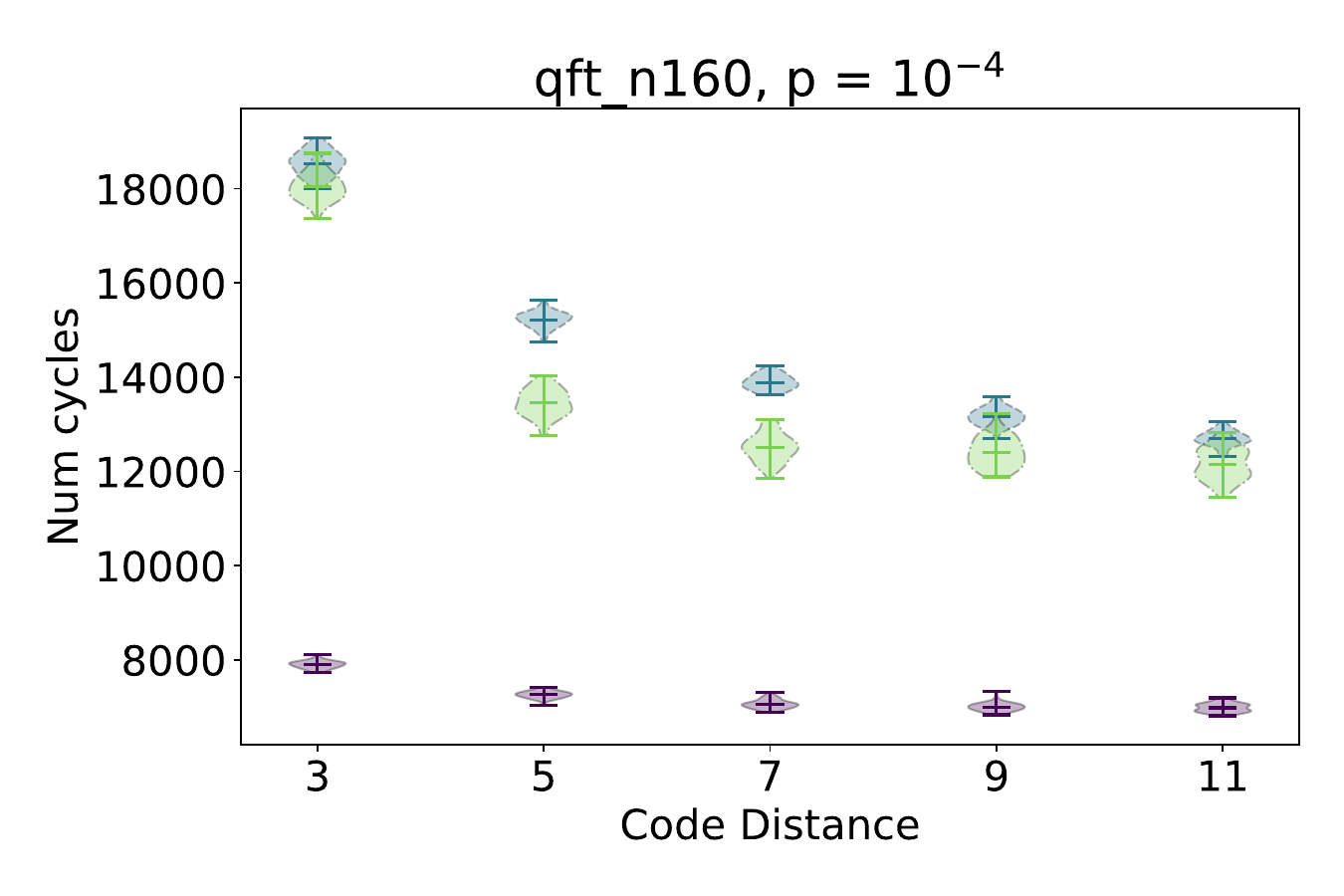}
    \end{subfigure}
    \begin{subfigure}{0.24\textwidth}
        \centering
        \resizebox{\linewidth}{!}{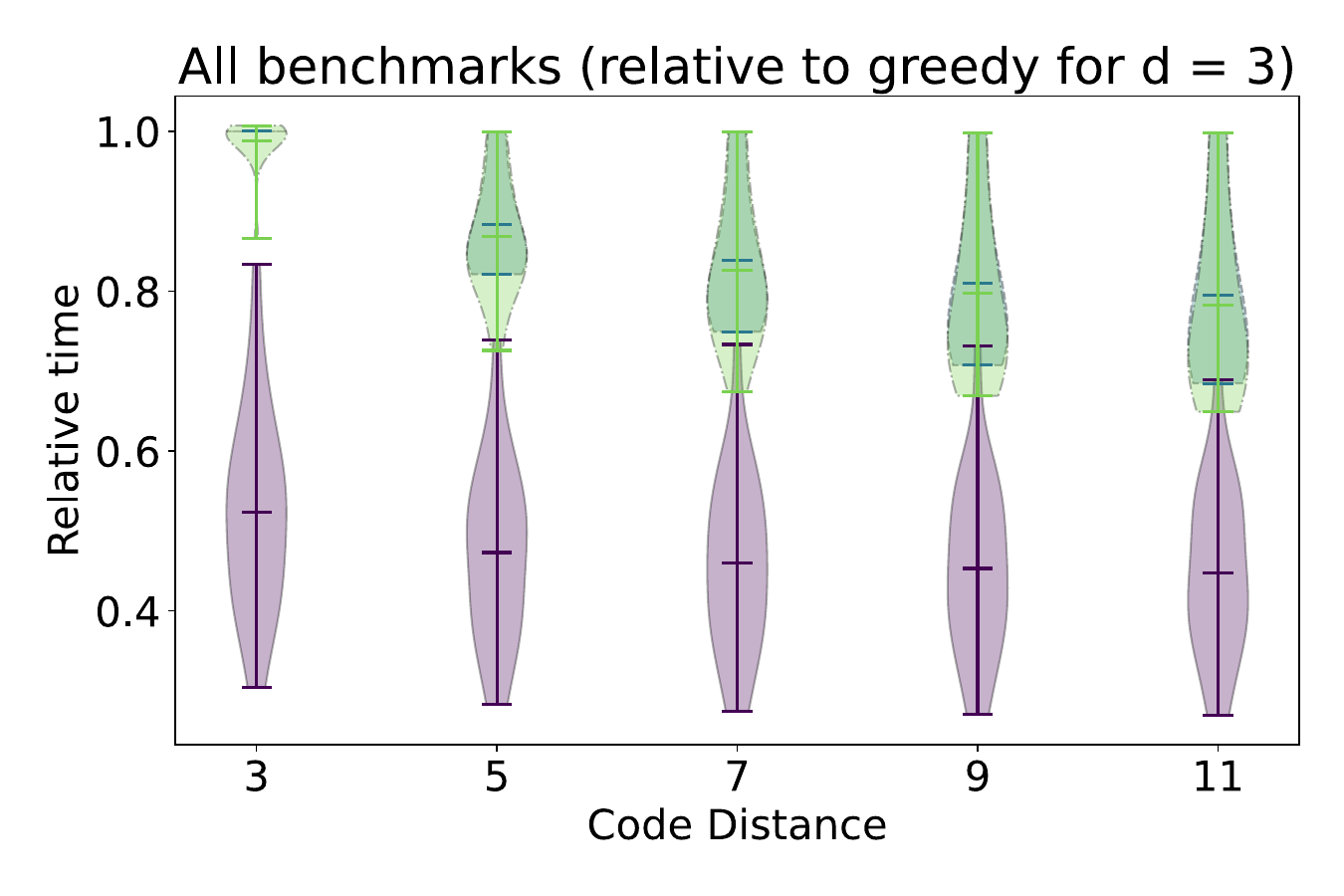}
    \end{subfigure}
    \begin{subfigure}{0.24\textwidth}
        \centering
        \resizebox{\linewidth}{!}{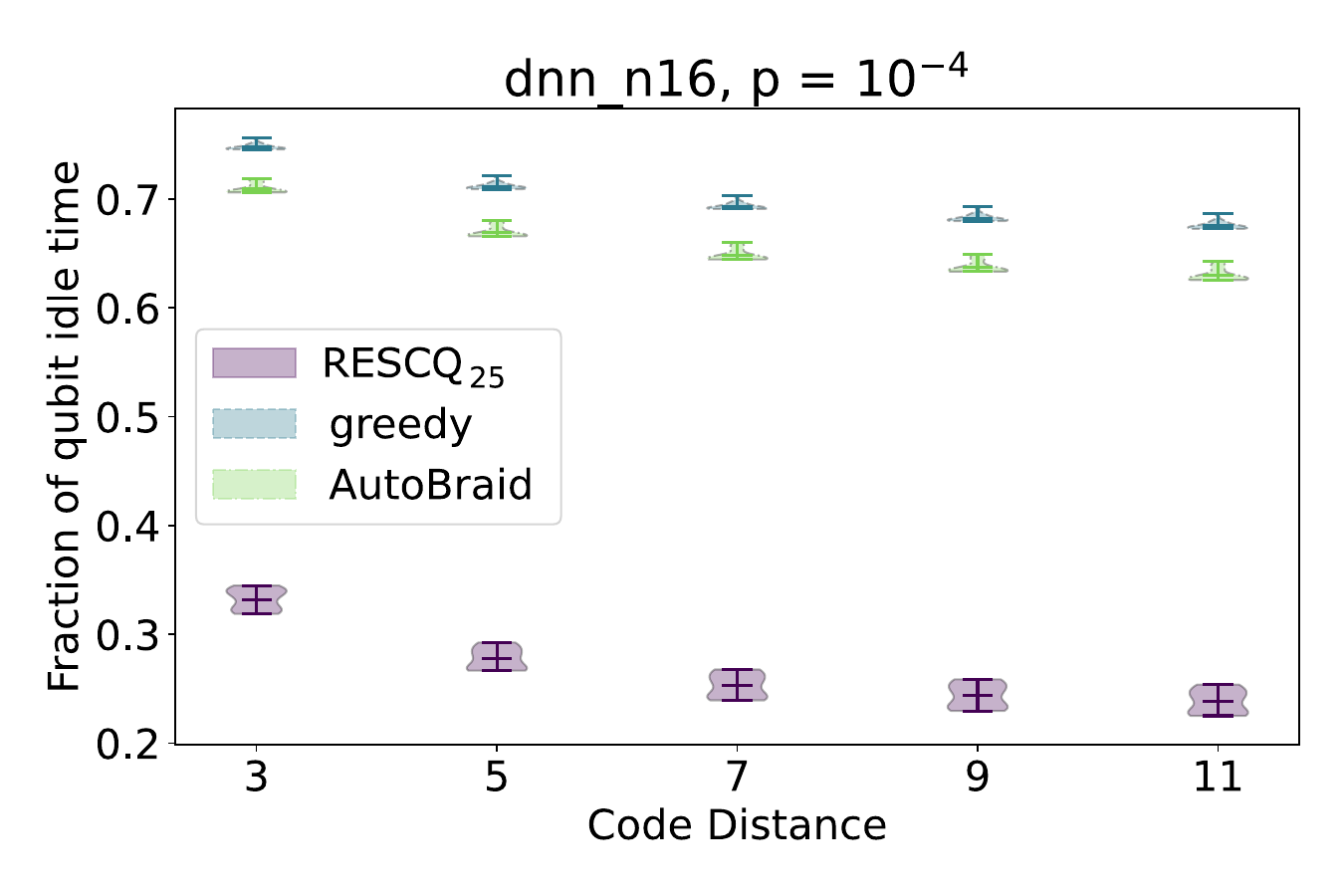}
    \end{subfigure}
    \begin{subfigure}{0.24\textwidth}
        \centering
        \resizebox{\linewidth}{!}{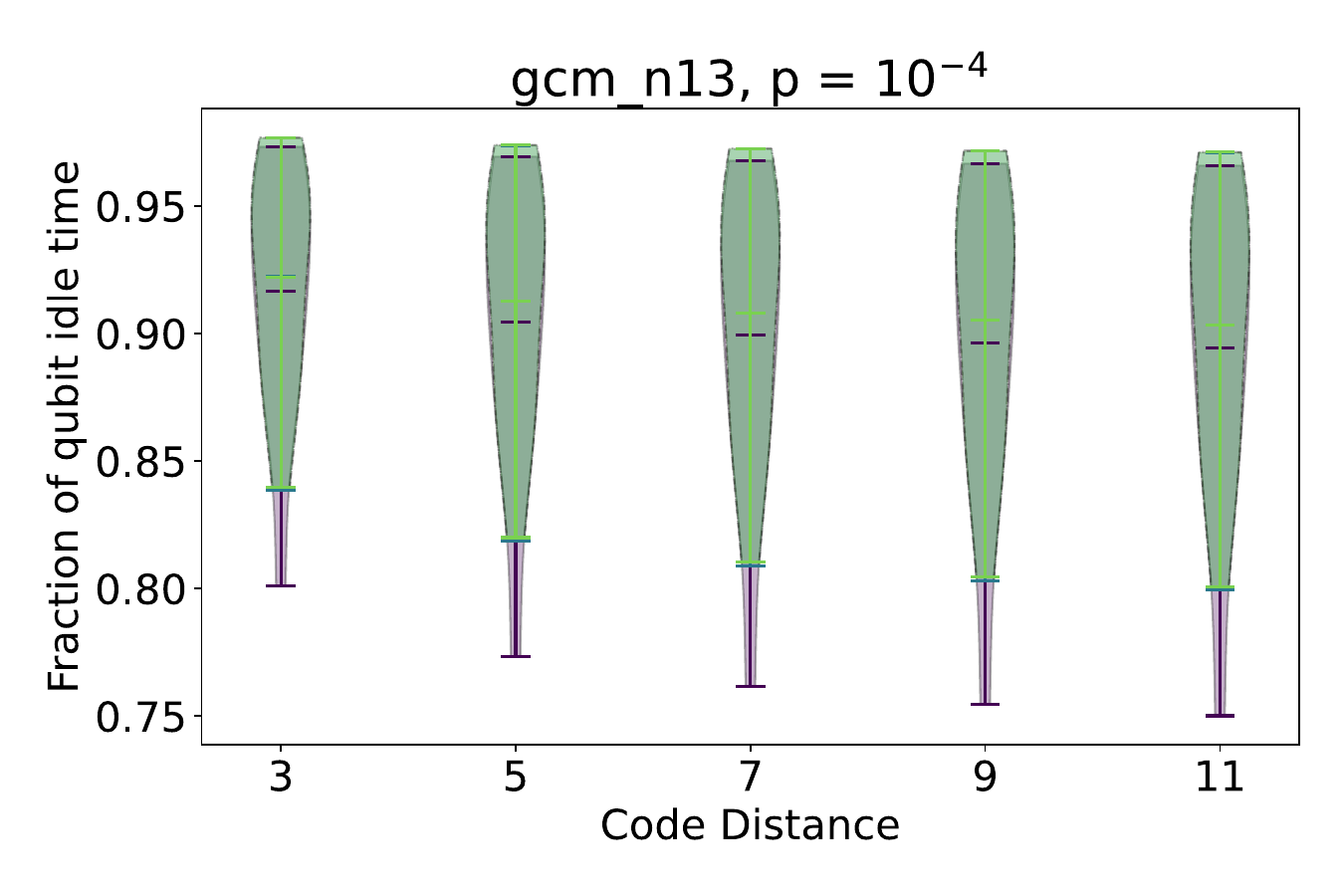}
    \end{subfigure}
    \begin{subfigure}{0.24\textwidth}
        \centering
        \resizebox{\linewidth}{!}{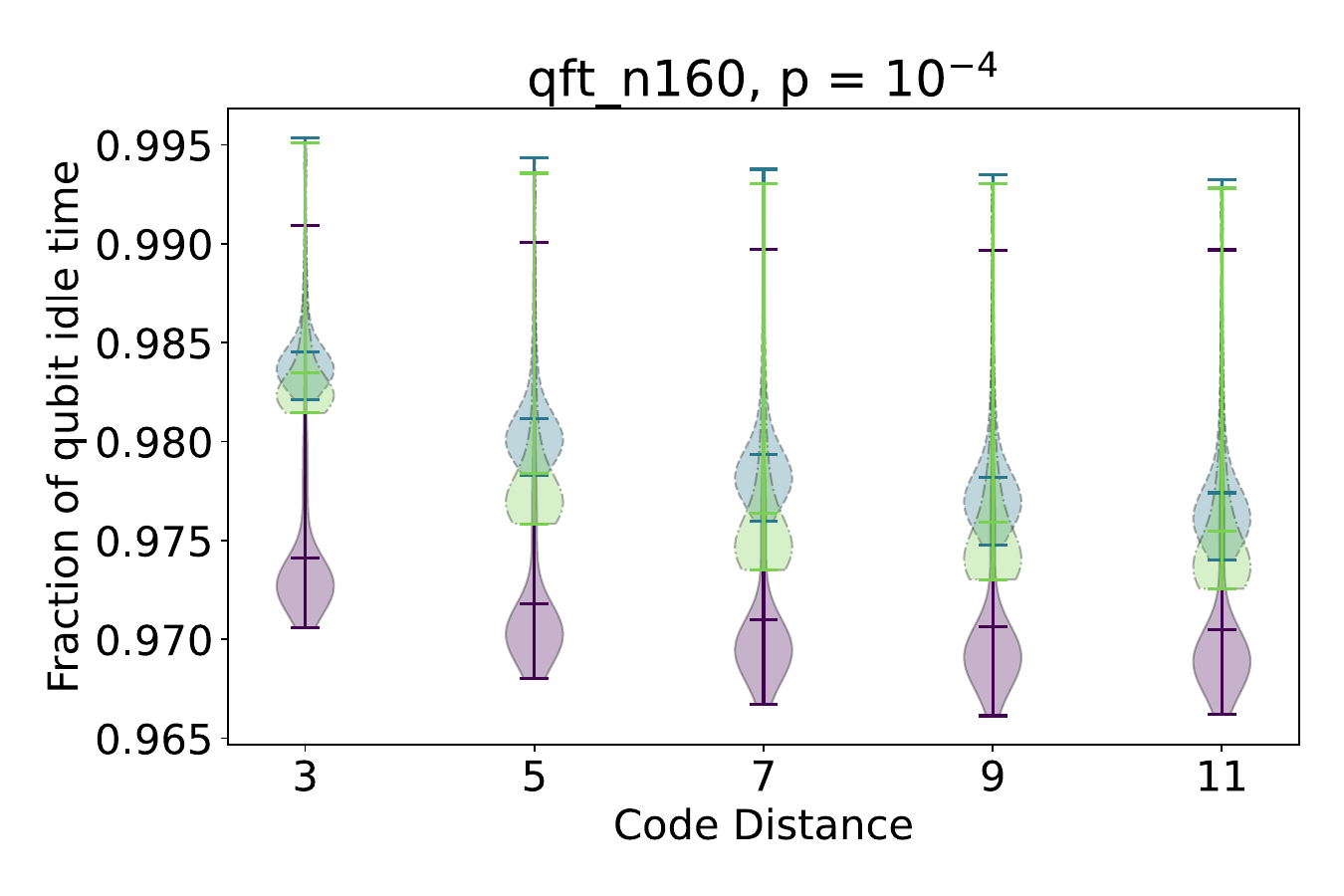}
    \end{subfigure}
    \begin{subfigure}{0.24\textwidth}
        \centering
        \resizebox{\linewidth}{!}{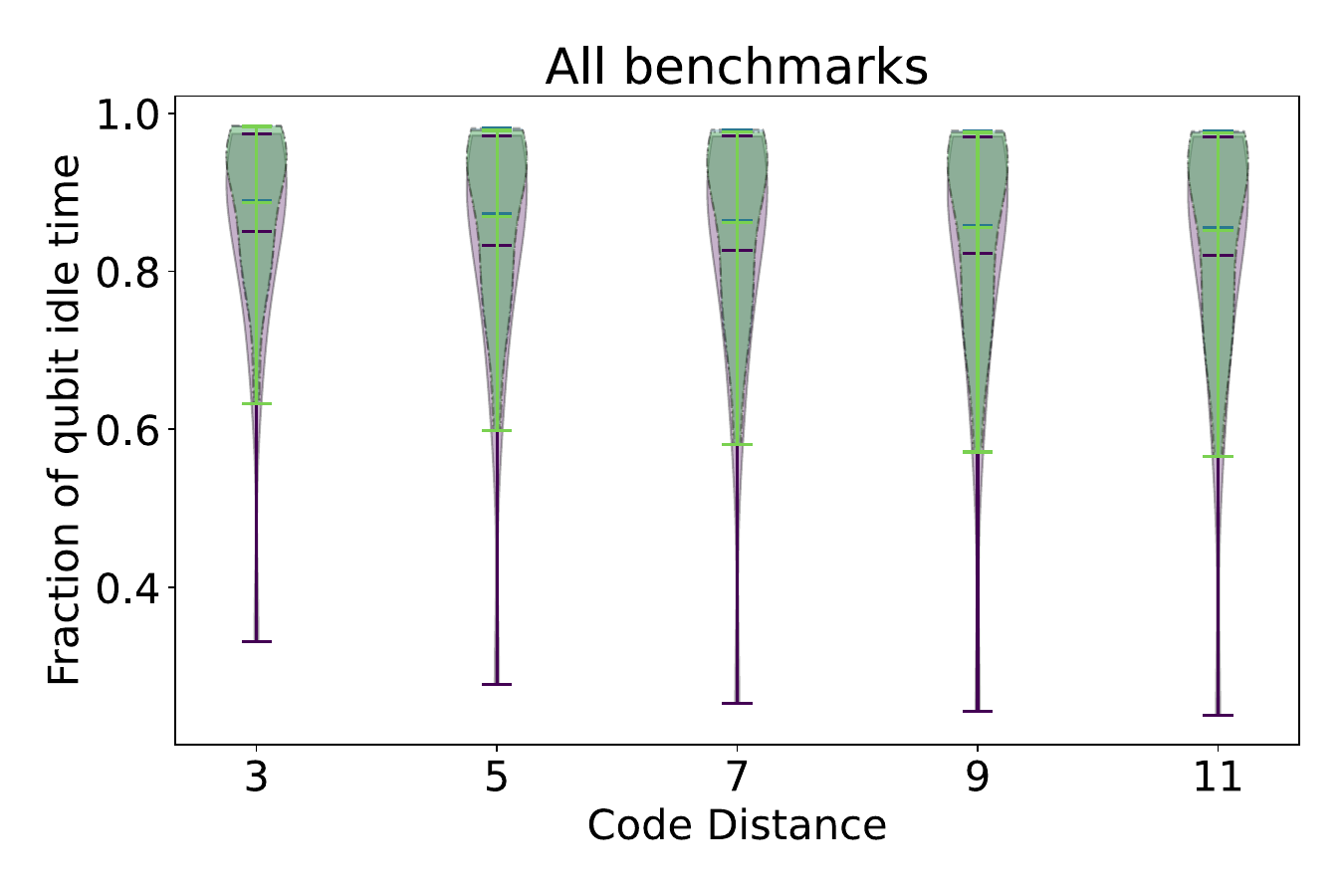}
    \end{subfigure}
    \caption{Sensitivity of different schedulers to varying the code distance}
    \label{fig:sensitivity_d}
\end{figure*}

\begin{figure*}
    \centering
    \begin{subfigure}{0.24\textwidth}
        \centering
        \resizebox{\linewidth}{!}{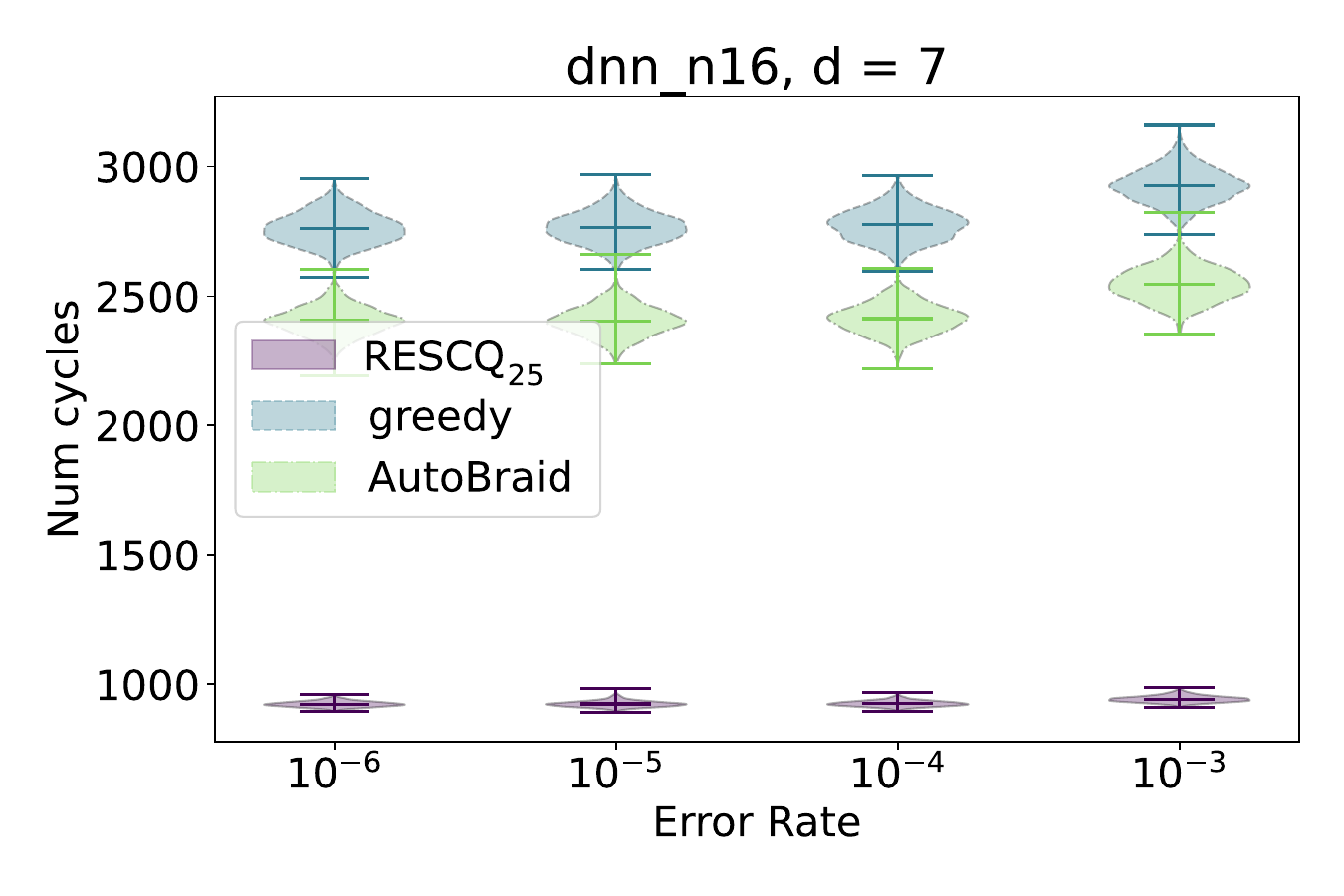}
    \end{subfigure}
    \begin{subfigure}{0.24\textwidth}
        \centering
        \resizebox{\linewidth}{!}{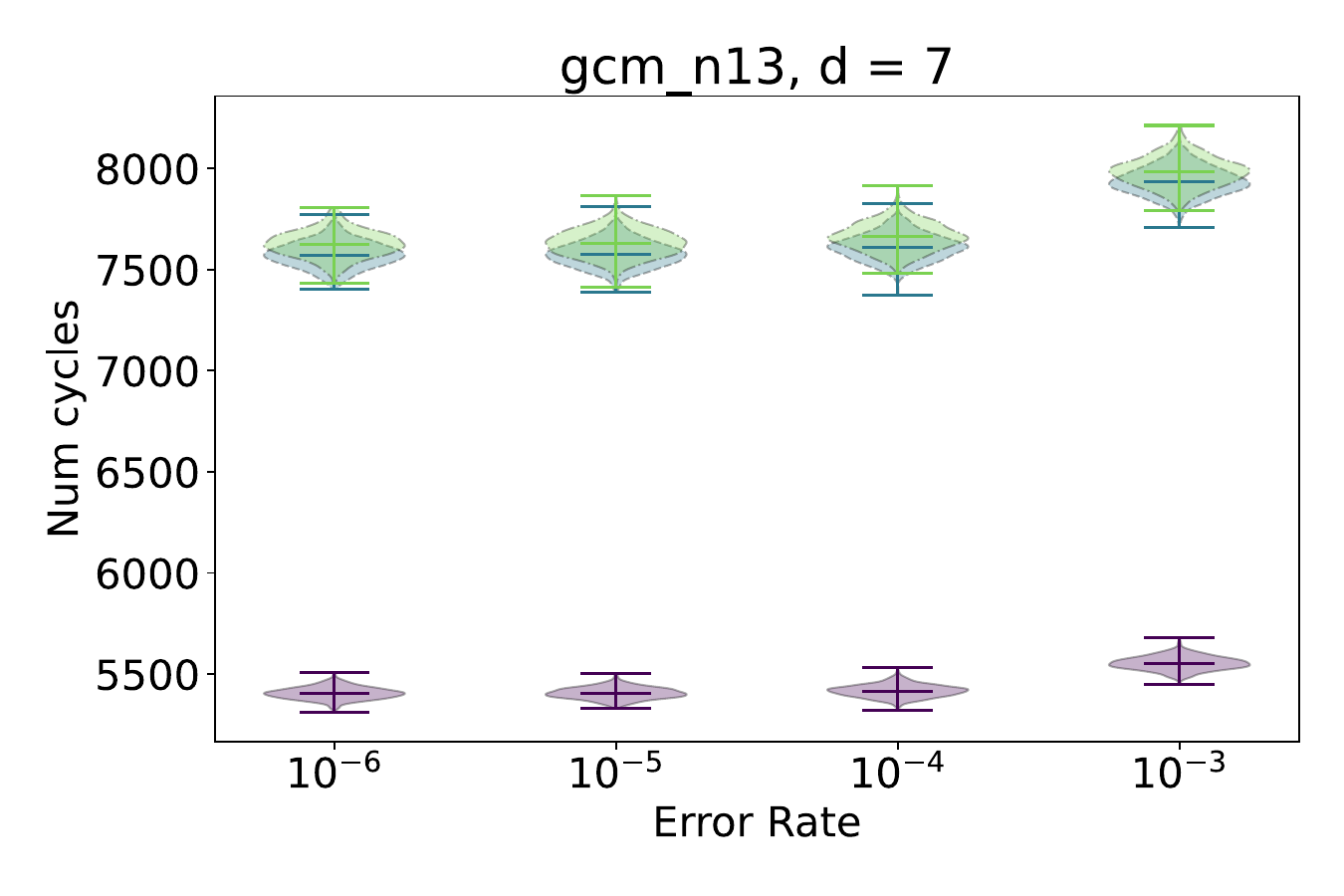}
    \end{subfigure}
    \begin{subfigure}{0.24\textwidth}
        \centering
        \resizebox{\linewidth}{!}{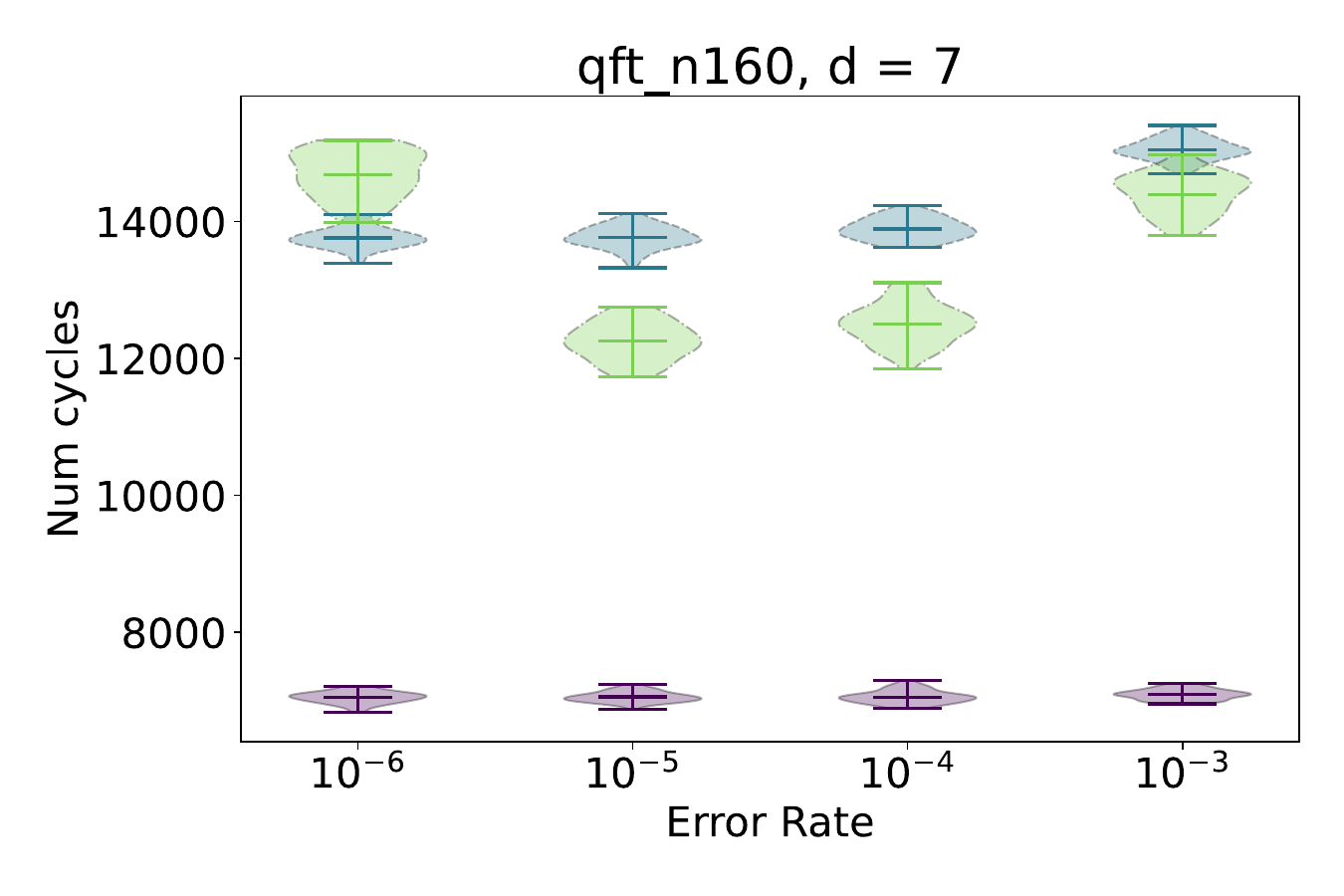}
    \end{subfigure}
    \begin{subfigure}{0.24\textwidth}
        \centering
        \resizebox{\linewidth}{!}{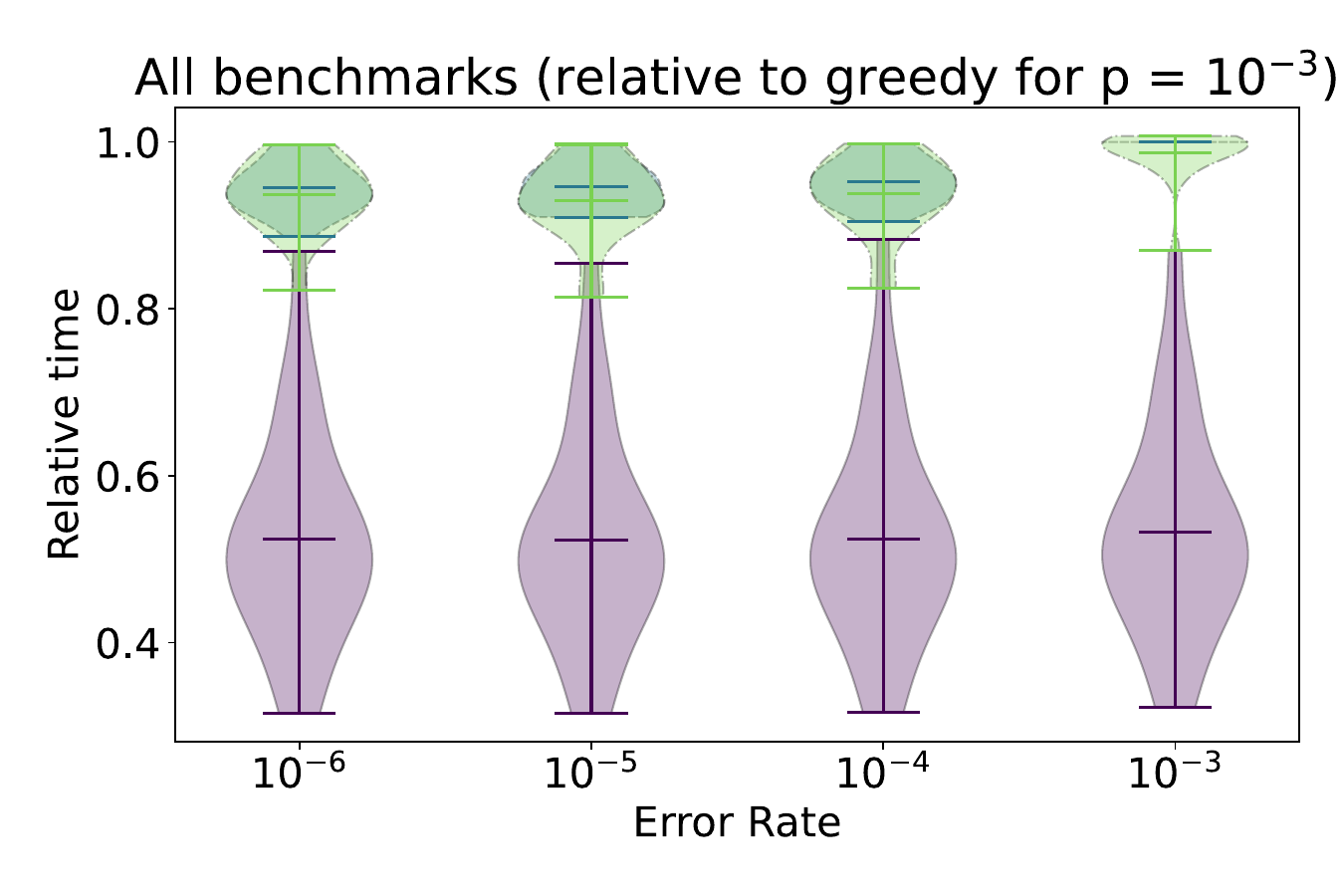}
    \end{subfigure}
    \begin{subfigure}{0.24\textwidth}
        \centering
        \resizebox{\linewidth}{!}{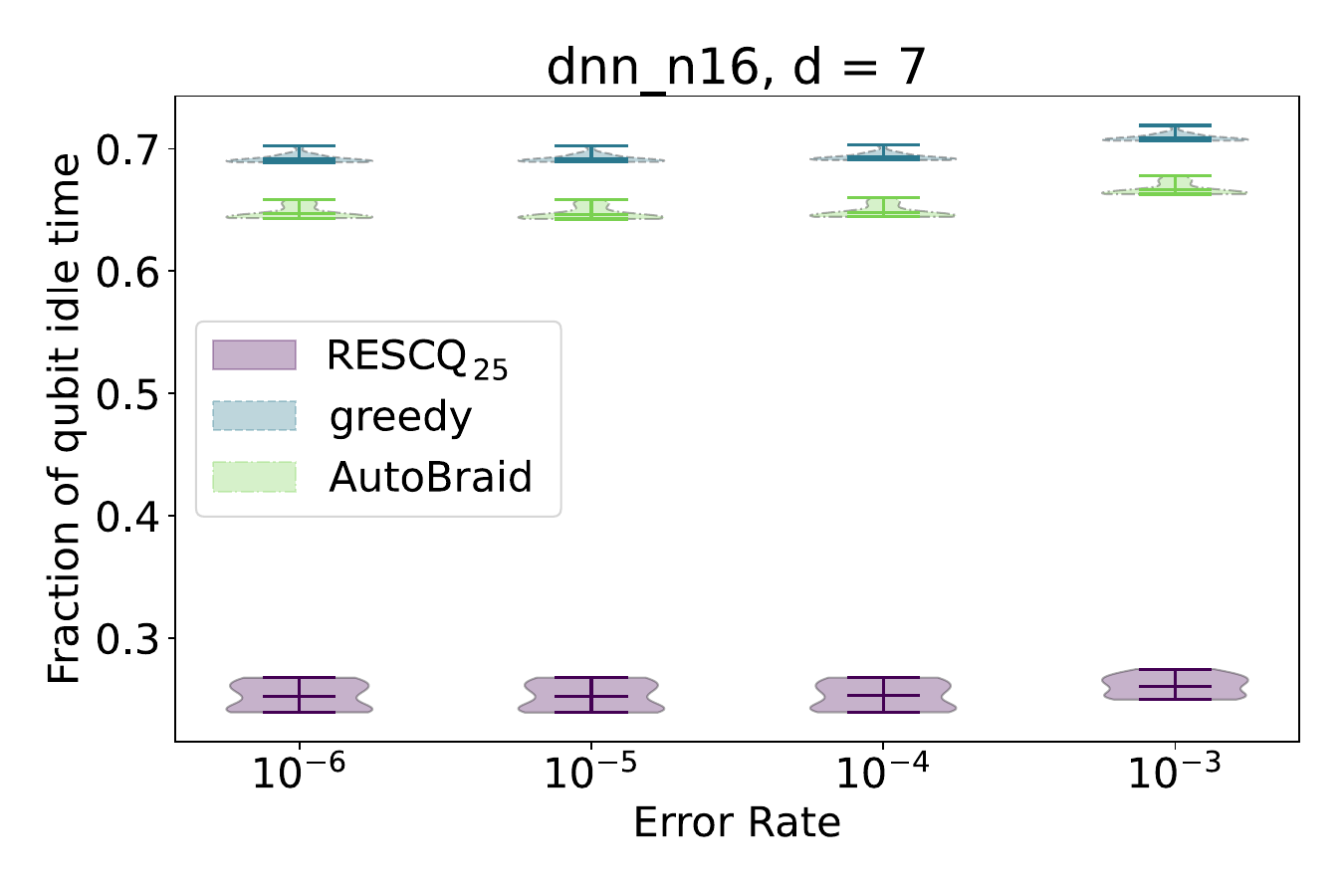}
    \end{subfigure}
    \begin{subfigure}{0.24\textwidth}
        \centering
        \resizebox{\linewidth}{!}{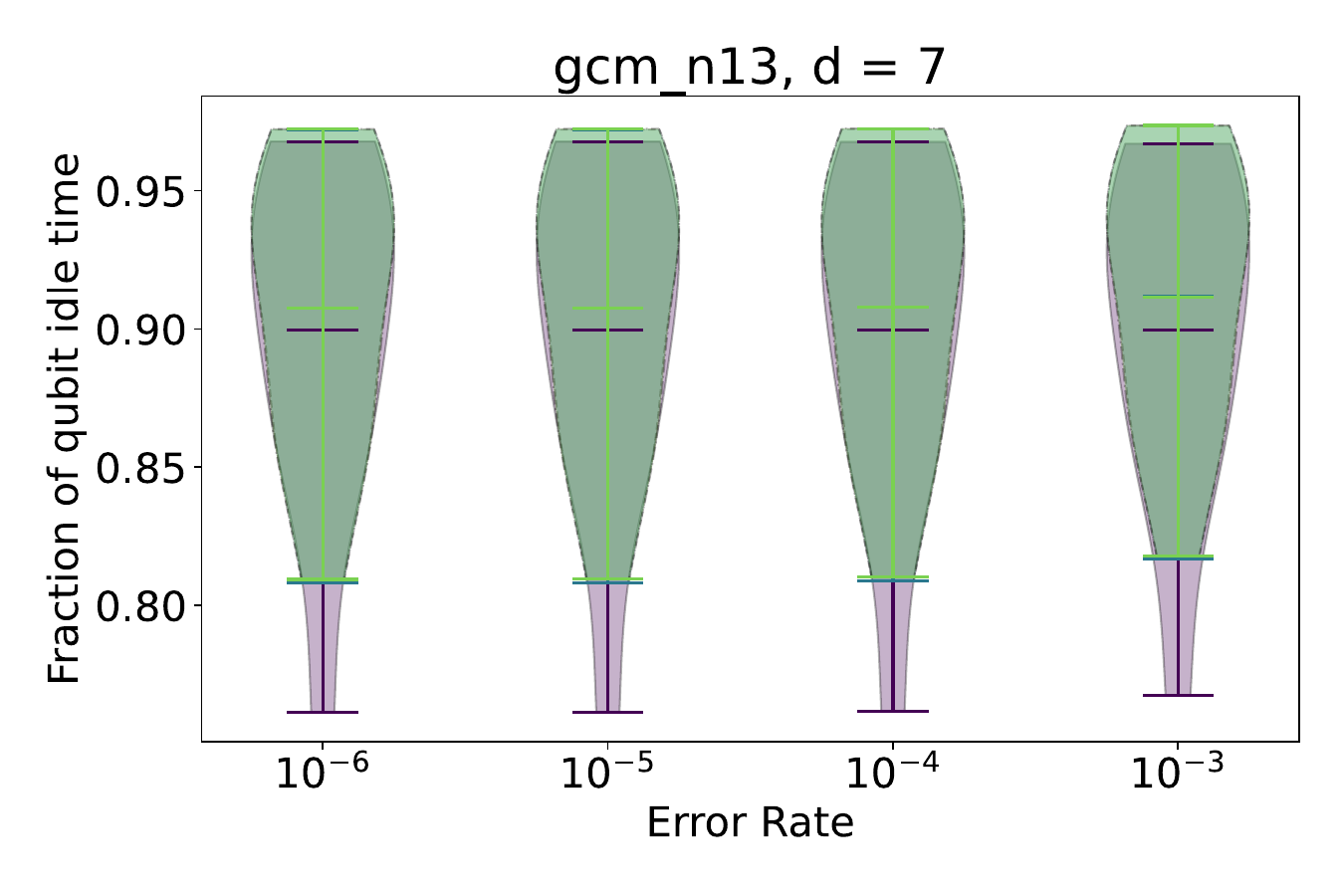}
    \end{subfigure}
    \begin{subfigure}{0.24\textwidth}
        \centering
        \resizebox{\linewidth}{!}{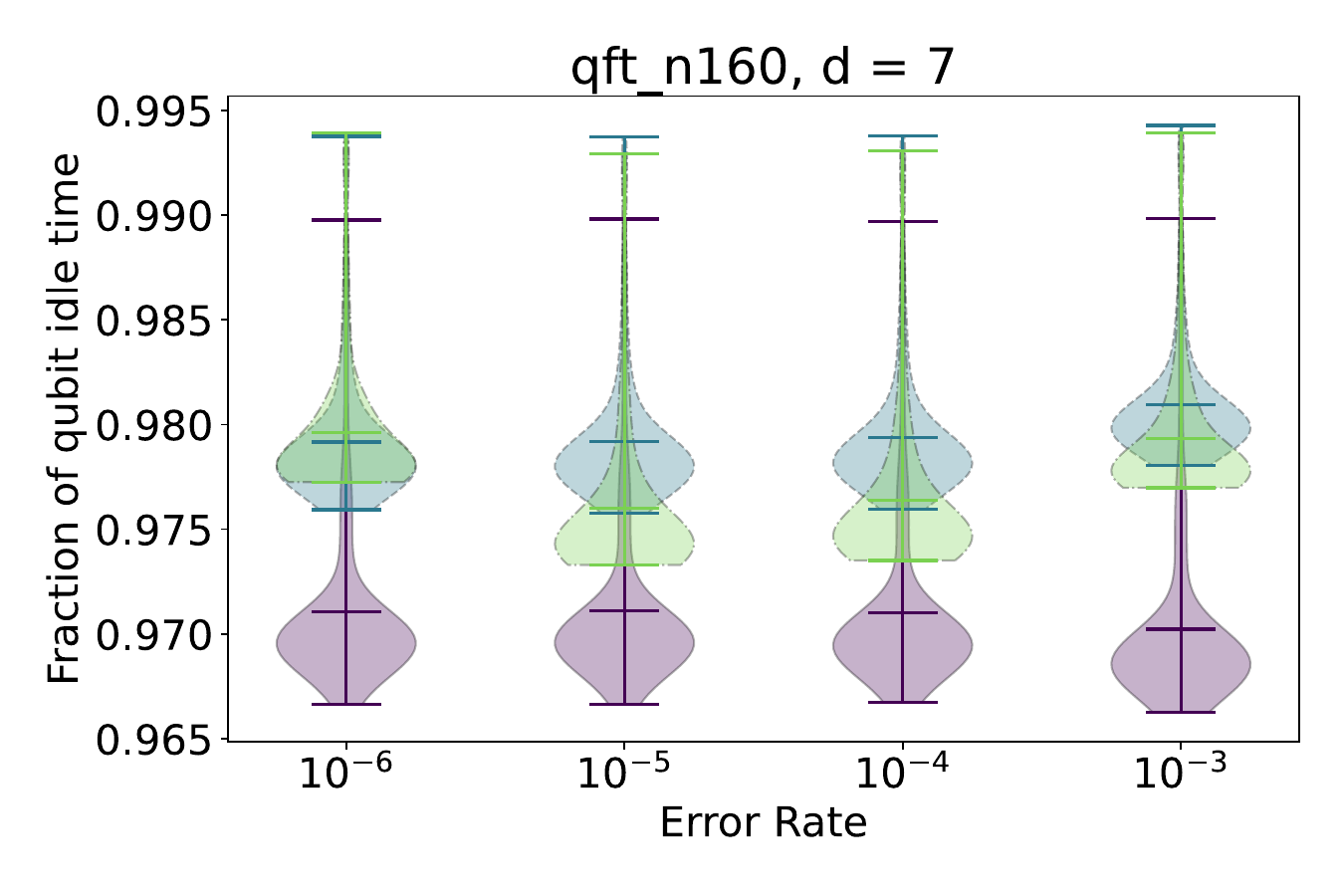}
    \end{subfigure}
    \begin{subfigure}{0.24\textwidth}
        \centering
        \resizebox{\linewidth}{!}{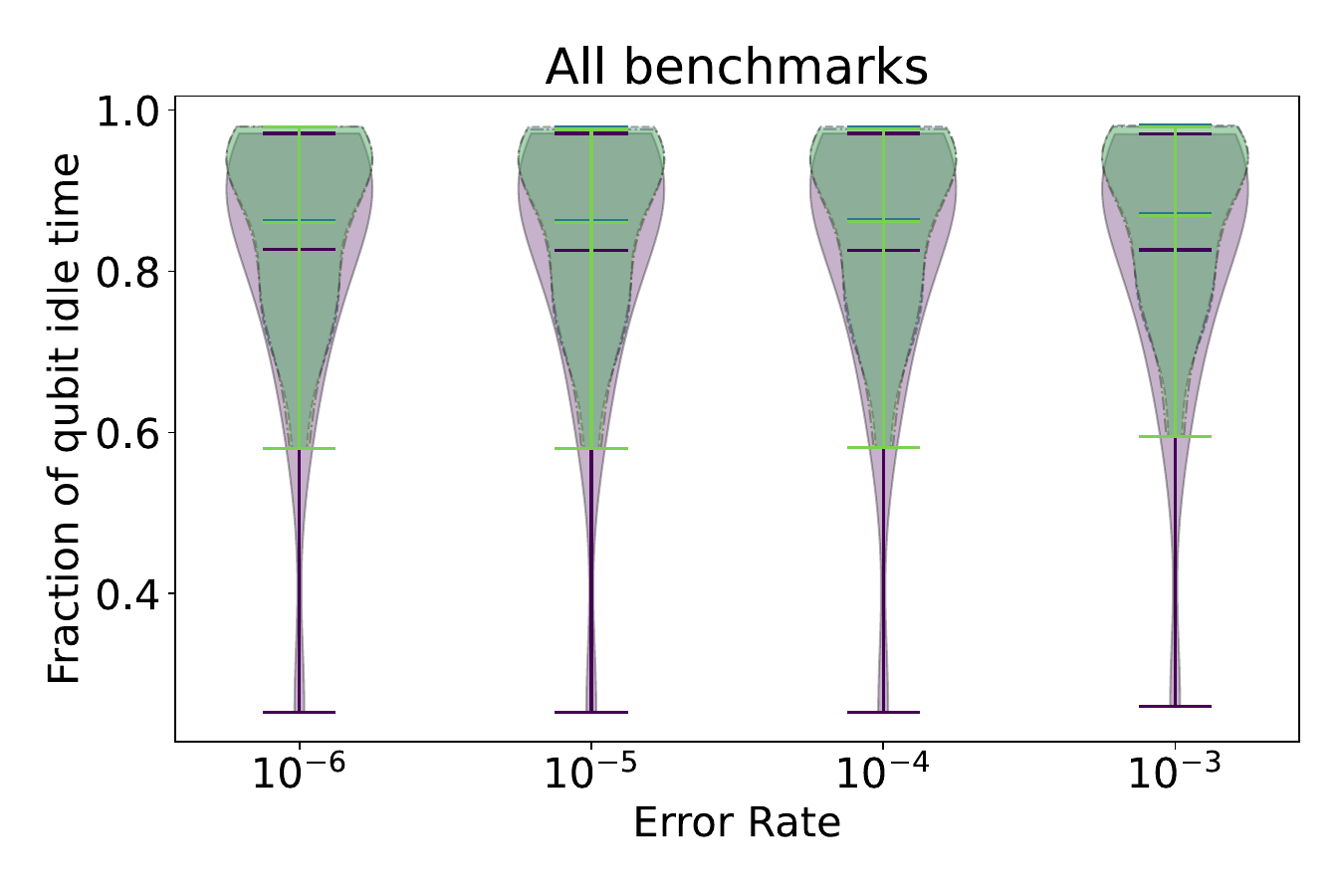}
    \end{subfigure}
    \caption{Sensitivity of different schedulers to varying the physical qubit error rate}
    \label{fig:sensitivity_p}
\end{figure*}

We evaluate \acronym{} against the baseline schemes for a large number of different code distances (Figure~\ref{fig:sensitivity_d}), physical qubit error rates (Figure~\ref{fig:sensitivity_p}) and also analyze how performance for \acronym{} is affected with different $k$ on changing p and $d$ (Figure~\ref{fig:sensitivity_k}). We report the execution times and the fraction of time each data qubit remains idle for all benchmarks, and separately plot an example benchmark. We choose dnn\_n16, gcm\_n13 and qft\_n160 as representative benchmarks. These benchmarks are chosen because of the density of $Rz$ gates. dnn\_n16 has about $6$ $Rz$ gates for each CNOT gate, the largest among all benchmarks. gcm\_n13 has about $2$ $Rz$'s per CNOT and qft\_n160 has an equal number of $Rz$'s and CNOTs. Finally, qft\_n160 has $160$ qubits, therefore, we also show how \acronym{} scales with more qubits.
For Figure~\ref{fig:sensitivity_d} and Figure~\ref{fig:sensitivity_p}, we set $k = 25$, indicated by $RESCQ_{25}$.

\subsubsection{Sensitivity to the Code Distance}

Figure~\ref{fig:sensitivity_d} shows the performance of \acronym{} against the baselines under varying code distance. We fix p $= 10^{-4}$. The execution time improves as $d$ is increased for all benchmarks and all schedulers. %
We perform $d$ rounds of syndrome measurements in every lattice surgery cycle. Therefore, the time taken for one measurement is $1/d$ and the time taken for a single RUS preparation attempt is $O(\alpha/d)$, where $\alpha$ is a function of p. The number of RUS attempts per lattice surgery cycle increases with code distance, improving the execution time. Our dynamic scheme is largely insensitive to the code distance (for both execution and idling time), because in \acronym{}, successful preparation of the $\ket{m_\theta}$ state is almost guaranteed within the first parallelized attempt. Eager preparation of $\ket{m_{2\theta}}$ ensures we do not stall the data qubit in the case of injection failure.%

\subsubsection{Sensitivity to the Physical Qubit Error Rate}

We plot the performance of the schedulers relative to different physical qubit error rates in Figure~\ref{fig:sensitivity_p}. Unlike the sensitivity to verying code distances, all schemes are relatively insensitive to verying physical qubit error rates. Smaller physical error rates decrease the expected number of RUS attempts needed to obtain the $\ket{m_\theta}$ state, but this decrease is relatively insignificant.%

\begin{figure*}
    \centering
    \begin{subfigure}{0.24\textwidth}
        \centering
        \resizebox{\linewidth}{!}{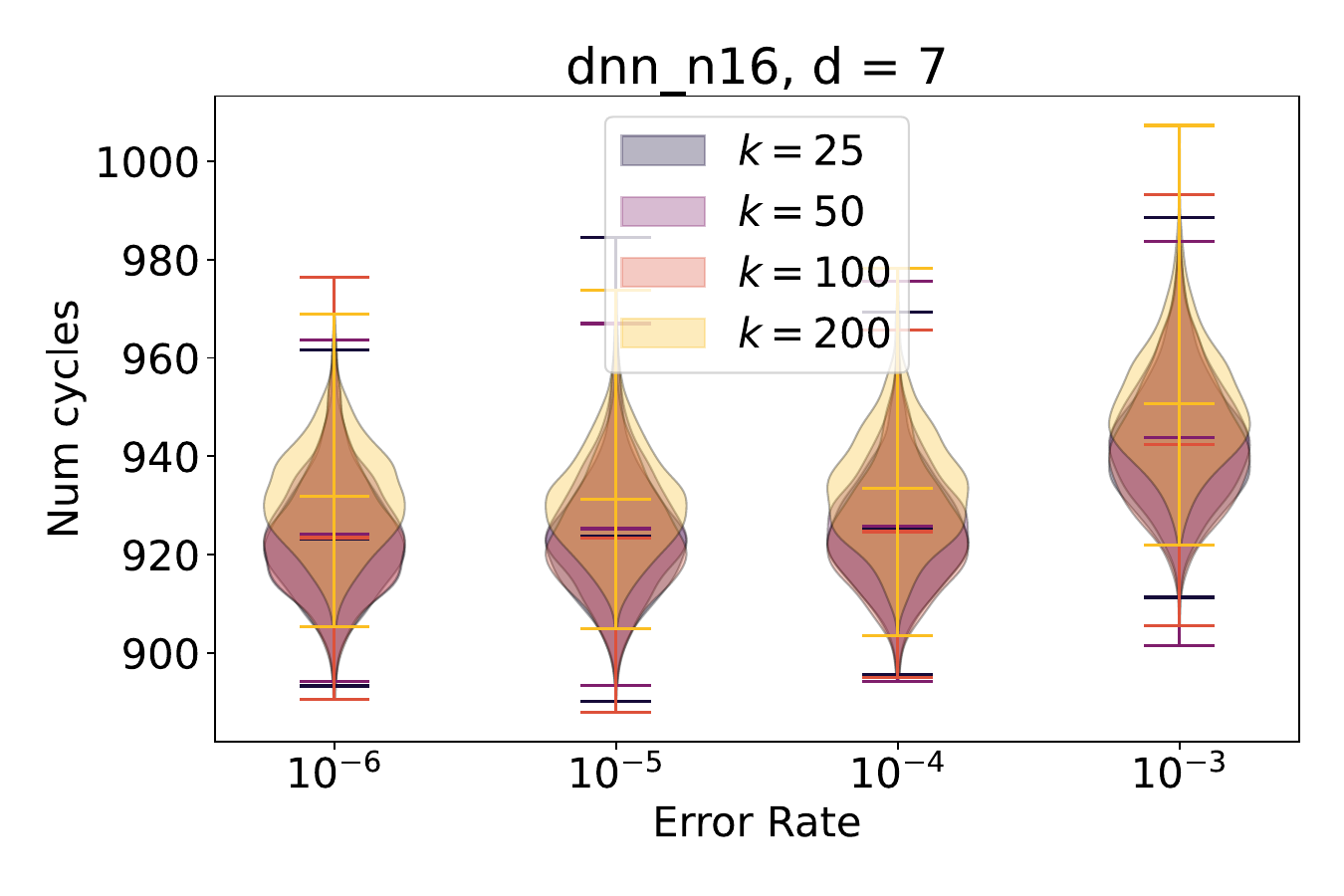}
    \end{subfigure}
    \begin{subfigure}{0.24\textwidth}
        \centering
        \resizebox{\linewidth}{!}{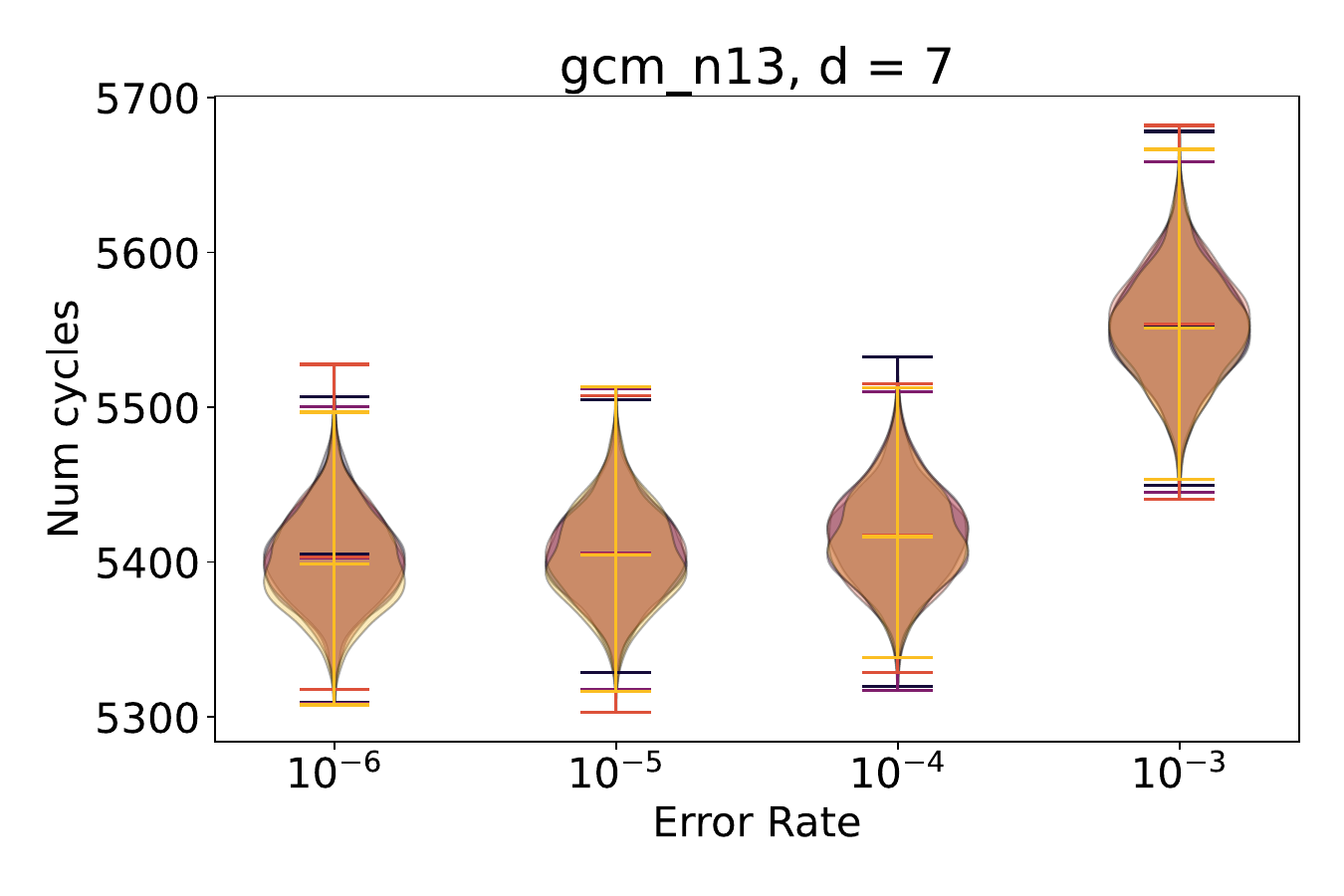}
    \end{subfigure}
    \begin{subfigure}{0.24\textwidth}
        \centering
        \resizebox{\linewidth}{!}{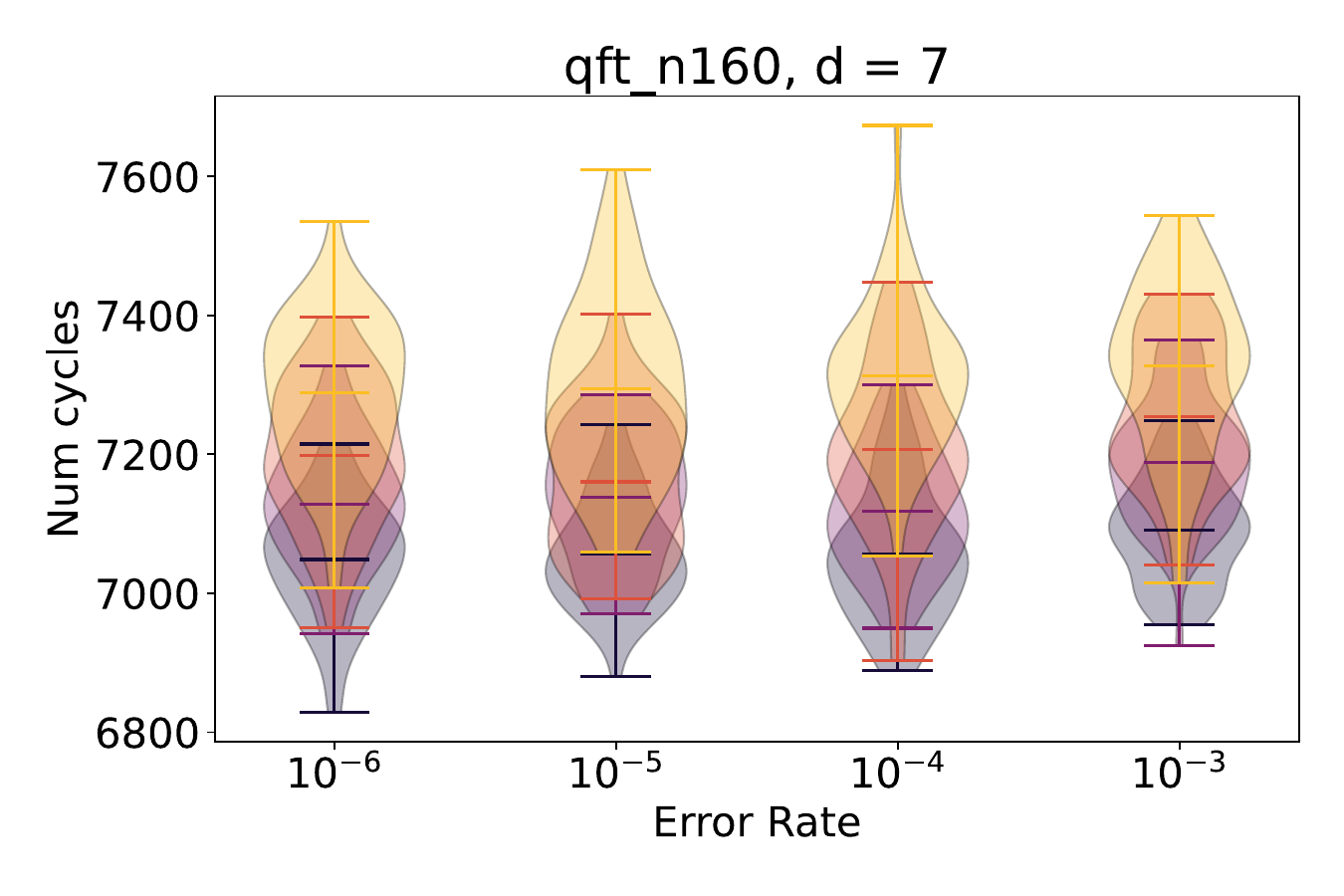}
    \end{subfigure}
    \begin{subfigure}{0.24\textwidth}
        \centering
        \resizebox{\linewidth}{!}{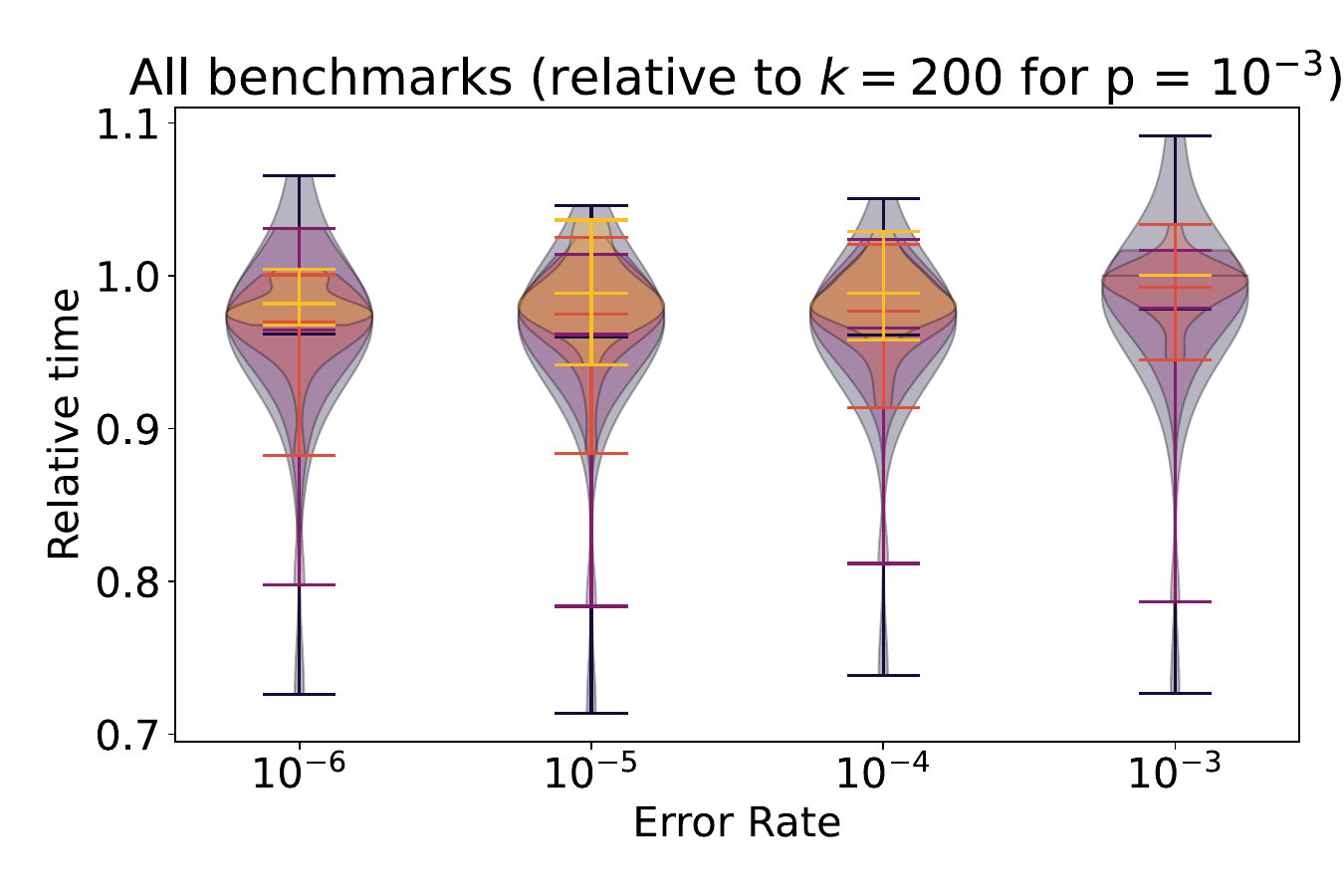}
    \end{subfigure}
    \begin{subfigure}{0.24\textwidth}
        \centering
        \resizebox{\linewidth}{!}{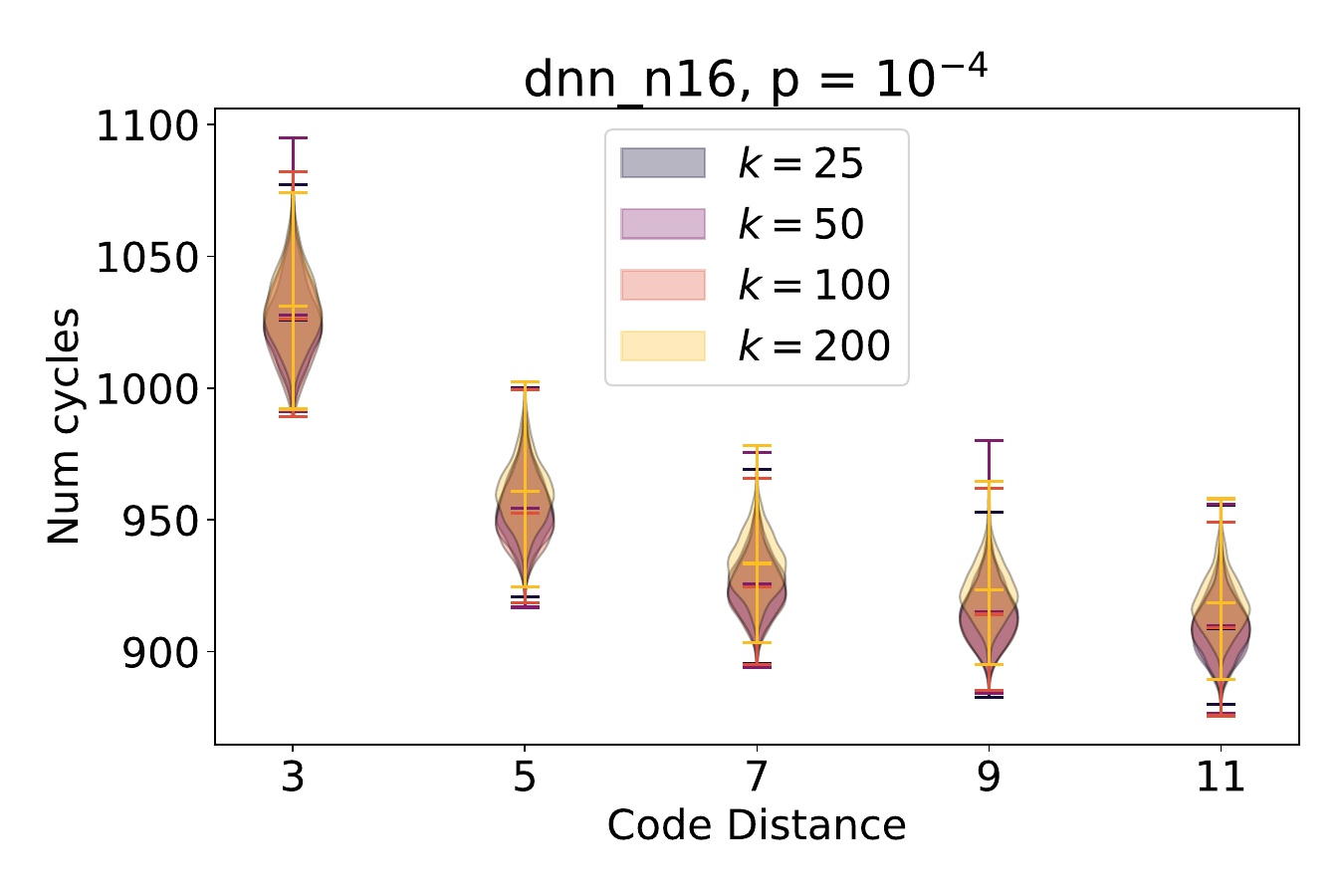}
    \end{subfigure}
    \begin{subfigure}{0.24\textwidth}
        \centering
        \resizebox{\linewidth}{!}{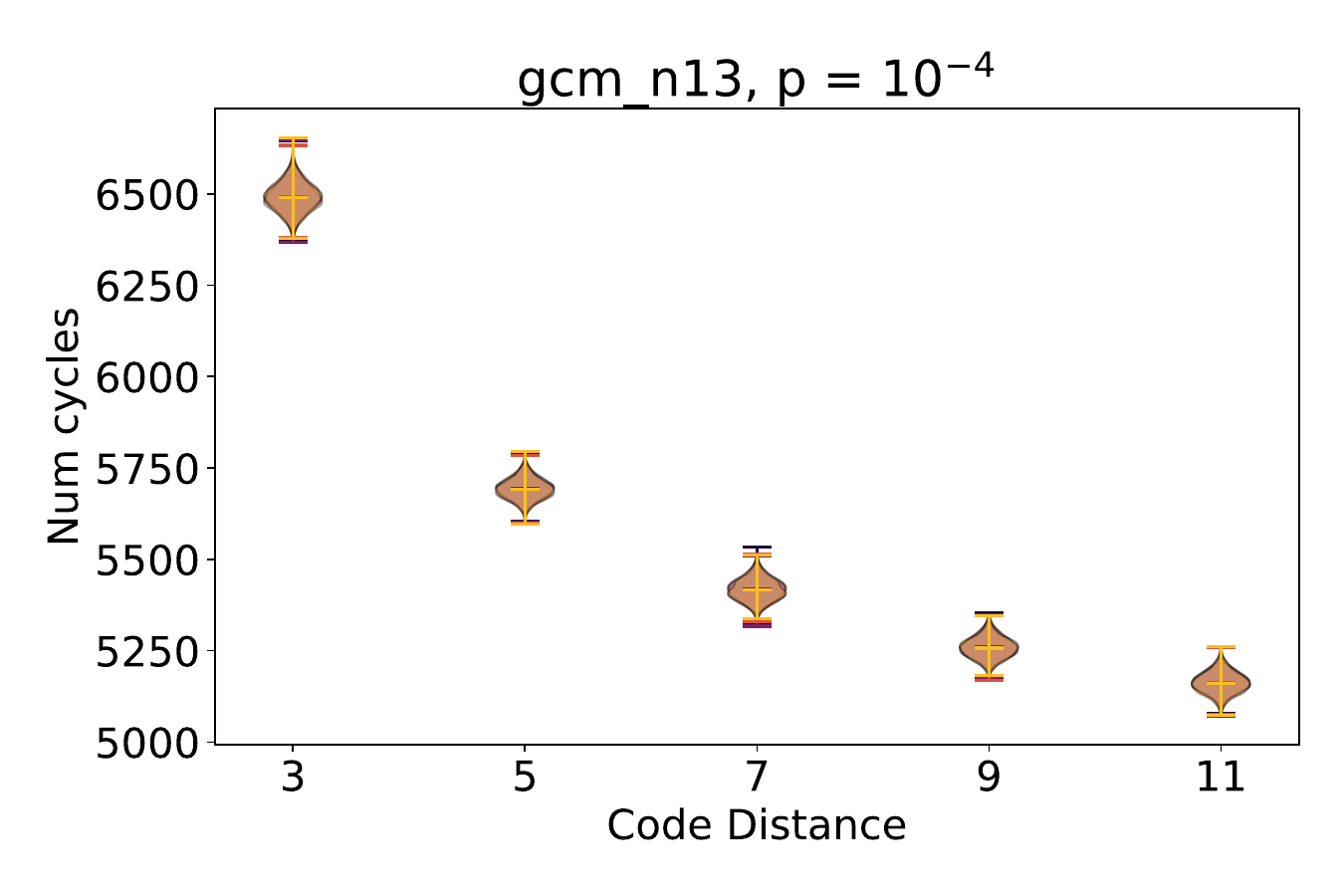}
    \end{subfigure}
    \begin{subfigure}{0.24\textwidth}
        \centering
        \resizebox{\linewidth}{!}{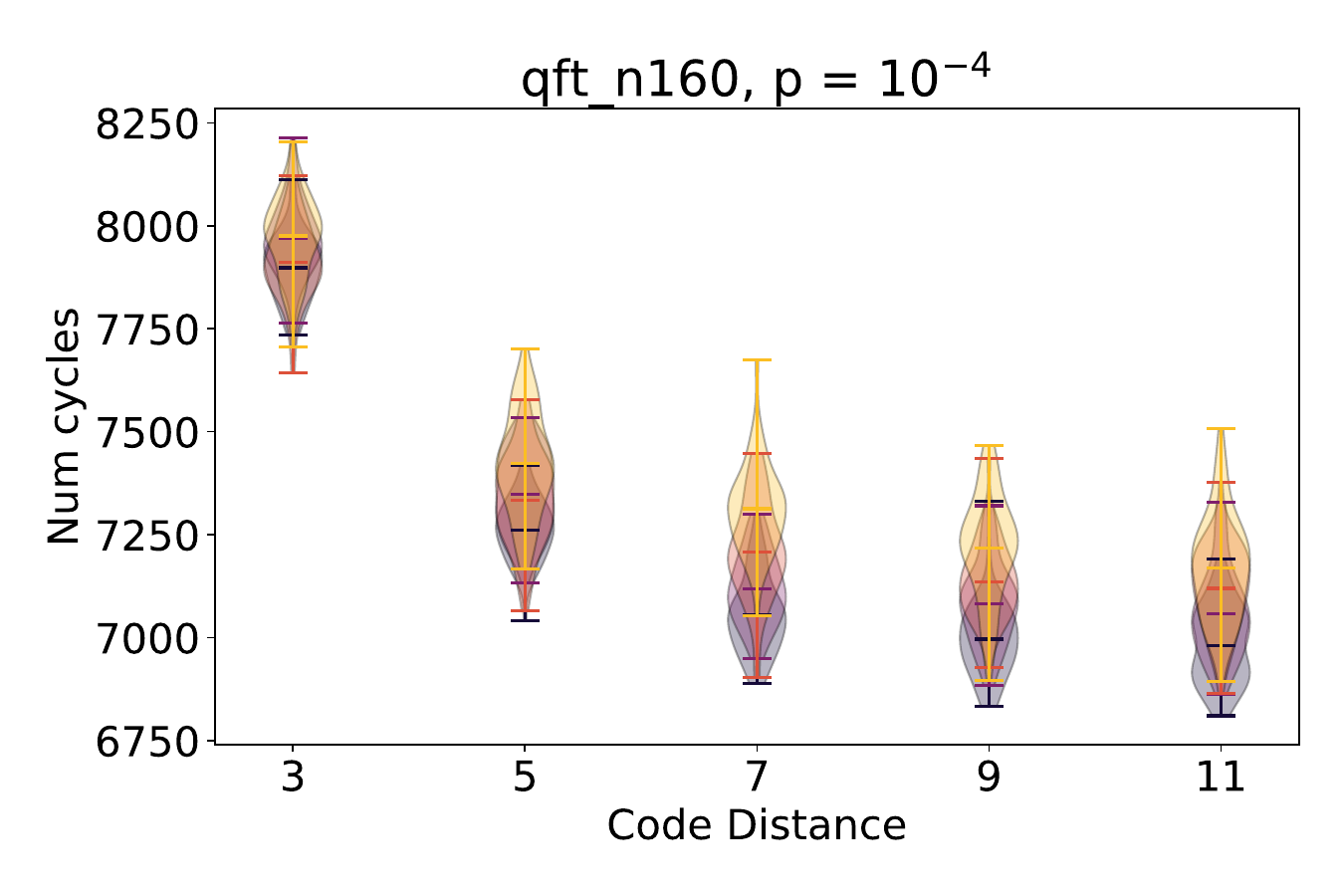}
    \end{subfigure}
    \begin{subfigure}{0.24\textwidth}
        \centering
        \resizebox{\linewidth}{!}{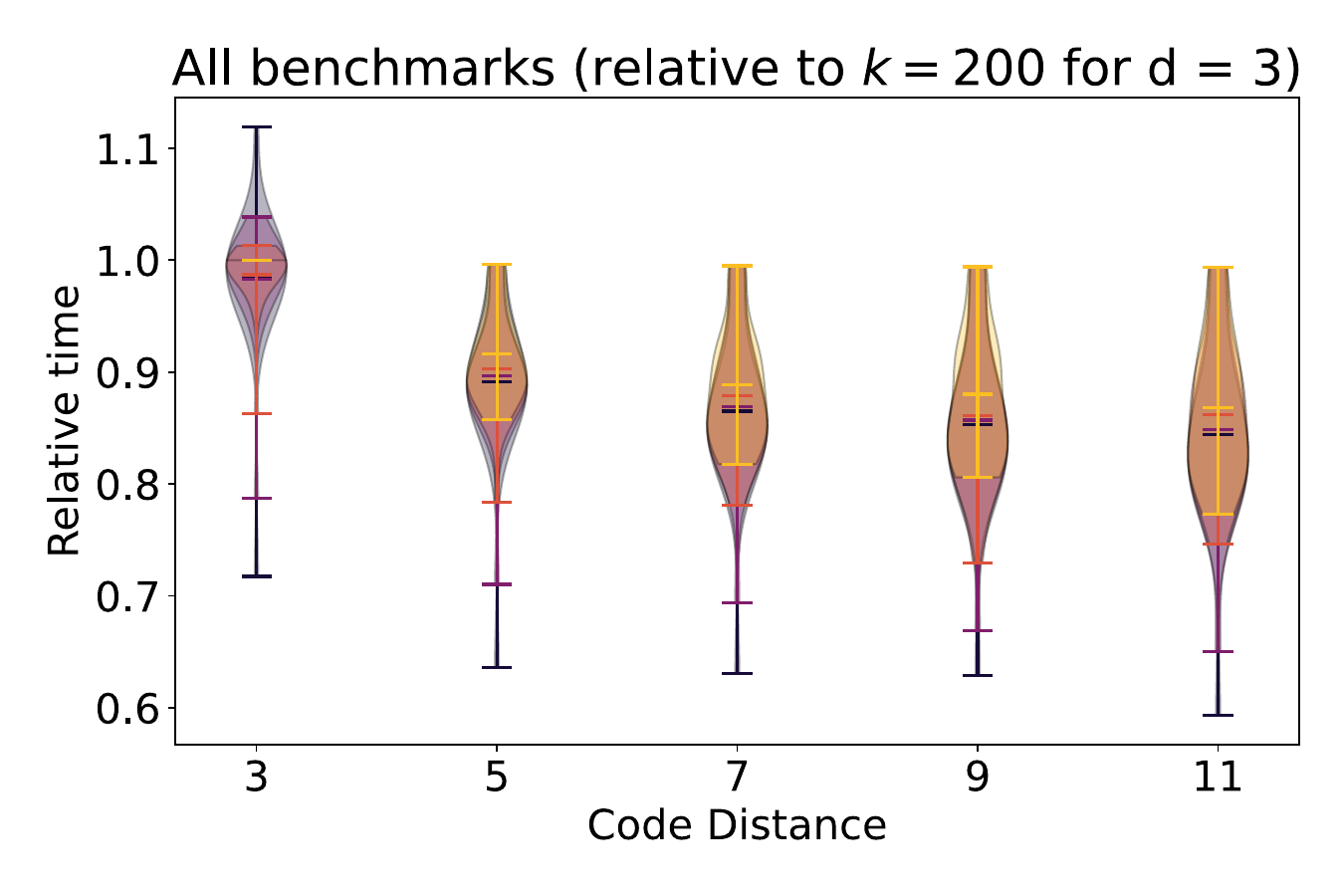}
    \end{subfigure}
    \caption{Sensitivity of \acronym{} to varying $d$ and p for different $k$ (frequency of MST computation)}
    \label{fig:sensitivity_k}
\end{figure*}

\subsubsection{Sensitivity to MST Computation Frequency}\label{section:evaluationk}
Figure~\ref{fig:sensitivity_k} shows the sensitivity of \acronym{} to the code distance and physical qubit error rate for $k\in\{25, 50, 100, 200\}$. For all benchmarks, the sensitivity to both p and $d$ is mostly unaffected by $k$. The performance is optimal when $k = 25$ and deteriorates negligibly as $k$ increases. %
Computing the MST less frequently does not directly translate to a decrease in performance since we still have to choose between the $16$ different routing in Algorithm~\ref{algorithm:cnotexecution}, distributing the routing congestion when we have multiple CNOTs between the same pairs of qubits. In addition to this, since the same MST is used for $k$ consecutive cycles, it is likely that some set of edges with low activity in the past will be used for multiple CNOT gates in the next $k$ cycles. However, this leads to the overused edges having a higher activity and thus being absent from the updated MST, balancing out the load across all ancilla qubits. This maximises performance gains even for higher values of $k$.

\begin{figure*}
    \centering
    \begin{subfigure}{0.24\textwidth}
        \centering
        \resizebox{\linewidth}{!}{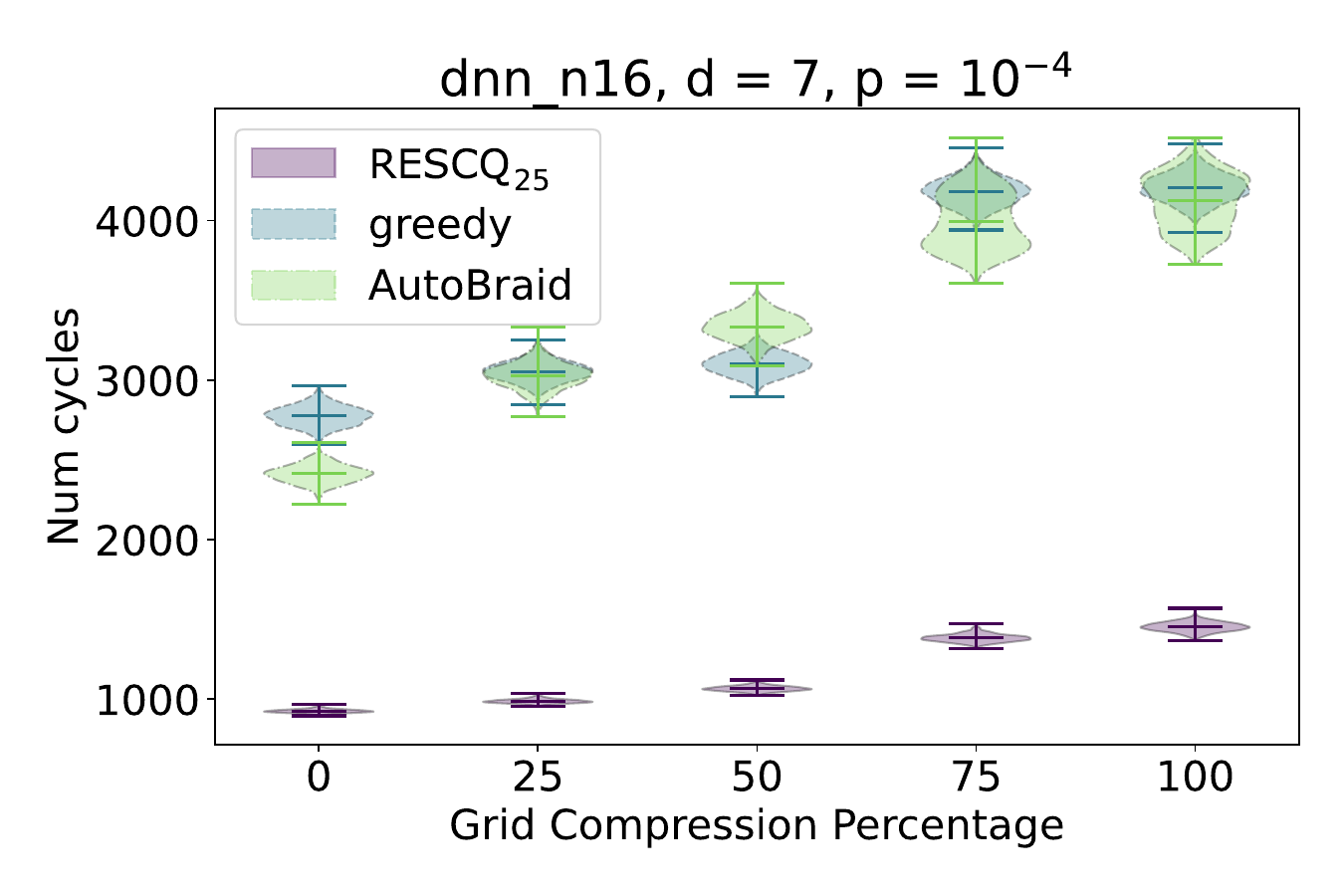}
    \end{subfigure}
    \begin{subfigure}{0.24\textwidth}
        \centering
        \resizebox{\linewidth}{!}{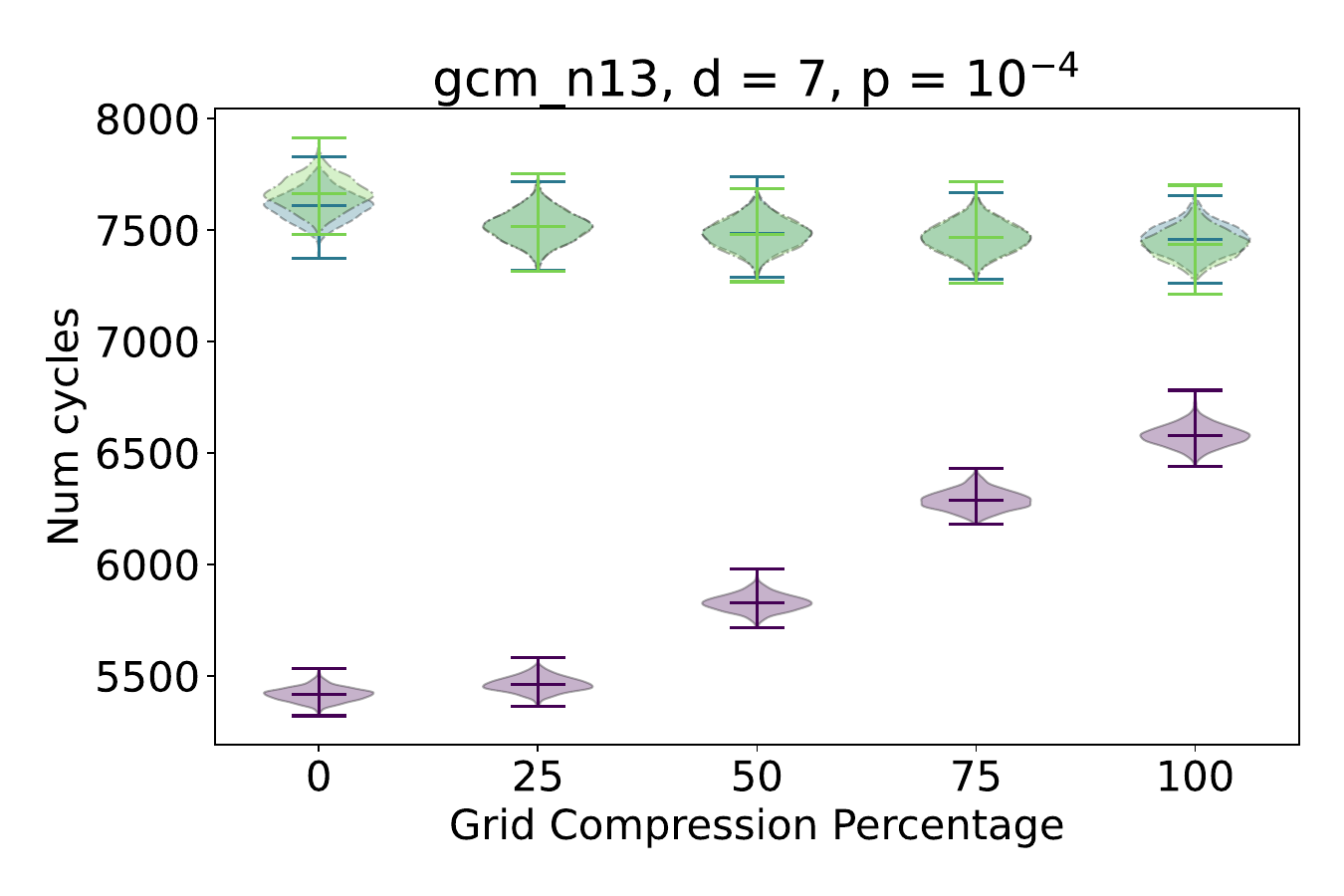}
    \end{subfigure}
    \begin{subfigure}{0.24\textwidth}
        \centering
        \resizebox{\linewidth}{!}{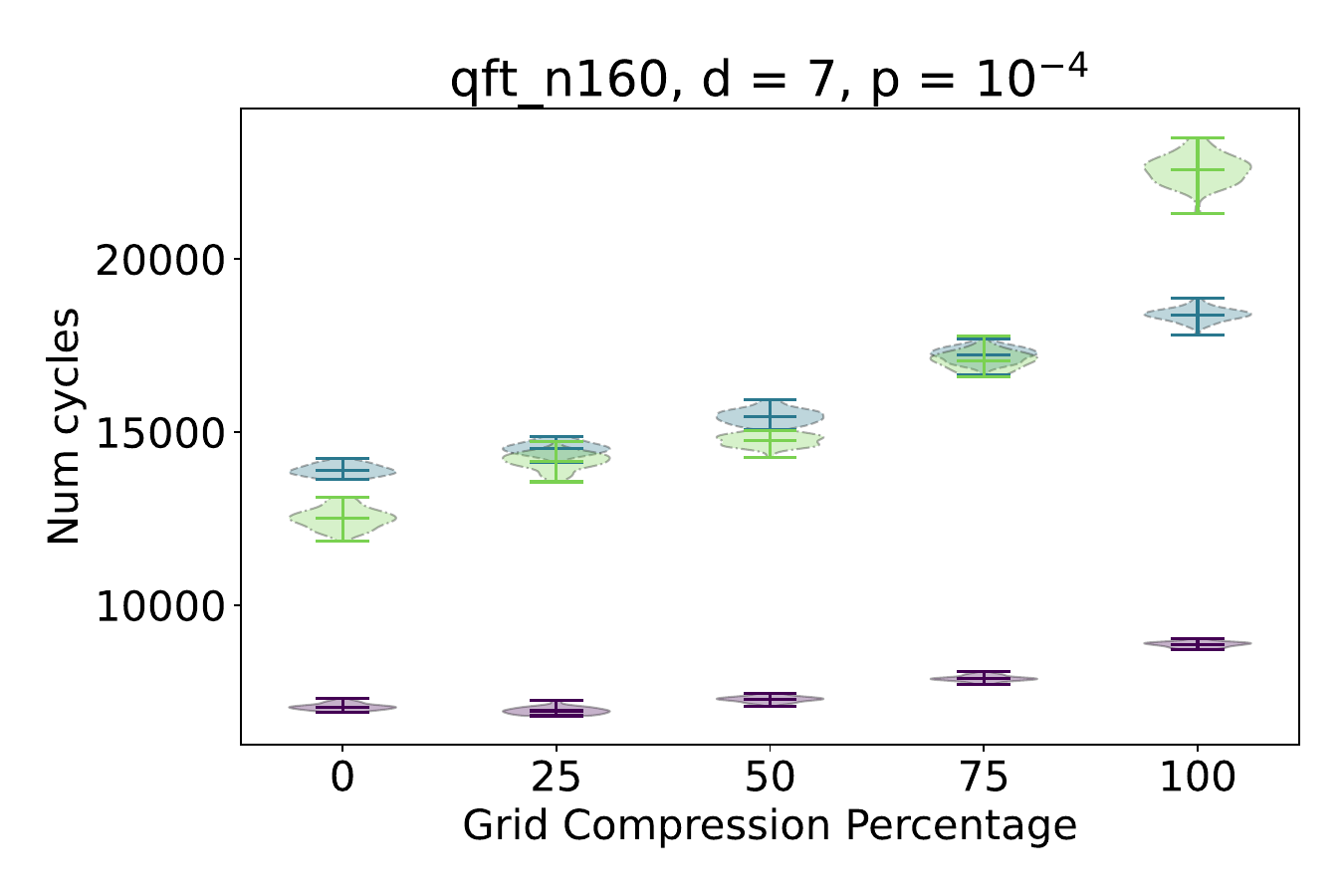}
    \end{subfigure}
    \begin{subfigure}{0.24\textwidth}
        \centering
        \resizebox{\linewidth}{!}{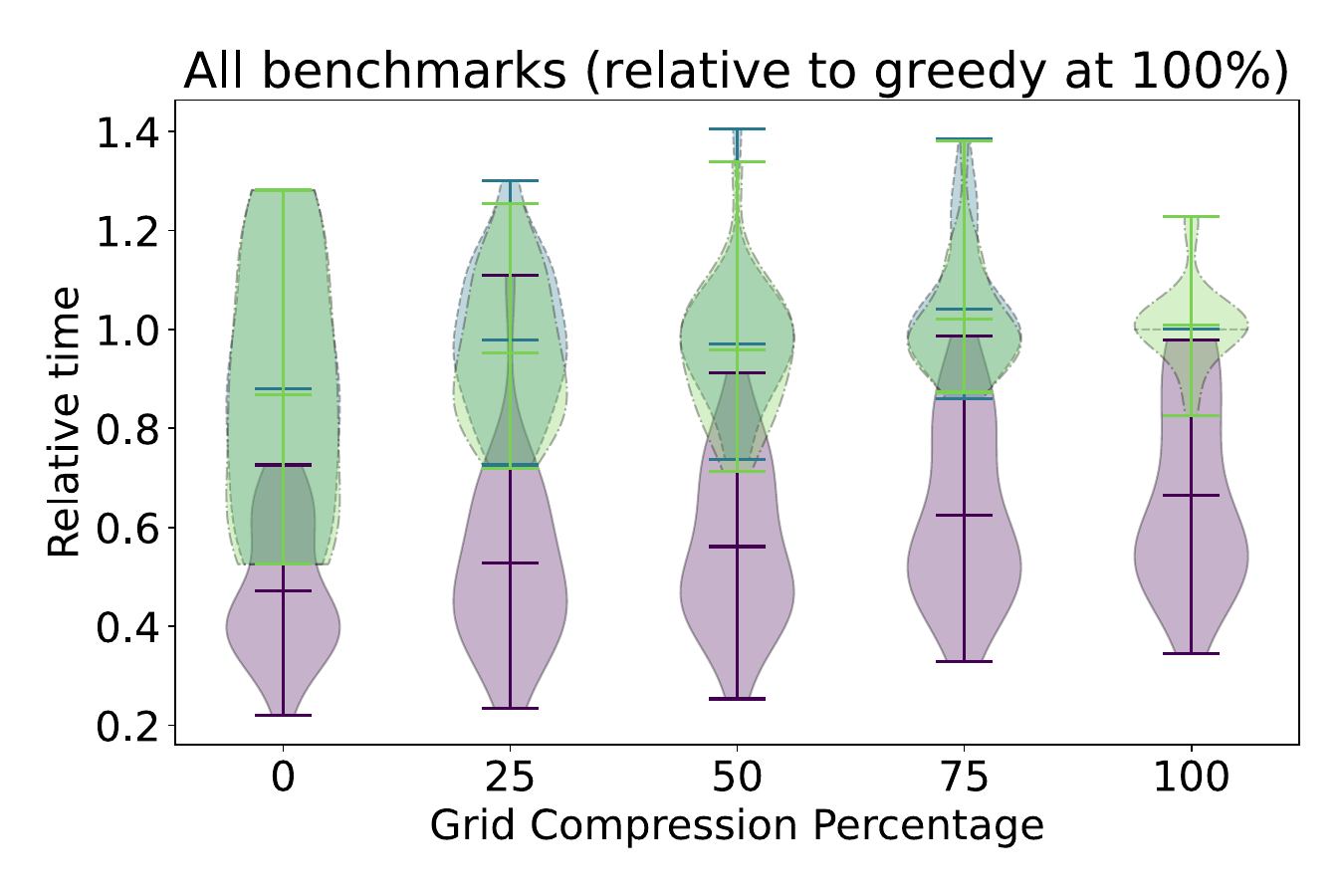}
    \end{subfigure}
    \caption{Sensitivity of different schedulers to the ancilla availability (grid compression)}
    \label{fig:sensitivity_compression}
\end{figure*}

\subsection{Hardware-Software Co-Design}
\begin{figure}
    \centering
    \resizebox{\columnwidth}{!}{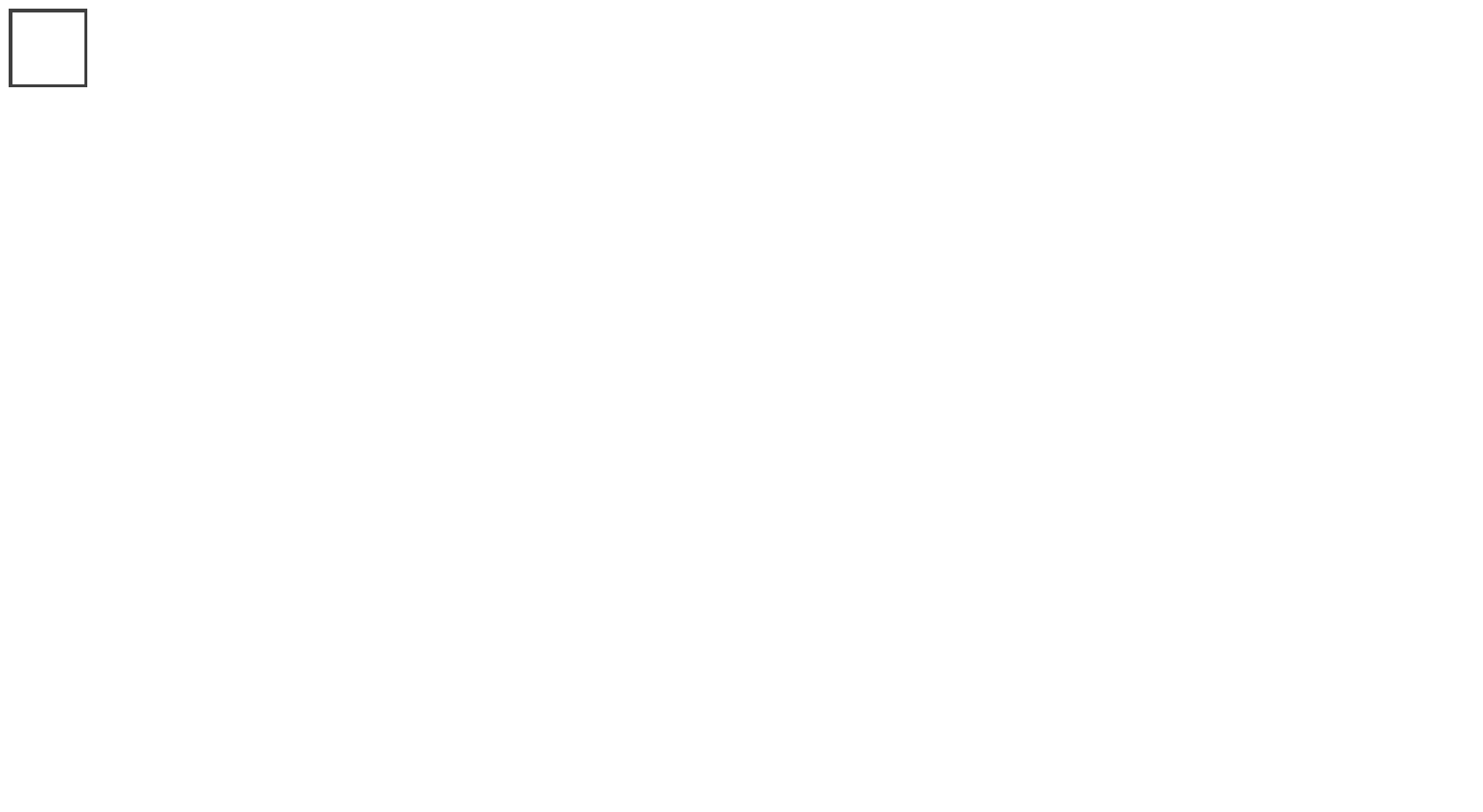}
    \caption{Grids of $8$ data qubits (grey) at different compressions. Increasing compression limits the availability of ancilla qubits (white) per data qubit (grey), causing congestion. Results corresponding to different compressions are shown in Figure~\ref{fig:sensitivity_compression}.}
    \label{fig:grid_compression}
\end{figure}

As discussed in Section~\ref{section:relatedwork}, the default STAR grid uses $2\times 2$ blocks for each data qubit, and thus there are $3$ ancilla qubits for each data qubit. This space overhead is significant, especially in the near-term FTQC regime wherein the availability of physical qubits is limited. We explore the trade-off for reducing the number of ancillas in the grid by incrementally \textit{compressing} the grid. We choose a data qubit at random and modifying its patch from a $2\times 2$ to a $2\times 1$ patch until all data qubits have been \textit{compressed} (while still ensuring the grid remains connected). We then evaluate the schedulers between $0\%$ and $100\%$ compression, which correspond to 3 ancilla per data qubit and a single ancilla per data qubit, respectively. An example grid at different compressions is shown in Figure~\ref{fig:grid_compression}. We report the results obtained in Figure~\ref{fig:sensitivity_compression}.%

Grid compression is significantly detrimental to the performance of the baseline schemes. While the reduced ancilla availability is on its own impactful, the baseline schemes are further crippled since they do not account for congestion or the introduction of edge-rotations while routing. This can sometimes cancel out the effects of resource contention, e.g., in benchmarks like gcm\_n13, when the grid is compressed, the probability of the baseline schemes to choose paths that do not require edge-rotation gates becomes more likely, reducing the average per-gate execution time. %
On the other hand, for the dnn\_n16 and qft\_n160 benchmarks (and for almost all of the other benchmarks), the increased stalls due to congestion dominates, which leads to increased execution times on grid compression. This congestion problem is mitigated in the case of \acronym{} due to the efficient schedules generated by the queues. The dynamic nature of \acronym{} is able to alleviate a large portion of stalls caused by reduced ancilla resources, only experiencing a relatively minor increase in execution time with grid compression.

\subsection{Overheads of Classical Computation}

\subsubsection{MST Computation Complexity}\label{section:mstcomplexityanalysis}
The MST is computed asynchronously and does not stall quantum execution (Figure~\ref{fig:mst_timeline}). Computing the MST on an arbitrary graph with $n$ vertices takes $O(n\log n)$ time \cite{clrs2009algorithms}. However, we only deal with a grid architecture and build the graph from ancilla qubits, further simplifying the structure. This simplification makes it easier for us to instead update the MST rather than recomputing the entire MST. We have $4$ cases for every edge $a_{ij}$ between ancilla qubits $i$ and $j$, only two of which require MST updates:
\begin{enumerate}
    \item $a_{ij}$ is not on the MST and its activity decreases -- we insert this edge to the MST and remove the maximum weight edge from the formed cycle. This will take $O(1)$ time since all cycles are of $O(1)$ size on a 2D grid.
    \item \sloppy $a_{ij}$ is on the MST and its activity increases -- we remove this MST edge and find the minimal weight edge between the two forests. This step will take $O(\max(rows, columns))$ since the edge removal will be along the horizontal or vertical direction.
\end{enumerate}
Therefore, the time taken to update the MST for each edge update will be $O(\max(rows, columns))$. Since we have a square grid of $n$ qubits and compute the MST every $k$ cycles, there are $O(k)$ edge updates and the overall complexity simplifies to $O(k\sqrt{n})$. Based on our simulations on modern processors, it takes $\approx 92\mu s$ for a $100\times 100$ grid and $\approx 330\mu s$ for a $1000\times 1000$ grid when $k = 200$.

\subsubsection{Path-Finding Algorithm Complexity}
The path-finding algorithm (Algorithm~\ref{algorithm:cnotexecution}) uses  and maintains a pointer to the lastest MST. All CNOT execution paths for the next $k$ cycles are computed from the latest MST and stored since they are independent of ancilla activity. Similarly, \texttt{startTime} is updated dynamically whenever any ancilla on the path is used in some gate. This leads to only $O(1)$ computation when the CNOT gate is added to the queues: deciding the best path by comparing \texttt{startTime} of the pre-computed paths.

\section{Conclusion}
Fault-tolerant quantum architectures are still far from being realized on physical systems. Various error correcting codes have been proposed, such as the surface code. The surface code architecture does not natively support non-Clifford gates and thus requires specialized support for such operations. Continuous rotation angle architectures have been proposed that allow for localized, low-latency ancilla preparation in the required states. However, these proposals have lacked realtime ancilla allocation support, incurring unecessary time overhead. We tackle this problem and propose \acronym{}, a realtime scheduler that utilizes variable ancilla availability and activity to perform efficient allocation of different gate operations in parallel, thus minimizing total program execution time. We improve significantly over the baseline proposals while simultaneously minimizing classical overhead for realtime recomputation.

\appendix
\section{Analysis of Repeat-Until-Success}\label{sec:appendix_rus}
In this appendix section, we discuss the success probability and the expected number of rounds needed for executing a single $Rz(\theta)$ gate. We will also compare the execution time with executing the same gate in the Clifford + T compilation.

\subsection{Preparation of \texorpdfstring{$\ket{m_\theta}$}{|m\_theta>}}
As described in \cite{akahoshi2023partially}, which we summarized in Section~\ref{section:relatedwork}, the $\ket{m_\theta}$ state is initially prepared in a [[4, 1, 1, 2]] error-detection subsystem code (Figure~\ref{fig:parallel_prepare}). On successful preparation, the code is expanded to the entire logical qubit (of distance $d$) and error-detection measurements are performed. In total, we have two rounds of error-detection, and both should succeed for successful state preparation. This is also called post-selection in literature.
\par We can embed multiple [[4, 1, 1, 2]] subsystem codes within an ancilla qubit (Figure~\ref{fig:parallel_prepare}), scaling with increasing code distance as $(d^2 - 1)/2$. Therefore, we prepare $\ket{m_\theta}$ in parallel on all $O(d^2)$ embedded subsystem codes. When any of the parallel preparations (i.e., the first error-detection round) succeed, we expand and perform the second error-detection round, destroying other parallel subsystem attempts. If the second round fails, we have to start all over by preparing $\ket{m_\theta}$ in multiple subsystem codes, and expanding again. We call both steps of error-detection along with the expansion as a single `attempt'.

\begin{figure}
    \centering
    \begin{subfigure}{\linewidth}
        \centering
        \resizebox{\linewidth}{!}{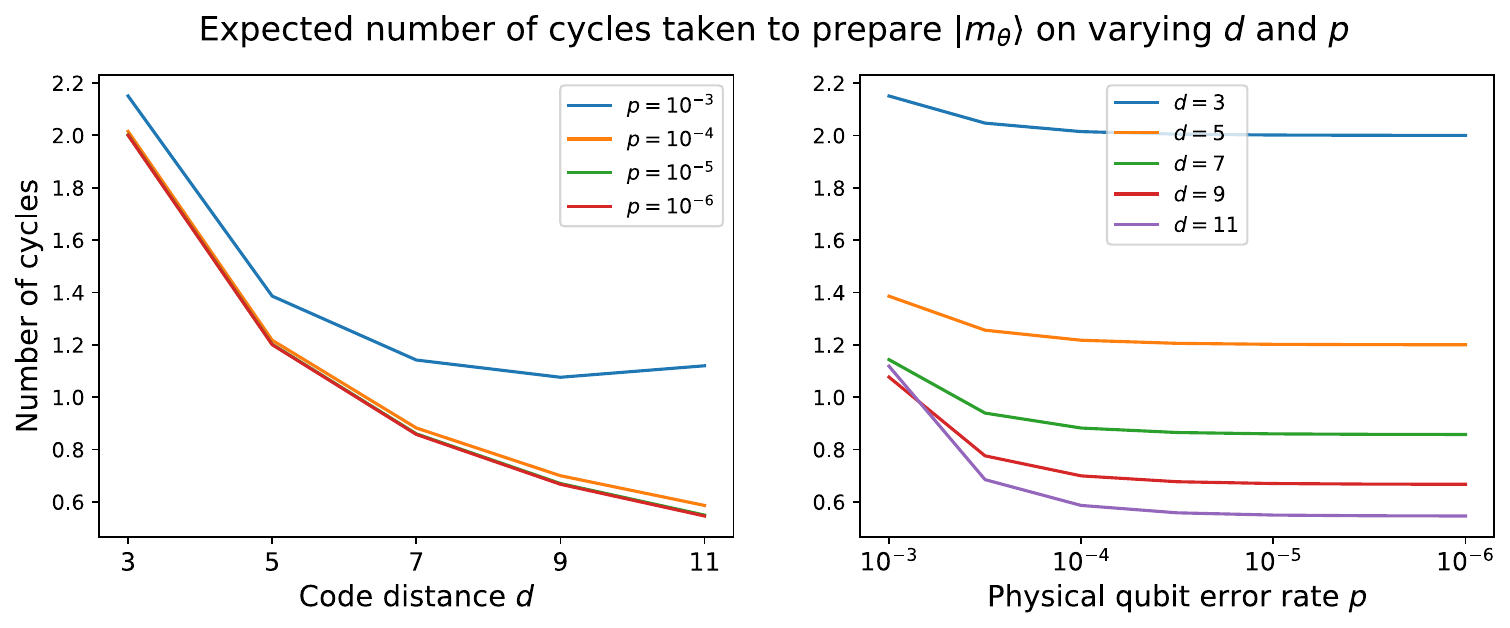}
        \label{fig:cycles_prob_analysis}
    \end{subfigure}
    \centering
    \begin{subfigure}{\linewidth}
        \centering
        \resizebox{\linewidth}{!}{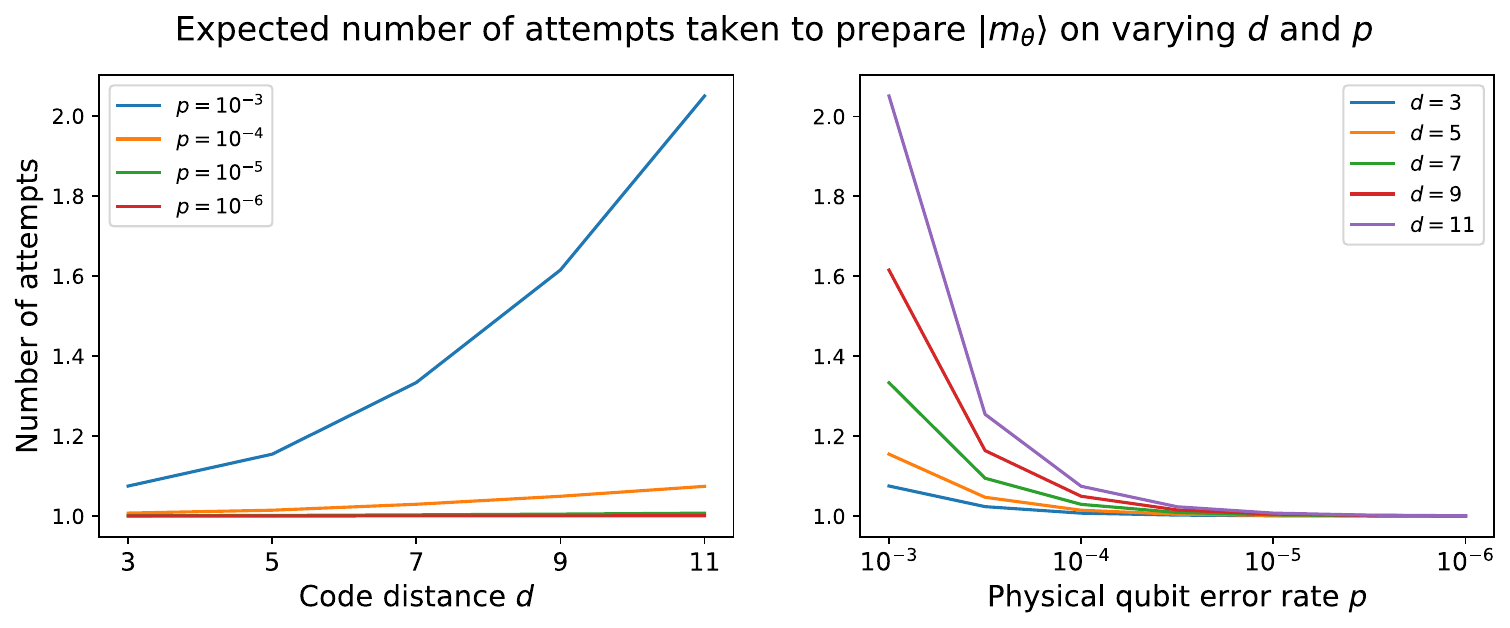}
        \label{fig:attempts_prob_analysis}
    \end{subfigure}
    \caption{Plots depicting the effect of different code distances ($d$) and physical qubit error rates ($p$) on the expected cycles and attempts it takes to prepare the $\ket{m_\theta}$ state.}
    \label{fig:prob_analysis}
\end{figure}

\par Figure~\ref{fig:prob_analysis} shows the effect of increasing code distance or decreasing the physical qubit error rate on the number of cycles and attempts taken to prepare the $\ket{m_\theta}$ state. As intuition suggests, the expected cycle time decreases as $d$ is increased or $p$ is decreased since both changes decrease the probability of errors in logical qubits.
\par However, something to note is the increase in the expected number of attempts as the code distance is increased. An increase in the the code distance increases the number of syndrome bits of the second error-detection step, increasing the fidelity of the prepared $\ket{m_\theta}$ at the cost of increased attempts. Since the cycle time is proportional to $1/d$, the increase in the expected number of attempts is not reflected in the expected number of cycles.
\par Even though the expected attempts are close to $1$ for most combinations of $d$ and $p$, the overall protocol is still non-deterministic since injection of $\ket{m_\theta}$ has a fixed success rate of $50\%$ (Section~\ref{section:rzexecution}). However, a faster preparation helps reduce program execution time by minimising the time each data qubit spends idling while waiting for preparation of $\ket{m_\theta}$. The ideas proposed in \acronym{} (Section~\ref{section:ancillamanage}) further alleviate the stalls by parallel and eager preparation.

\subsection{$\ket{m_\theta}$ Injection vs \texorpdfstring{$T$}{T} Injection}
As we have shown in Equation~\ref{eq:exp_inj_attempts}, it takes $2$ `steps' (in expectation) for the execution of each $Rz(\theta)$ gate. We define each `step' as a successful preparation `attempt' followed by a ZZ or CNOT injection (Figure~\ref{fig:injstrats}). Assuming a baseline scheduling policy and worst case execution times from Figure~\ref{fig:prob_analysis}, this requires an expected $2\text{ steps} \times(\text{preparation} + \text{injection cycles}) = 2\times (2.2 + 2) = 8.4$ cycles.
\par In contrast, assuming a single $T$ factory and using the analysis provided in \cite{litinski2019game}, it requires $11$ cycles (with $99.9\%$ probability error-detection succeeds) to prepare a $T$ state. Since factories are typically present at the boundaries of the grid, we need to route the $T$ state to the data qubit for injection. Since we will have concurrent operations going on, such as CNOT gates and $T$ injection onto other data qubits, this will impose routing constraints (Section~\ref{section:cnotexecution}) and factory contention (Section~\ref{section:ancillamanage}). Assuming that a valid path always exists for $T$ injection and we have a dedicated factory for each data qubit (both assumptions are impractical due to immense space overheads), we require between $2-13$ cycles for the execution of a single $T$ gate ($2$ cycles for injection + $0-11$ cycles for $T$ preparation depending on when the $T$ factory started preparation). Since the execution of a single $Rz(\theta)$ gate requires more than $100\times$ $T$ gates (along with additional $H$ and $S$ gates)\cite{camp1}, we require $200-1300$ cycles for the execution of a single $Rz(\theta)$ gate.
\par Therefore, even with some simplifying assumptions, we have an overhead of more than $20-150\times$ for the execution of a single $Rz(\theta)$ gate in the Clifford+T gate set.

\section{Artifact Appendix}

\subsection{Abstract}

This artifact contains details on the source code for simulating different scheduling policies (\acronym{}, greedy, AutoBraid) on the variety of benchmarks used in this work. This artifact also contains details about configuring the simulator to test the schedulers on different parameters, how to run benchmarks and generate the plots presented in this work.

\subsection{Artifact check-list (meta-information)}

{\small
\begin{itemize}
  \item {\bf Algorithm: }\acronym{} and other baselines
  \item {\bf Program: }\texttt{sim} (benchmarks used\cite{li2022qasmbench, tomesh2022supermarq} for plots are processed and available)
  \item {\bf Compilation: }Using \texttt{gcc} or \texttt{clang} via \texttt{cmake} (\texttt{boost} is required)
  \item {\bf Metrics: }Total execution time, time taken for each gate execution
  \item {\bf Output: }Log files and plots containing violin plots of sensitivity analysis, histograms of execution metrics and heatmaps of grid activity
  \item {\bf Experiments: }Using \texttt{cmake} (tested on minimum version 3.22) and a config file (probability computations are seeded and multiple runs are executed to account for spread and non-determinism)
  \item {\bf How much disk space required (approximately)?: }About 120 MB for a single choice of $d, p$ totalled across all benchmarks (1-2 GB to obtain logs for all sensitivity plots)
  \item {\bf How much time is needed to prepare workflow (approximately)?: }Under 5 minutes
  \item {\bf How much time is needed to complete experiments (approximately)?: }About 0.5-1 hour to generate all plots (when run using 16 threads)
  \item {\bf Publicly available?: }Yes
  \item {\bf Code licenses (if publicly available)?: }MIT License
  \item {\bf Archived (provide DOI)?: }\href{https://doi.org/10.5281/zenodo.14769159}{10.5281/zenodo.14769159}
\end{itemize}
}

\subsection{Description}

\subsubsection{How to access}

Download or clone the simulator files available at \href{https://doi.org/10.5281/zenodo.14769159}{10.5281/zenodo.14769159}.

\subsubsection{Software dependencies}
\sloppy Compilation is done using \texttt{cmake} (tested on minimum version v3.22); \texttt{boost} is required for compilation. Instructions to install \texttt{boost} are available in the \texttt{README}. Python packages used for postprocessing (generating plots from log files) are listed in \texttt{requirements.txt} file inside the \texttt{postprocess} directory.

\subsubsection{Data sets}
Benchmarks\cite{li2022qasmbench, tomesh2022supermarq} used for testing and generating plots have been compiled to the Clifford+Rz gate set using Qiskit\cite{Qiskit} and are available as ready to use in the \texttt{circuits} directory.

\subsection{Installation}

All installation steps are described in the \texttt{README} of the artifact. Requirements include \texttt{cmake} ($\geq$ v3.22) and the \texttt{boost} library to compile the simulator, and a Python virtual environment to run the postprocessing scripts.

\subsubsection{Basic Test}
After the installation is successful, it can be tested by running the following commands:\\
\texttt{\$ ./scripts/basic\_test.sh}\\
This should create a directory \texttt{outputs} if it did not exist earlier. The logs and some initial plots (generated by \texttt{postprocess.py}) can be found inside the \texttt{outputs/basic\_test} directory.

\subsection{Experiment workflow}
The experiment workflow involves running the simulator with different config files which are then postprocessed using Python scripts in the the \texttt{postprocess} directory. All shell scripts are provided in the \texttt{scripts} directory. We use the \texttt{gen\_configs.sh} script to generate the config files relevant for the plots in the paper. The \texttt{sim} executable is then run on these configurations and are post-processed using \texttt{postprocess.py}. We use \texttt{run\_all.sh} (which calls the scripts \texttt{run\_scheduler.sh} and \texttt{run\_postprocess.sh}) to run the simulator in parallel on different configurations by creating sub-processes. The maximum number of these sub-processes is determined by the \texttt{MAX\_PROCESSES} argument required by this script. Note that it takes about 0.5-1 hour when using 16 threads.

\subsection{Evaluation and expected results}

To generate all logs and plots that are used in the paper, run the following:\\
\texttt{\$ ./scripts/run\_all.sh <MAX\_PROCESSES>}\\
\texttt{\$ python postprocess/final.py outputs plots}\\
\texttt{\$ python postprocess/histograms.py outputs plots}\\
\texttt{\$ python postprocess/sensitivity.py outputs plots}

Each of the three Python scripts generate plots for Figure~\ref{fig:plot}, Figure~\ref{fig:histograms}, and Figures~\ref{fig:sensitivity_d},~\ref{fig:sensitivity_p},~\ref{fig:sensitivity_k},~\ref{fig:sensitivity_compression}, respectively. \texttt{plots/execution/50\_4\_7.svg} is used to generate Figure~\ref{fig:plot}. \texttt{plots/sensitivity\_0/autobraid\_4\_7.svg} and \texttt{plots/sensitivity\_0/rescq\_4\_7.svg} are used to generate Figure~\ref{fig:histograms}.
\par For the sensitivity figures, the format used to describe the figures in the \texttt{plots/senstivity\_0} directory is \texttt{benchmark} followed by \texttt{idling} if the plot is for the qubit idling time, \texttt{mst} if the plot is for the sensitivity to $k$ (MST computation frequency) or simply \texttt{benchmark} if the plot is for execution times. This is followed by either \texttt{d<number>} indicating the code distance $d$ or \texttt{p<number>} indicating the physical error rate $p = 10\textsuperscript{-number}$. This is the parameter that is fixed for the plot and the other parameter is varied for the sesntivity plots. For example, \texttt{gcm\_n13\_p4.svg} is the plot that depicts the sensitivity of different scheduling schemes to varying code distance when $p = 10^{-4}$ for the \texttt{gcm\_n13} benchmark.\\
\textit{Note: To reduce the time it takes to run the simulator on all configurations, the \texttt{number\_of\_runs} parameter in the config files has been reduced to 10 (from 50 in \texttt{large} benchmarks, and 1000 in \texttt{medium} and \texttt{supermarq} benchmarks). Since the simulator's execution is seeded, this will lead to increased variations in execution time, however, the trends and relative performance of \acronym{} should remain the same.}

\subsection{Experiment customization}
The config file can be modified to play around with different parameters. The postprocessing scripts can also be modified to visualize different performance metrics. To test \acronym{} on benchmarks not included in the repository, the programs can be compiled to the Clifford+Rz gate set using any open-source quantum software (such as Qiskit\cite{Qiskit}) and then parsed into a format that the total number of gates on the first line, followed by the following on each line:\\
\texttt{<gate name> <list of qubit(s) involved> <rotation angle for Rz gates>}

\subsection{Methodology}

Submission, reviewing and badging methodology:

\begin{itemize}
  \item \url{https://www.acm.org/publications/policies/artifact-review-badging}
  \item \url{http://cTuning.org/ae/submission-20201122.html}
  \item \url{http://cTuning.org/ae/reviewing-20201122.html}
\end{itemize}

\balance{}



\begin{thebibliography}{37}


\ifx \showCODEN    \undefined \def \showCODEN     #1{\unskip}     \fi
\ifx \showDOI      \undefined \def \showDOI       #1{#1}\fi
\ifx \showISBNx    \undefined \def \showISBNx     #1{\unskip}     \fi
\ifx \showISBNxiii \undefined \def \showISBNxiii  #1{\unskip}     \fi
\ifx \showISSN     \undefined \def \showISSN      #1{\unskip}     \fi
\ifx \showLCCN     \undefined \def \showLCCN      #1{\unskip}     \fi
\ifx \shownote     \undefined \def \shownote      #1{#1}          \fi
\ifx \showarticletitle \undefined \def \showarticletitle #1{#1}   \fi
\ifx \showURL      \undefined \def \showURL       {\relax}        \fi
\providecommand\bibfield[2]{#2}
\providecommand\bibinfo[2]{#2}
\providecommand\natexlab[1]{#1}
\providecommand\showeprint[2][]{arXiv:#2}

\bibitem[Akahoshi et~al\mbox{.}(2024)]%
        {akahoshi2023partially}
\bibfield{author}{\bibinfo{person}{Yutaro Akahoshi}, \bibinfo{person}{Kazunori
  Maruyama}, \bibinfo{person}{Hirotaka Oshima}, \bibinfo{person}{Shintaro
  Sato}, {and} \bibinfo{person}{Keisuke Fujii}.}
  \bibinfo{year}{2024}\natexlab{}.
\newblock \showarticletitle{Partially Fault-Tolerant Quantum Computing
  Architecture with Error-Corrected Clifford Gates and Space-Time Efficient
  Analog Rotations}.
\newblock \bibinfo{journal}{\emph{PRX Quantum}}  \bibinfo{volume}{5}
  (\bibinfo{date}{Mar} \bibinfo{year}{2024}), \bibinfo{pages}{010337}.
\newblock
Issue 1.
\urldef\tempurl%
\url{https://doi.org/10.1103/PRXQuantum.5.010337}
\showDOI{\tempurl}


\bibitem[Anderson et~al\mbox{.}(2014)]%
        {anderson2014fault}
\bibfield{author}{\bibinfo{person}{Jonas~T. Anderson},
  \bibinfo{person}{Guillaume Duclos-Cianci}, {and} \bibinfo{person}{David
  Poulin}.} \bibinfo{year}{2014}\natexlab{}.
\newblock \showarticletitle{Fault-Tolerant Conversion between the Steane and
  Reed-Muller Quantum Codes}.
\newblock \bibinfo{journal}{\emph{Phys. Rev. Lett.}}  \bibinfo{volume}{113}
  (\bibinfo{date}{Aug} \bibinfo{year}{2014}), \bibinfo{pages}{080501}.
\newblock
Issue 8.
\urldef\tempurl%
\url{https://doi.org/10.1103/PhysRevLett.113.080501}
\showDOI{\tempurl}


\bibitem[Beverland et~al\mbox{.}(2022)]%
        {comp2}
\bibfield{author}{\bibinfo{person}{Michael Beverland}, \bibinfo{person}{Vadym
  Kliuchnikov}, {and} \bibinfo{person}{Eddie Schoute}.}
  \bibinfo{year}{2022}\natexlab{}.
\newblock \showarticletitle{Surface Code Compilation via Edge-Disjoint Paths}.
\newblock \bibinfo{journal}{\emph{PRX Quantum}}  \bibinfo{volume}{3}
  (\bibinfo{date}{May} \bibinfo{year}{2022}), \bibinfo{pages}{020342}.
\newblock
Issue 2.
\urldef\tempurl%
\url{https://doi.org/10.1103/PRXQuantum.3.020342}
\showDOI{\tempurl}


\bibitem[Campbell and Howard(2017a)]%
        {camp2}
\bibfield{author}{\bibinfo{person}{Earl~T. Campbell} {and}
  \bibinfo{person}{Mark Howard}.} \bibinfo{year}{2017}\natexlab{a}.
\newblock \showarticletitle{Unified framework for magic state distillation and
  multiqubit gate synthesis with reduced resource cost}.
\newblock \bibinfo{journal}{\emph{Phys. Rev. A}}  \bibinfo{volume}{95}
  (\bibinfo{date}{Feb} \bibinfo{year}{2017}), \bibinfo{pages}{022316}.
\newblock
Issue 2.
\urldef\tempurl%
\url{https://doi.org/10.1103/PhysRevA.95.022316}
\showDOI{\tempurl}


\bibitem[Campbell and Howard(2017b)]%
        {camp1}
\bibfield{author}{\bibinfo{person}{Earl~T. Campbell} {and}
  \bibinfo{person}{Mark Howard}.} \bibinfo{year}{2017}\natexlab{b}.
\newblock \showarticletitle{Unifying Gate Synthesis and Magic State
  Distillation}.
\newblock \bibinfo{journal}{\emph{Phys. Rev. Lett.}}  \bibinfo{volume}{118}
  (\bibinfo{date}{Feb} \bibinfo{year}{2017}), \bibinfo{pages}{060501}.
\newblock
Issue 6.
\urldef\tempurl%
\url{https://doi.org/10.1103/PhysRevLett.118.060501}
\showDOI{\tempurl}


\bibitem[Choi et~al\mbox{.}(2023)]%
        {choi2023fault}
\bibfield{author}{\bibinfo{person}{Hyeongrak Choi}, \bibinfo{person}{Frederic~T
  Chong}, \bibinfo{person}{Dirk Englund}, {and} \bibinfo{person}{Yongshan
  Ding}.} \bibinfo{year}{2023}\natexlab{}.
\newblock \showarticletitle{Fault Tolerant Non-Clifford State Preparation for
  Arbitrary Rotations}.
\newblock \bibinfo{journal}{\emph{arXiv preprint arXiv:2303.17380}}
  (\bibinfo{year}{2023}).
\newblock


\bibitem[Cormen et~al\mbox{.}(2009)]%
        {clrs2009algorithms}
\bibfield{author}{\bibinfo{person}{Thomas~H. Cormen},
  \bibinfo{person}{Charles~E. Leiserson}, \bibinfo{person}{Ronald~L. Rivest},
  {and} \bibinfo{person}{Clifford Stein}.} \bibinfo{year}{2009}\natexlab{}.
\newblock \bibinfo{booktitle}{\emph{Introduction to Algorithms, Third Edition}
  (\bibinfo{edition}{3rd} ed.)}.
\newblock \bibinfo{publisher}{The MIT Press}.
\newblock
\showISBNx{0262033844}


\bibitem[Ding et~al\mbox{.}(2018)]%
        {ding2018magic}
\bibfield{author}{\bibinfo{person}{Yongshan Ding}, \bibinfo{person}{Adam
  Holmes}, \bibinfo{person}{Ali Javadi-Abhari}, \bibinfo{person}{Diana
  Franklin}, \bibinfo{person}{Margaret Martonosi}, {and}
  \bibinfo{person}{Frederic Chong}.} \bibinfo{year}{2018}\natexlab{}.
\newblock \showarticletitle{Magic-state functional units: Mapping and
  scheduling multi-level distillation circuits for fault-tolerant quantum
  architectures}. In \bibinfo{booktitle}{\emph{2018 51st Annual IEEE/ACM
  International Symposium on Microarchitecture (MICRO)}}. IEEE,
  \bibinfo{pages}{828--840}.
\newblock


\bibitem[Ding et~al\mbox{.}(2020)]%
        {ding2020square}
\bibfield{author}{\bibinfo{person}{Yongshan Ding}, \bibinfo{person}{Xin-Chuan
  Wu}, \bibinfo{person}{Adam Holmes}, \bibinfo{person}{Ash Wiseth},
  \bibinfo{person}{Diana Franklin}, \bibinfo{person}{Margaret Martonosi}, {and}
  \bibinfo{person}{Frederic~T Chong}.} \bibinfo{year}{2020}\natexlab{}.
\newblock \showarticletitle{Square: Strategic quantum ancilla reuse for modular
  quantum programs via cost-effective uncomputation}. In
  \bibinfo{booktitle}{\emph{2020 ACM/IEEE 47th Annual International Symposium
  on Computer Architecture (ISCA)}}. IEEE, \bibinfo{pages}{570--583}.
\newblock


\bibitem[Eastin and Knill(2009)]%
        {eastin2009restrictions}
\bibfield{author}{\bibinfo{person}{Bryan Eastin} {and} \bibinfo{person}{Emanuel
  Knill}.} \bibinfo{year}{2009}\natexlab{}.
\newblock \showarticletitle{Restrictions on transversal encoded quantum gate
  sets}.
\newblock \bibinfo{journal}{\emph{Physical review letters}}
  \bibinfo{volume}{102}, \bibinfo{number}{11} (\bibinfo{year}{2009}),
  \bibinfo{pages}{110502}.
\newblock


\bibitem[Fowler et~al\mbox{.}(2012)]%
        {Fowler_2012}
\bibfield{author}{\bibinfo{person}{Austin~G. Fowler}, \bibinfo{person}{Matteo
  Mariantoni}, \bibinfo{person}{John~M. Martinis}, {and}
  \bibinfo{person}{Andrew~N. Cleland}.} \bibinfo{year}{2012}\natexlab{}.
\newblock \showarticletitle{Surface codes: Towards practical large-scale
  quantum computation}.
\newblock \bibinfo{journal}{\emph{Physical Review A}} \bibinfo{volume}{86},
  \bibinfo{number}{3} (\bibinfo{date}{sep} \bibinfo{year}{2012}).
\newblock
\urldef\tempurl%
\url{https://doi.org/10.1103/physreva.86.032324}
\showDOI{\tempurl}


\bibitem[Gidney and Eker{\aa}(2021)]%
        {gidney2021factor}
\bibfield{author}{\bibinfo{person}{Craig Gidney} {and} \bibinfo{person}{Martin
  Eker{\aa}}.} \bibinfo{year}{2021}\natexlab{}.
\newblock \showarticletitle{How to factor 2048 bit RSA integers in 8 hours
  using 20 million noisy qubits}.
\newblock \bibinfo{journal}{\emph{Quantum}}  \bibinfo{volume}{5}
  (\bibinfo{year}{2021}), \bibinfo{pages}{433}.
\newblock


\bibitem[Higgott(2022)]%
        {higgott2022pymatching}
\bibfield{author}{\bibinfo{person}{Oscar Higgott}.}
  \bibinfo{year}{2022}\natexlab{}.
\newblock \showarticletitle{PyMatching: A Python package for decoding quantum
  codes with minimum-weight perfect matching}.
\newblock \bibinfo{journal}{\emph{ACM Transactions on Quantum Computing}}
  \bibinfo{volume}{3}, \bibinfo{number}{3} (\bibinfo{year}{2022}),
  \bibinfo{pages}{1--16}.
\newblock


\bibitem[Holmes et~al\mbox{.}(2020)]%
        {holmes2020nisq+}
\bibfield{author}{\bibinfo{person}{Adam Holmes}, \bibinfo{person}{Mohammad~Reza
  Jokar}, \bibinfo{person}{Ghasem Pasandi}, \bibinfo{person}{Yongshan Ding},
  \bibinfo{person}{Massoud Pedram}, {and} \bibinfo{person}{Frederic~T. Chong}.}
  \bibinfo{year}{2020}\natexlab{}.
\newblock \showarticletitle{NISQ+: Boosting quantum computing power by
  approximating quantum error correction}.
\newblock \bibinfo{journal}{\emph{2020 ACM/IEEE 47th Annual International
  Symposium on Computer Architecture (ISCA)}} (\bibinfo{year}{2020}).
\newblock
\urldef\tempurl%
\url{https://doi.org/10.1109/ISCA45697.2020.00053}
\showDOI{\tempurl}


\bibitem[Horsman et~al\mbox{.}(2012)]%
        {horsman2012surface}
\bibfield{author}{\bibinfo{person}{Dominic Horsman}, \bibinfo{person}{Austin~G
  Fowler}, \bibinfo{person}{Simon Devitt}, {and} \bibinfo{person}{Rodney
  Van~Meter}.} \bibinfo{year}{2012}\natexlab{}.
\newblock \showarticletitle{Surface code quantum computing by lattice surgery}.
\newblock \bibinfo{journal}{\emph{New Journal of Physics}}
  \bibinfo{volume}{14}, \bibinfo{number}{12} (\bibinfo{year}{2012}),
  \bibinfo{pages}{123011}.
\newblock


\bibitem[Hua et~al\mbox{.}(2021)]%
        {hua2021autobraid}
\bibfield{author}{\bibinfo{person}{Fei Hua}, \bibinfo{person}{Yanhao Chen},
  \bibinfo{person}{Yuwei Jin}, \bibinfo{person}{Chi Zhang},
  \bibinfo{person}{Ari Hayes}, \bibinfo{person}{Youtao Zhang}, {and}
  \bibinfo{person}{Eddy~Z Zhang}.} \bibinfo{year}{2021}\natexlab{}.
\newblock \showarticletitle{Autobraid: A framework for enabling efficient
  surface code communication in quantum computing}. In
  \bibinfo{booktitle}{\emph{MICRO-54: 54th Annual IEEE/ACM International
  Symposium on Microarchitecture}}. \bibinfo{pages}{925--936}.
\newblock


\bibitem[Jandura and Pupillo(2024)]%
        {jandura2024surface}
\bibfield{author}{\bibinfo{person}{Sven Jandura} {and} \bibinfo{person}{Guido
  Pupillo}.} \bibinfo{year}{2024}\natexlab{}.
\newblock \showarticletitle{Surface Code Stabilizer Measurements for Rydberg
  Atoms}.
\newblock \bibinfo{journal}{\emph{arXiv preprint arXiv:2405.16621}}
  (\bibinfo{year}{2024}).
\newblock


\bibitem[Javadi-Abhari et~al\mbox{.}(2017)]%
        {javadi-abhari2017optimized}
\bibfield{author}{\bibinfo{person}{Ali Javadi-Abhari}, \bibinfo{person}{Pranav
  Gokhale}, \bibinfo{person}{Adam Holmes}, \bibinfo{person}{Diana Franklin},
  \bibinfo{person}{Kenneth~R. Brown}, \bibinfo{person}{Margaret Martonosi},
  {and} \bibinfo{person}{Frederic~T. Chong}.} \bibinfo{year}{2017}\natexlab{}.
\newblock \showarticletitle{Optimized surface code communication in
  superconducting quantum computers}. In \bibinfo{booktitle}{\emph{Proceedings
  of the 50th Annual IEEE/ACM International Symposium on Microarchitecture}}
  (Cambridge, Massachusetts) \emph{(\bibinfo{series}{MICRO-50 '17})}.
  \bibinfo{publisher}{Association for Computing Machinery},
  \bibinfo{address}{New York, NY, USA}, \bibinfo{pages}{692–705}.
\newblock
\showISBNx{9781450349529}
\urldef\tempurl%
\url{https://doi.org/10.1145/3123939.3123949}
\showDOI{\tempurl}


\bibitem[Kubica(2018)]%
        {kubica2018abcs}
\bibfield{author}{\bibinfo{person}{Aleksander~Marek Kubica}.}
  \bibinfo{year}{2018}\natexlab{}.
\newblock \emph{\bibinfo{title}{The ABCs of the color code: A study of
  topological quantum codes as toy models for fault-tolerant quantum
  computation and quantum phases of matter}}.
\newblock \bibinfo{thesistype}{Ph.\,D. Dissertation}.
  \bibinfo{school}{California Institute of Technology}.
\newblock


\bibitem[LeBlond et~al\mbox{.}(2023)]%
        {leblond2023tiscc}
\bibfield{author}{\bibinfo{person}{Tyler LeBlond}, \bibinfo{person}{Ryan~S
  Bennink}, \bibinfo{person}{Justin~G Lietz}, {and}
  \bibinfo{person}{Christopher~M Seck}.} \bibinfo{year}{2023}\natexlab{}.
\newblock \showarticletitle{Tiscc: A surface code compiler and resource
  estimator for trapped-ion processors}. In
  \bibinfo{booktitle}{\emph{Proceedings of the SC'23 Workshops of The
  International Conference on High Performance Computing, Network, Storage, and
  Analysis}}. \bibinfo{pages}{1426--1435}.
\newblock


\bibitem[Li et~al\mbox{.}(2022)]%
        {li2022qasmbench}
\bibfield{author}{\bibinfo{person}{Ang Li}, \bibinfo{person}{Samuel Stein},
  \bibinfo{person}{Sriram Krishnamoorthy}, {and} \bibinfo{person}{James Ang}.}
  \bibinfo{year}{2022}\natexlab{}.
\newblock \bibinfo{title}{QASMBench: A Low-level QASM Benchmark Suite for NISQ
  Evaluation and Simulation}.
\newblock
\newblock
\showeprint[arxiv]{2005.13018}~[quant-ph]


\bibitem[Li et~al\mbox{.}(2019)]%
        {li2019tackling}
\bibfield{author}{\bibinfo{person}{Gushu Li}, \bibinfo{person}{Yufei Ding},
  {and} \bibinfo{person}{Yuan Xie}.} \bibinfo{year}{2019}\natexlab{}.
\newblock \showarticletitle{Tackling the qubit mapping problem for NISQ-era
  quantum devices}. In \bibinfo{booktitle}{\emph{Proceedings of the
  twenty-fourth international conference on architectural support for
  programming languages and operating systems}}. \bibinfo{pages}{1001--1014}.
\newblock


\bibitem[Litinski(2019a)]%
        {litinski2019game}
\bibfield{author}{\bibinfo{person}{Daniel Litinski}.}
  \bibinfo{year}{2019}\natexlab{a}.
\newblock \showarticletitle{A game of surface codes: Large-scale quantum
  computing with lattice surgery}.
\newblock \bibinfo{journal}{\emph{Quantum}}  \bibinfo{volume}{3}
  (\bibinfo{year}{2019}), \bibinfo{pages}{128}.
\newblock


\bibitem[Litinski(2019b)]%
        {litinski2019magic}
\bibfield{author}{\bibinfo{person}{Daniel Litinski}.}
  \bibinfo{year}{2019}\natexlab{b}.
\newblock \showarticletitle{Magic state distillation: Not as costly as you
  think}.
\newblock \bibinfo{journal}{\emph{Quantum}}  \bibinfo{volume}{3}
  (\bibinfo{year}{2019}), \bibinfo{pages}{205}.
\newblock


\bibitem[Molavi et~al\mbox{.}(2023)]%
        {comp3}
\bibfield{author}{\bibinfo{person}{Abtin Molavi}, \bibinfo{person}{Amanda Xu},
  \bibinfo{person}{Swamit Tannu}, {and} \bibinfo{person}{Aws Albarghouthi}.}
  \bibinfo{year}{2023}\natexlab{}.
\newblock \showarticletitle{Compilation for Surface Code Quantum Computers}.
\newblock \bibinfo{journal}{\emph{arXiv preprint arXiv:2311.18042}}
  (\bibinfo{year}{2023}).
\newblock


\bibitem[Murali et~al\mbox{.}(2019)]%
        {murali2019noise}
\bibfield{author}{\bibinfo{person}{Prakash Murali}, \bibinfo{person}{Jonathan~M
  Baker}, \bibinfo{person}{Ali Javadi-Abhari}, \bibinfo{person}{Frederic~T
  Chong}, {and} \bibinfo{person}{Margaret Martonosi}.}
  \bibinfo{year}{2019}\natexlab{}.
\newblock \showarticletitle{Noise-adaptive compiler mappings for noisy
  intermediate-scale quantum computers}. In
  \bibinfo{booktitle}{\emph{Proceedings of the twenty-fourth international
  conference on architectural support for programming languages and operating
  systems}}. \bibinfo{pages}{1015--1029}.
\newblock


\bibitem[{Qiskit contributors}(2023)]%
        {Qiskit}
\bibfield{author}{\bibinfo{person}{{Qiskit contributors}}.}
  \bibinfo{year}{2023}\natexlab{}.
\newblock \bibinfo{title}{Qiskit: An Open-source Framework for Quantum
  Computing}.
\newblock
\newblock
\urldef\tempurl%
\url{https://doi.org/10.5281/zenodo.2573505}
\showDOI{\tempurl}


\bibitem[Ravi et~al\mbox{.}(2023)]%
        {ravi2023better}
\bibfield{author}{\bibinfo{person}{Gokul~Subramanian Ravi},
  \bibinfo{person}{Jonathan~M Baker}, \bibinfo{person}{Arash Fayyazi},
  \bibinfo{person}{Sophia~Fuhui Lin}, \bibinfo{person}{Ali Javadi-Abhari},
  \bibinfo{person}{Massoud Pedram}, {and} \bibinfo{person}{Frederic~T Chong}.}
  \bibinfo{year}{2023}\natexlab{}.
\newblock \showarticletitle{Better than worst-case decoding for quantum error
  correction}. In \bibinfo{booktitle}{\emph{Proceedings of the 28th ACM
  International Conference on Architectural Support for Programming Languages
  and Operating Systems, Volume 2}}. \bibinfo{pages}{88--102}.
\newblock


\bibitem[Ross and Selinger(2014)]%
        {ross2014optimal}
\bibfield{author}{\bibinfo{person}{Neil~J Ross} {and} \bibinfo{person}{Peter
  Selinger}.} \bibinfo{year}{2014}\natexlab{}.
\newblock \showarticletitle{Optimal ancilla-free Clifford+ T approximation of
  z-rotations}.
\newblock \bibinfo{journal}{\emph{arXiv preprint arXiv:1403.2975}}
  (\bibinfo{year}{2014}).
\newblock


\bibitem[Ross and Selinger(2016)]%
        {ross2016gridsynth}
\bibfield{author}{\bibinfo{person}{Neil~J. Ross} {and} \bibinfo{person}{Peter
  Selinger}.} \bibinfo{year}{2016}\natexlab{}.
\newblock \showarticletitle{Optimal ancilla-free Clifford+T approximation of
  z-rotations}.
\newblock \bibinfo{journal}{\emph{Quantum Info. Comput.}} \bibinfo{volume}{16},
  \bibinfo{number}{11–12} (\bibinfo{date}{sep} \bibinfo{year}{2016}),
  \bibinfo{pages}{901–953}.
\newblock
\showISSN{1533-7146}


\bibitem[Sethi and Baker(2025)]%
        {this_work}
\bibfield{author}{\bibinfo{person}{Sayam Sethi} {and} \bibinfo{person}{Jonathan
  Baker}.} \bibinfo{year}{2025}\natexlab{}.
\newblock
  \bibinfo{booktitle}{\emph{5ayam5/Realtime-Scheduling-for-Continuous-Angle-QEC-Architectures}}.
\newblock
\urldef\tempurl%
\url{https://doi.org/10.5281/zenodo.14769159}
\showDOI{\tempurl}


\bibitem[Tannu and Qureshi(2019)]%
        {tannu2019not}
\bibfield{author}{\bibinfo{person}{Swamit~S Tannu} {and}
  \bibinfo{person}{Moinuddin~K Qureshi}.} \bibinfo{year}{2019}\natexlab{}.
\newblock \showarticletitle{Not all qubits are created equal: A case for
  variability-aware policies for NISQ-era quantum computers}. In
  \bibinfo{booktitle}{\emph{Proceedings of the twenty-fourth international
  conference on architectural support for programming languages and operating
  systems}}. \bibinfo{pages}{987--999}.
\newblock


\bibitem[Tomesh et~al\mbox{.}(2022)]%
        {tomesh2022supermarq}
\bibfield{author}{\bibinfo{person}{T. Tomesh}, \bibinfo{person}{P. Gokhale},
  \bibinfo{person}{V. Omole}, \bibinfo{person}{G. Ravi}, \bibinfo{person}{K.~N.
  Smith}, \bibinfo{person}{J. Viszlai}, \bibinfo{person}{X. Wu},
  \bibinfo{person}{N. Hardavellas}, \bibinfo{person}{M.~R. Martonosi}, {and}
  \bibinfo{person}{F.~T. Chong}.} \bibinfo{year}{2022}\natexlab{}.
\newblock \showarticletitle{SupermarQ: A Scalable Quantum Benchmark Suite}. In
  \bibinfo{booktitle}{\emph{2022 IEEE International Symposium on
  High-Performance Computer Architecture (HPCA)}}. \bibinfo{publisher}{IEEE
  Computer Society}, \bibinfo{address}{Los Alamitos, CA, USA},
  \bibinfo{pages}{587--603}.
\newblock
\urldef\tempurl%
\url{https://doi.org/10.1109/HPCA53966.2022.00050}
\showDOI{\tempurl}


\bibitem[Toshio et~al\mbox{.}(2024)]%
        {toshio2024practicalquantumadvantagepartially}
\bibfield{author}{\bibinfo{person}{Riki Toshio}, \bibinfo{person}{Yutaro
  Akahoshi}, \bibinfo{person}{Jun Fujisaki}, \bibinfo{person}{Hirotaka Oshima},
  \bibinfo{person}{Shintaro Sato}, {and} \bibinfo{person}{Keisuke Fujii}.}
  \bibinfo{year}{2024}\natexlab{}.
\newblock \showarticletitle{Practical quantum advantage on partially
  fault-tolerant quantum computer}.
\newblock \bibinfo{journal}{\emph{arXiv preprint arXiv:2408.14848}}
  (\bibinfo{year}{2024}).
\newblock


\bibitem[Viszlai et~al\mbox{.}(2023)]%
        {viszlai2023architecture}
\bibfield{author}{\bibinfo{person}{Joshua Viszlai},
  \bibinfo{person}{Sophia~Fuhui Lin}, \bibinfo{person}{Siddharth Dangwal},
  \bibinfo{person}{Jonathan~M Baker}, {and} \bibinfo{person}{Frederic~T
  Chong}.} \bibinfo{year}{2023}\natexlab{}.
\newblock \showarticletitle{An architecture for improved surface code
  connectivity in neutral atoms}.
\newblock \bibinfo{journal}{\emph{arXiv preprint arXiv:2309.13507}}
  (\bibinfo{year}{2023}).
\newblock


\bibitem[Watkins et~al\mbox{.}(2024)]%
        {comp1}
\bibfield{author}{\bibinfo{person}{George Watkins}, \bibinfo{person}{Hoang~Minh
  Nguyen}, \bibinfo{person}{Keelan Watkins}, \bibinfo{person}{Steven Pearce},
  \bibinfo{person}{Hoi-Kwan Lau}, {and} \bibinfo{person}{Alexandru Paler}.}
  \bibinfo{year}{2024}\natexlab{}.
\newblock \showarticletitle{A high performance compiler for very large scale
  surface code computations}.
\newblock \bibinfo{journal}{\emph{Quantum}}  \bibinfo{volume}{8}
  (\bibinfo{year}{2024}), \bibinfo{pages}{1354}.
\newblock


\bibitem[Wu et~al\mbox{.}(2022)]%
        {wu2022synthesis}
\bibfield{author}{\bibinfo{person}{Anbang Wu}, \bibinfo{person}{Gushu Li},
  \bibinfo{person}{Hezi Zhang}, \bibinfo{person}{Gian~Giacomo Guerreschi},
  \bibinfo{person}{Yufei Ding}, {and} \bibinfo{person}{Yuan Xie}.}
  \bibinfo{year}{2022}\natexlab{}.
\newblock \showarticletitle{A synthesis framework for stitching surface code
  with superconducting quantum devices}. In
  \bibinfo{booktitle}{\emph{Proceedings of the 49th Annual International
  Symposium on Computer Architecture}}. \bibinfo{pages}{337--350}.
\newblock


\end{thebibliography}
\end{document}